\documentclass[journal=ancac3,manuscript=article]{achemso}

\usepackage[T1]{fontenc} 
\usepackage{amsfonts}
\usepackage{amssymb}
\usepackage{amsmath}
\usepackage{amsthm}
\usepackage{color}

\author{Gerg\H{o} Kukucska}
\affiliation{Department of Biological Physics, E{\" o}tv{\" o}s Lor{\' a}nd University, P{\'a}zm{\'a}ny P{\'e}ter s{\'e}t{\'a}ny 1/A, 1117 Budapest, Hungary}
\author{Viktor Z{\'o}lyomi}
\affiliation{School of Physics and Astronomy, University of Manchester, Oxford Road, Manchester M13 9PL, UK}
\author{J{\'a}nos Koltai}
\affiliation{Department of Biological Physics, E{\" o}tv{\" o}s Lor{\' a}nd University, P{\'a}zm{\'a}ny P{\'e}ter s{\'e}t{\'a}ny 1/A, 1117 Budapest, Hungary}
\email{koltai@elte.hu}

\title[Resonance Raman spectroscopy of silicene and germanene]{Resonance Raman spectroscopy of silicene and germanene}

\begin{document}

\begin{tocentry}
\begin{center}

\includegraphics[scale=.098]{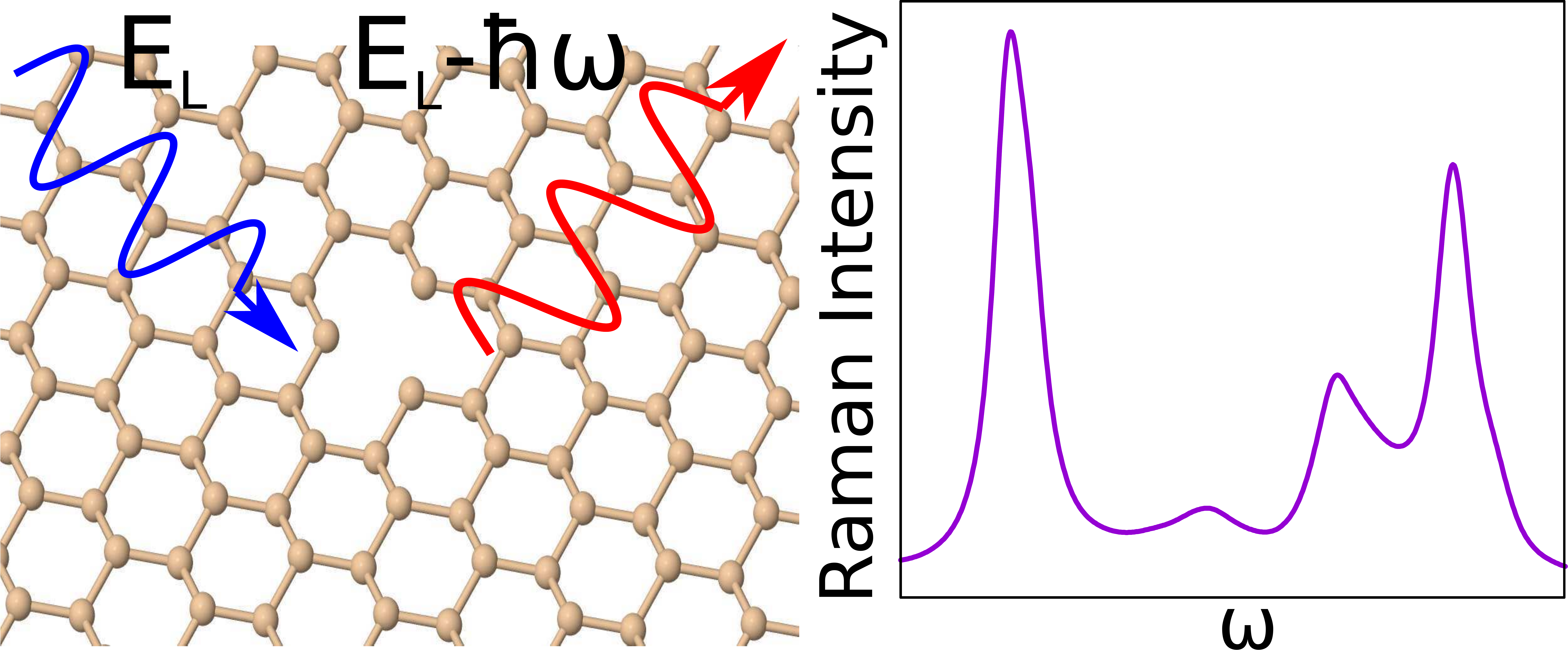}
\end{center}

\end{tocentry}

\begin{abstract}
We model Raman processes in silicene and germanene involving scattering of quasiparticles by, either, two phonons, or, one phonon and one point defect. We compute the resonance Raman intensities and lifetimes for laser excitations between 1 and 3$\,$eV using a newly developed third-nearest neighbour tight-binding model parametrized from first principles density functional theory. We identify features in the Raman spectra that are unique to the studied materials or the defects therein. We find that in silicene, a new Raman resonance arises from the $2.77\,\rm$eV $\pi-\sigma$ plasmon at the M point, measurably higher than the Raman resonance originating from the $2.12\,\rm$eV $\pi$ plasmon energy. We show that in germanene, the lifetimes of charge carriers, and thereby the linewidths of the Raman peaks, are influenced by spin-orbit splittings within the electronic structure. We use our model to predict scattering cross sections for defect induced Raman scattering involving adatoms, substitutional impurities, Stone-Wales pairs, and vacancies, and argue that the presence of each of these defects in silicene and germanene can be qualitatively matched to specific features in the Raman response.
\end{abstract}

\section{Introduction}

Graphene-like hexagonal materials composed of silicon or germanium are unique two-dimensional (2D) crystals with a promising future in nanoelectronics. These structures were predicted to be stable by density functional theory \cite{takeda_theoretical_1994,houssa_electronic_2010}, and fabricated on metallic substrates amid intensive experimental pursuit over the past decade \cite{aufray_graphene-like_2010,de_padova_evidence_2010,lalmi_epitaxial_2010,vogt_silicene:_2012,derivaz_continuous_2015,acun_germanene:_2015,curcella_multilayer_2017,gill_metallic_2017,borlido_ground_2018}. While synthesis of free-standing monolayers is still not accomplished, recent studies have shown that both silicene \cite{lian_dirac_2017} and germanene \cite{schroter_emergence_2017} monolayers synthesized on metallic surfaces exhibit Dirac-like bands. Moreover, these hybrid structures were already used to fabricate transistors from both silicene \cite{tao_silicene_2015} and germanene \cite{madhushankar_electronic_2017}.

\begin{figure}
\begin{center}
\includegraphics[scale=.095]{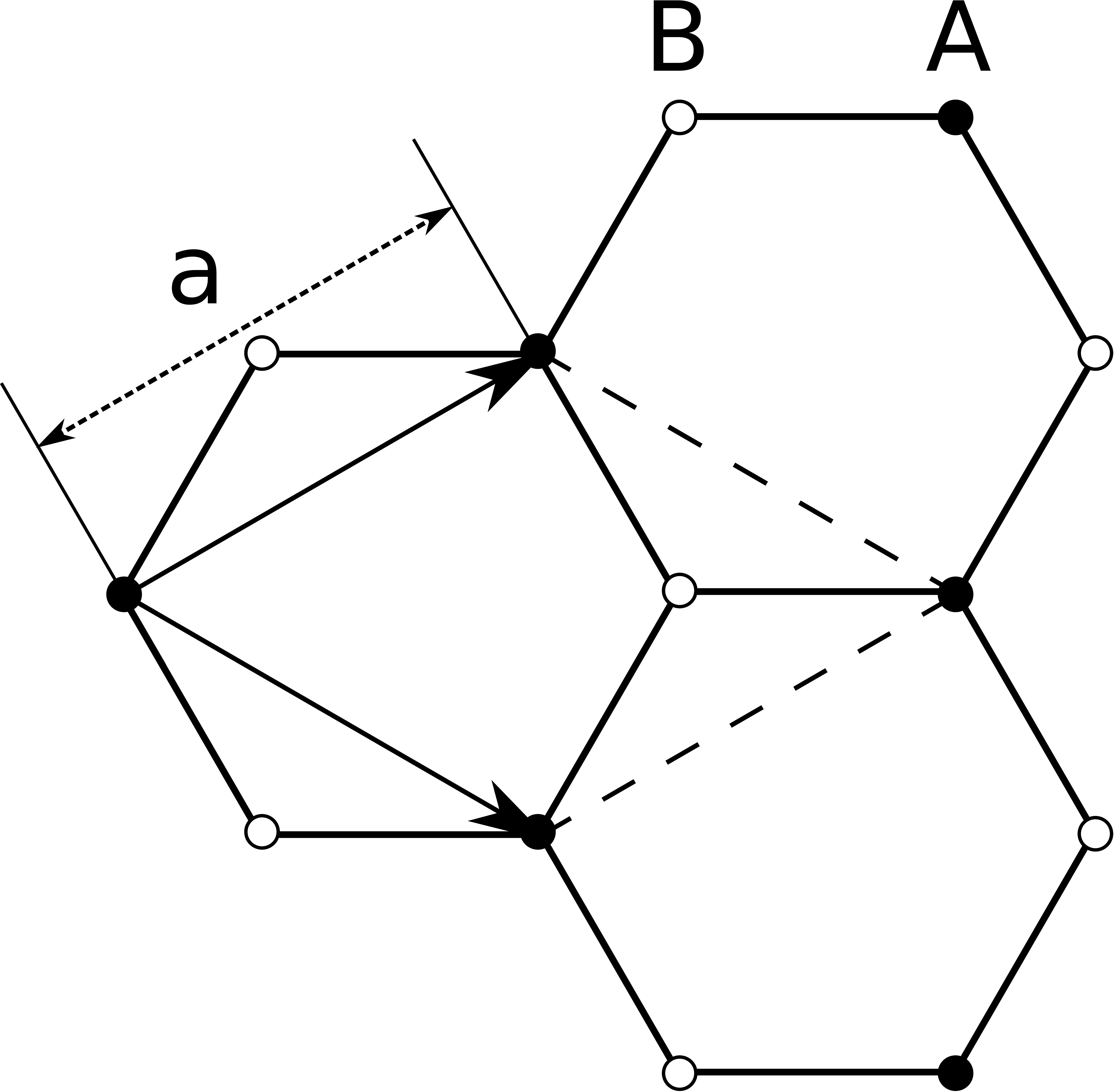}\\
\includegraphics[scale=.22]{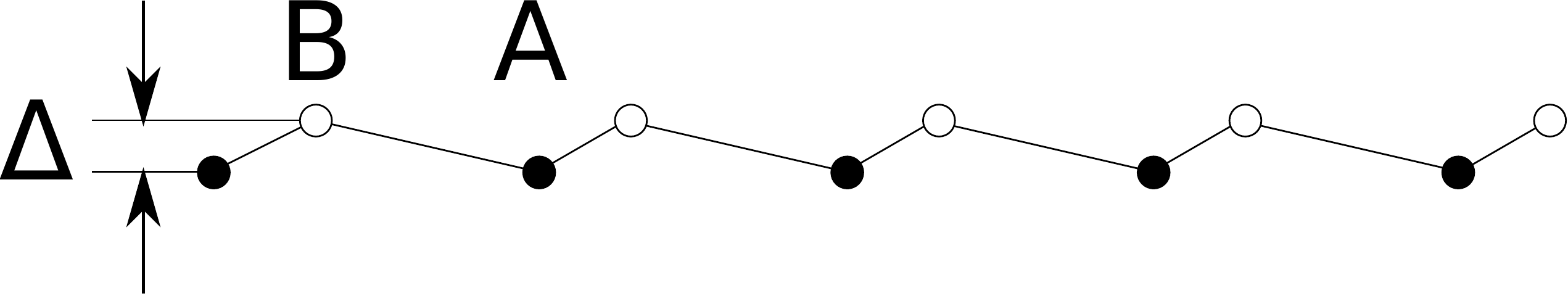}
\end{center}
\caption{\label{Buckling}Structure of silicene and germanene
}
\end{figure}

What sets silicene and germanene apart from their carbon counterpart \cite{novoselov_electric_2004} is that these structures exhibit a sublattice buckling, that is, the A and B sublattices of the honeycomb structure are vertically shifted relative to one another as shown in Fig. \ref{Buckling}. The buckled structure introduces new physics such as the opening of a spin-orbit induced band gap of 1--2$\,$meV and 24$\,$meV in silicene and germanene, respectively \cite{liu_low-energy_2011,drummond_electrically_2012,kaloni_quasi_2013,ezawa_monolayer_2015,zhuang_cooperative_2017}. Moreover, a topological phase transition can be induced in such materials by applying a perpendicular electric field which introduces a tunable band gap up to $\sim 100\,\rm meV$ \cite{drummond_electrically_2012,yan_tuning_2015}.

Raman spectroscopy offers a powerful way to analyze electronic and vibrational properties of these materials. This non-destructive technique relies on inelastic scattering of light to probe the phononic excitations in the lattice. It is also well suited to probe quantitative and qualitative properties of perturbations such as strain \cite{mohiuddin_uniaxial_2009,kukucska_theoretical_2017}, doping \cite{lee_optical_2012,kukucska_theoretical_2017} or lattice defects \cite{lucchese_quantifying_2010}. Defect scattering can activate otherwise forbidden peaks in the spectra, which can be used to identify the type of defect or edge orientation \cite{casiraghi_raman_2009,dresselhaus_defect_2010,venezuela_theory_2011}. Intensity of these peaks can be calculated within fourth-order time-dependent perturbation theory. These fourth-order processes also include scattering of defect-free systems by emission of two phonons.

In this paper, we present a tight-binding (TB) model parametrized directly from first principles density functional theory (DFT) to calculate i) two-phonon and ii) defect induced single-phonon Raman processes in freestanding monolayer silicene and germanene. We evaluate defect scattering matrix elements within the tight-binding formalism for various defects, and use these matrix elements to calculate the Raman spectra for defect induced processes. We argue, that the position and relative intensity ratio of the dominant peaks can be used to distinguish between different types of defects, implying that Raman characterization can be used to indicate the concentration of different point defects in silicene and germanene.

We demonstrate that in germanene spin-orbit coupling must be taken into account for an accurate description of excitations, whereas in silicene it can be safely neglected. We also compare our model for two-phonon processes to a previously developed non-orthogonal TB model \cite{popov_comparative_2004} of graphene and silicene, and point out a dependence of the predictions on the sublattice buckling parameter. In particular, by calculating the two phonon Raman spectra of germanene and comparing these spectra to the two phonon spectra of graphene we demonstrate a connection between the sublattice buckling and the amplitude of Raman peaks originating from out-of-plane vibrations.

\section{Results and discussion}
\begin{figure}
\begin{center}
\includegraphics[scale=.095]{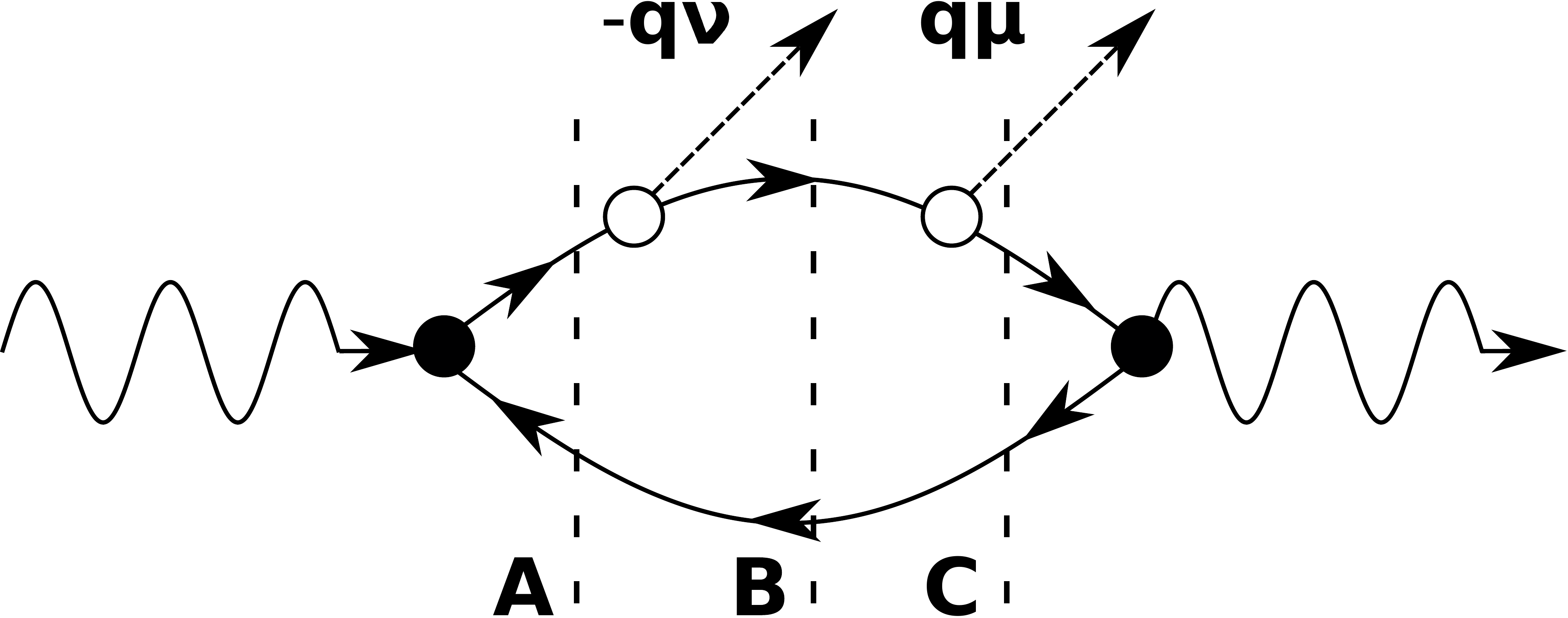}
\includegraphics[scale=.095]{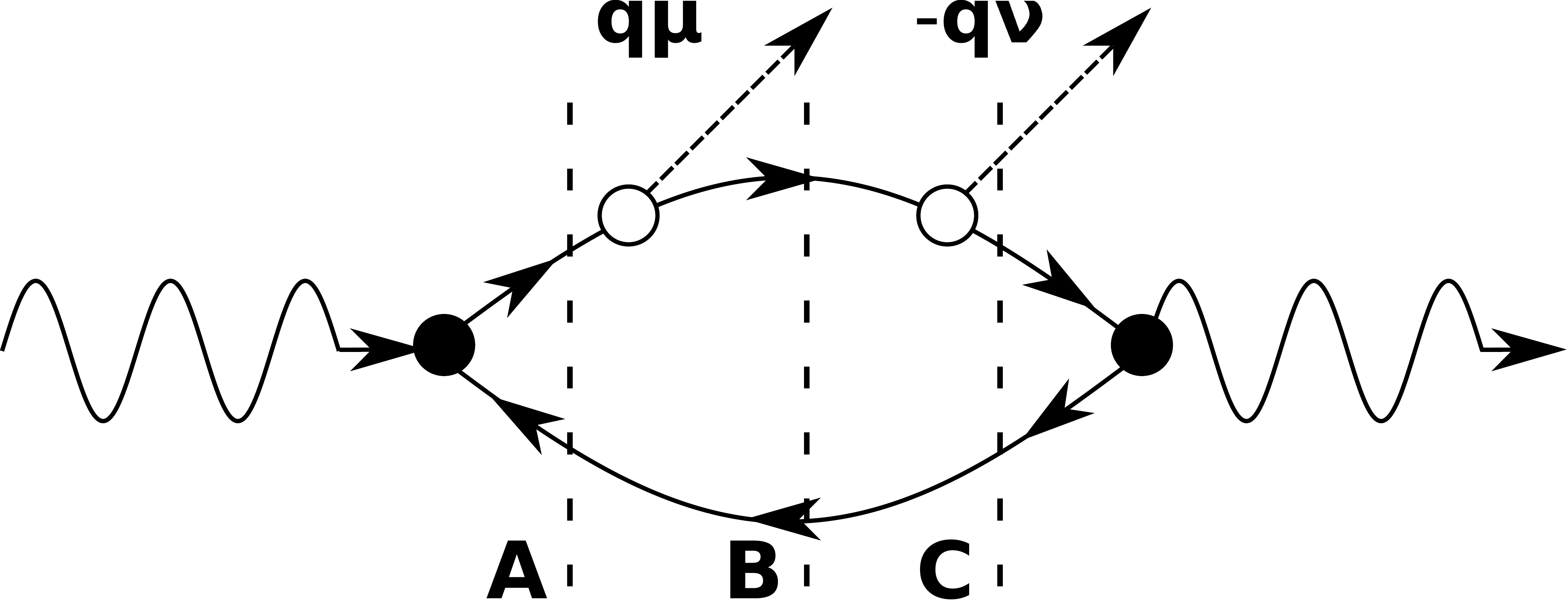}
\includegraphics[scale=.095]{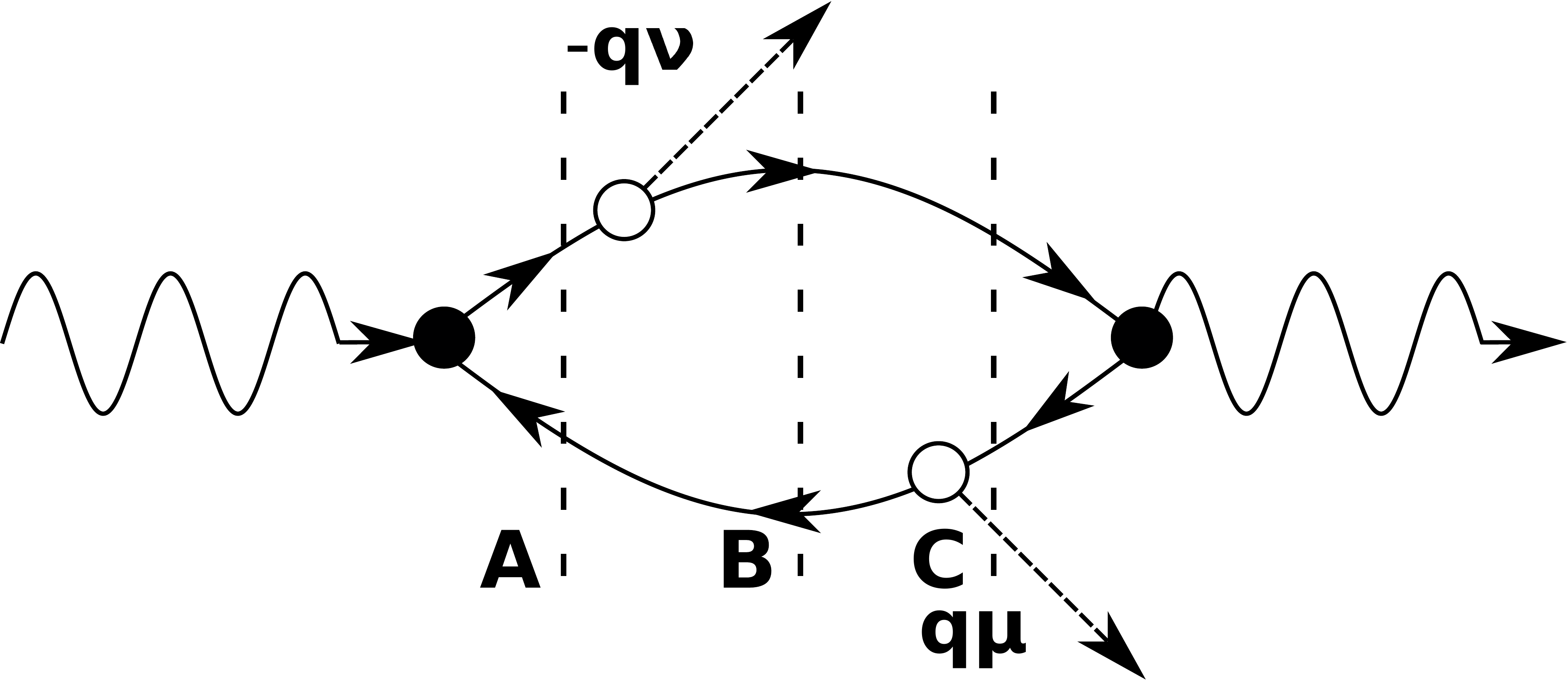}
\includegraphics[scale=.095]{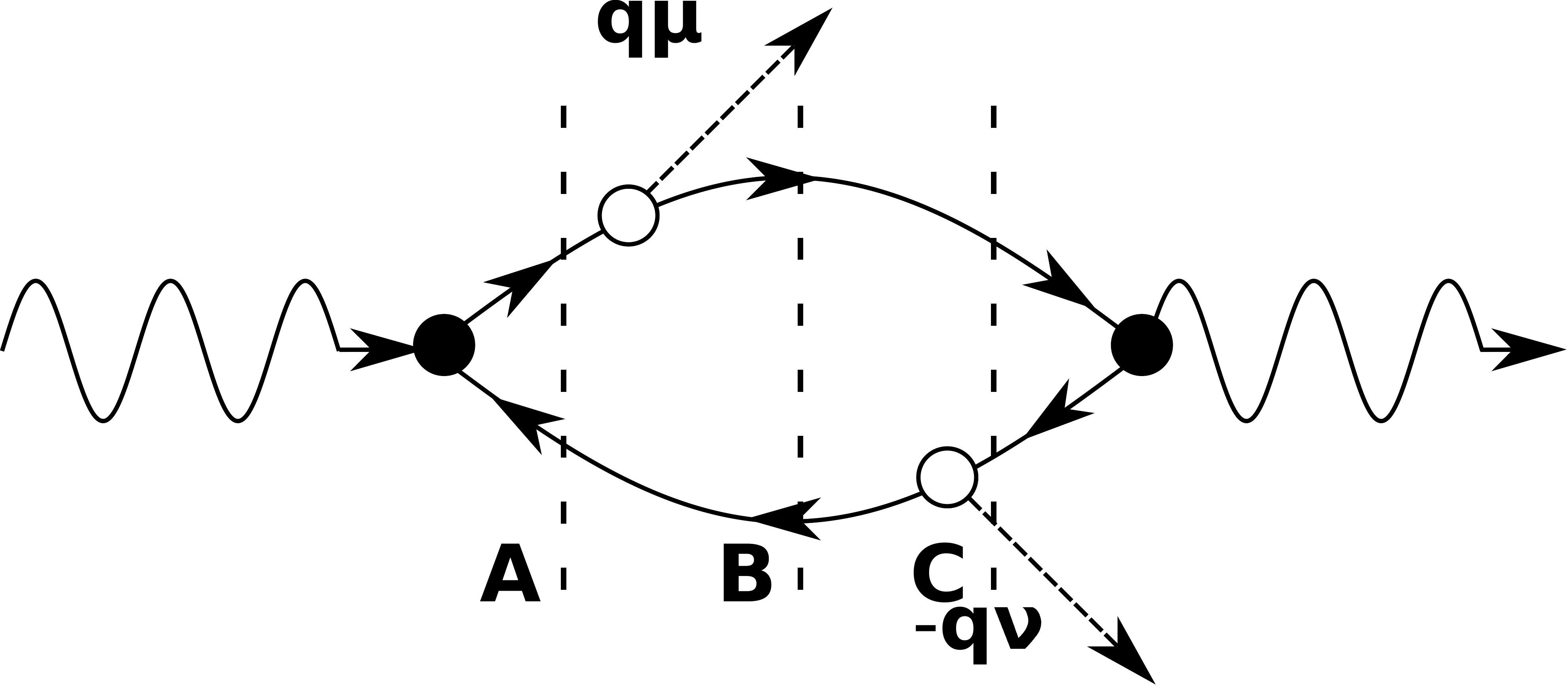}\\
\includegraphics[scale=.095]{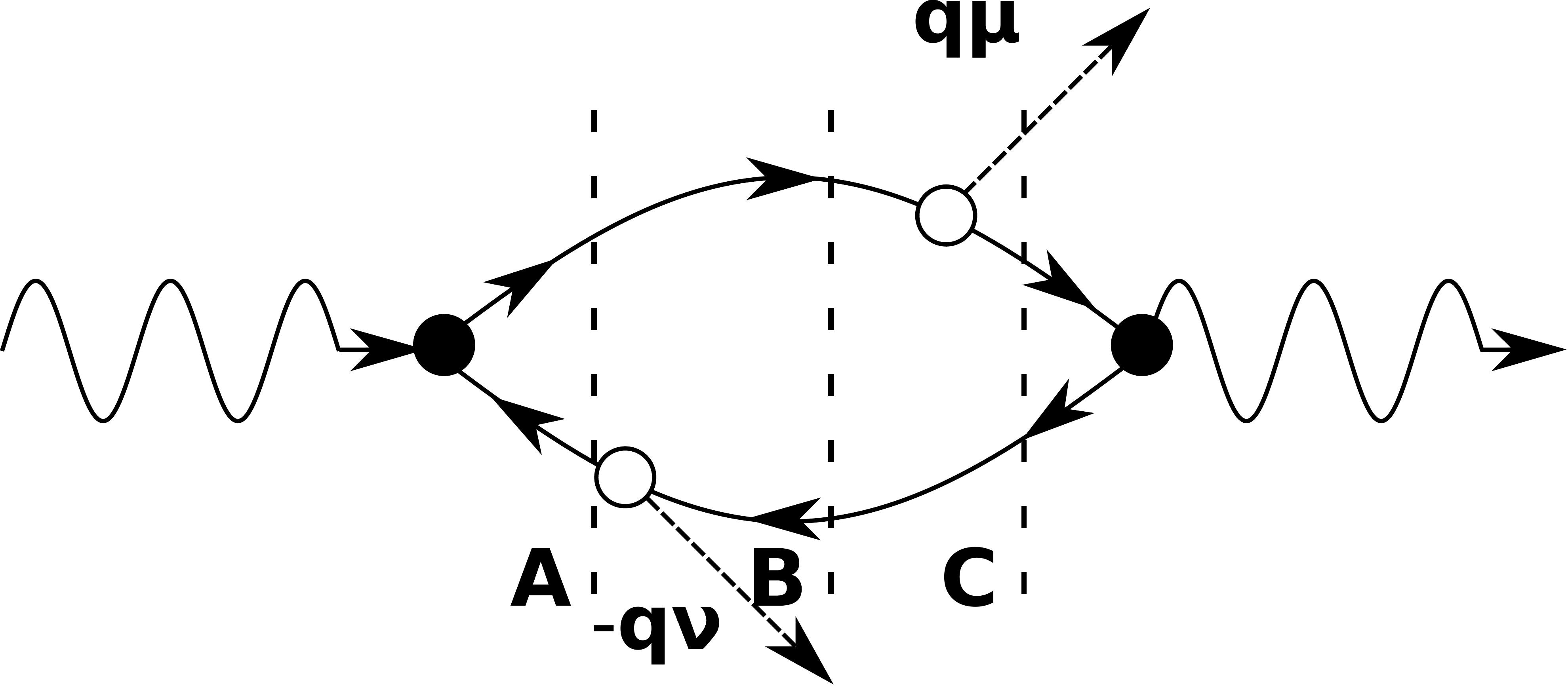}
\includegraphics[scale=.095]{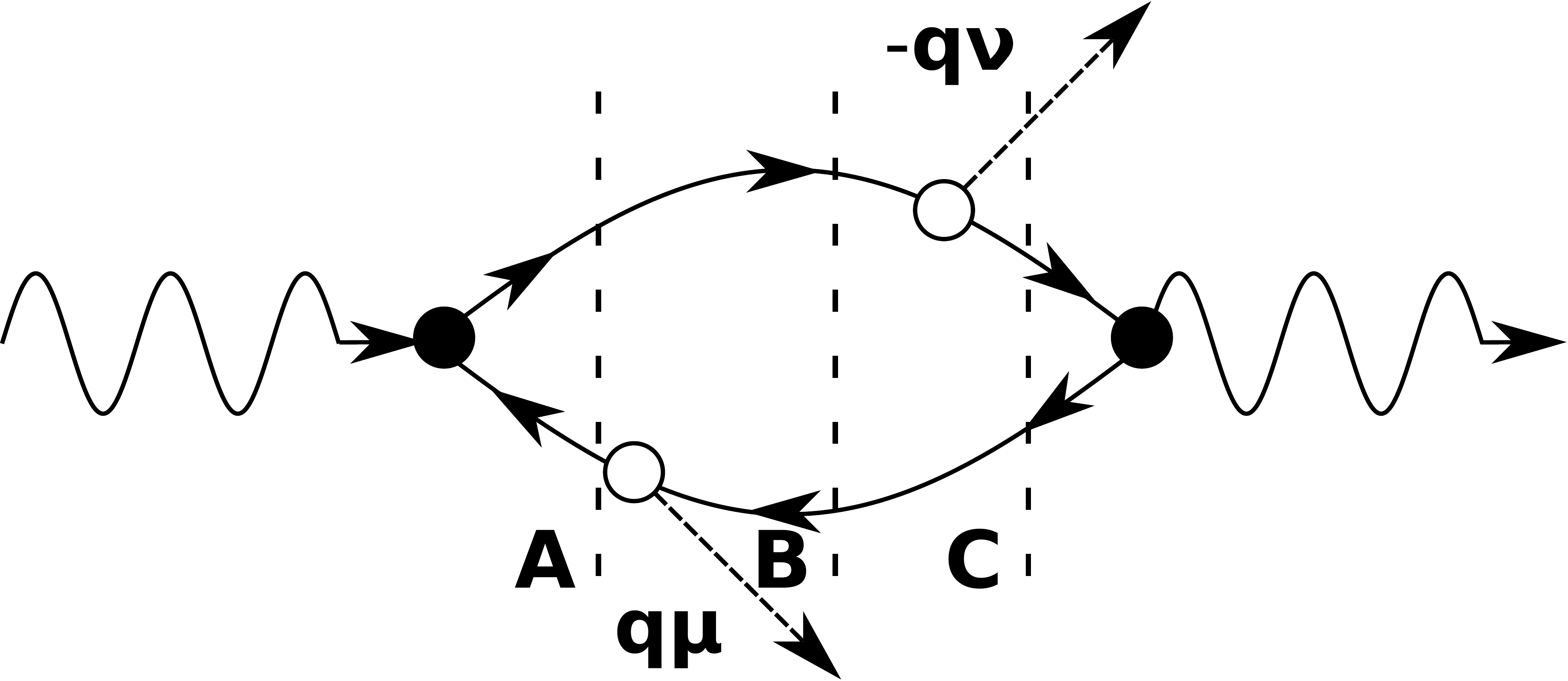}
\includegraphics[scale=.095]{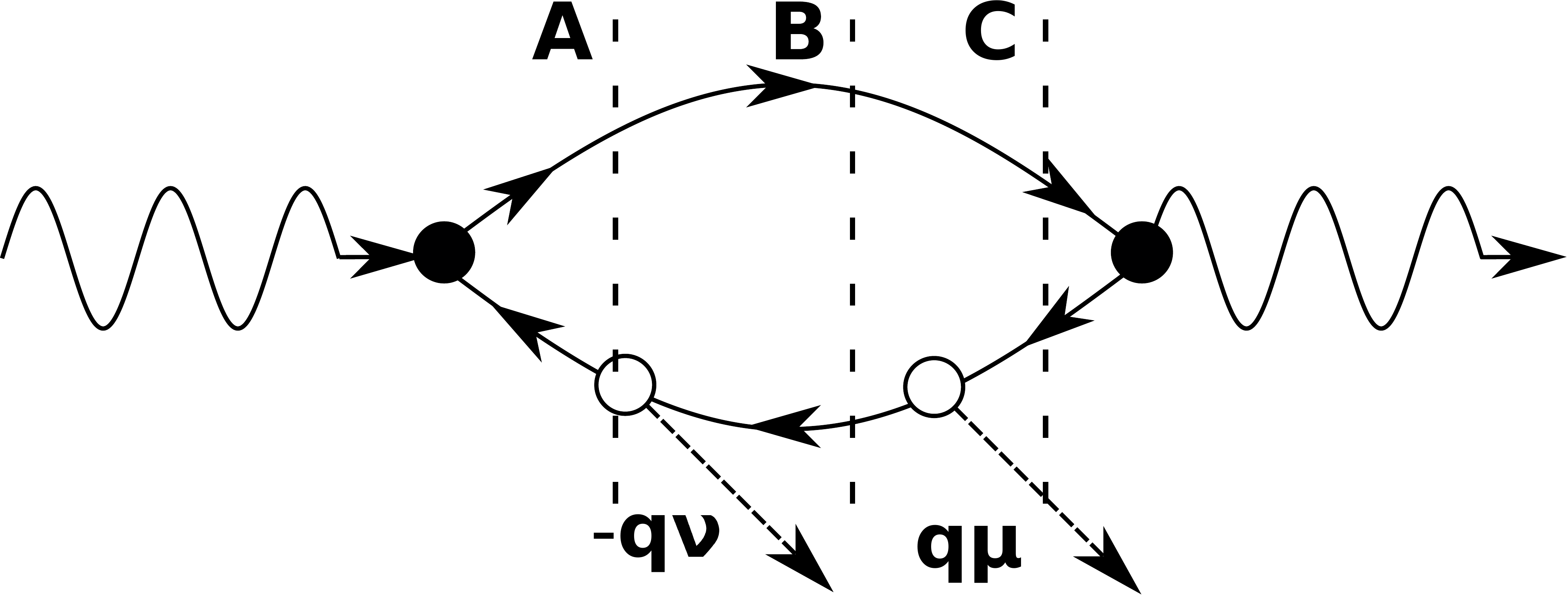}
\includegraphics[scale=.095]{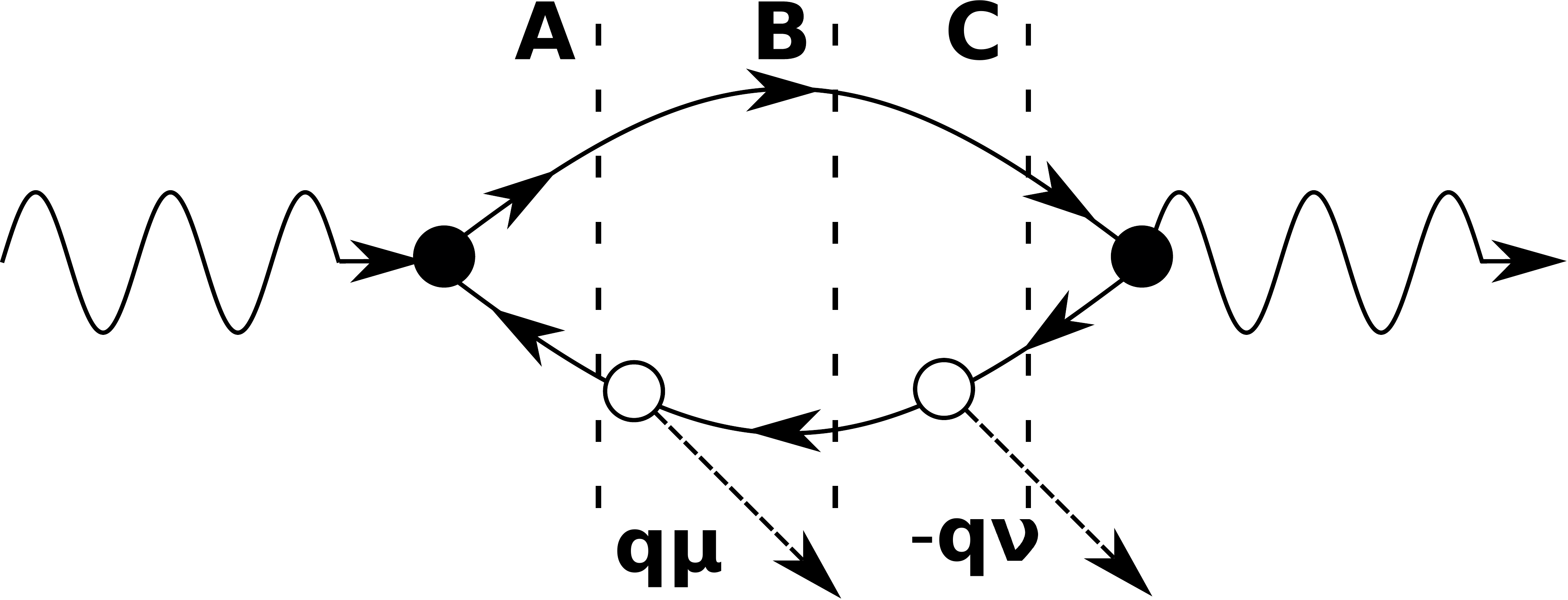}
\end{center}
\caption{\label{Feynman_2d}Feynman diagrams of two-phonon Raman processes
}
\end{figure}
\begin{figure}
\begin{center}
\includegraphics[scale=.095]{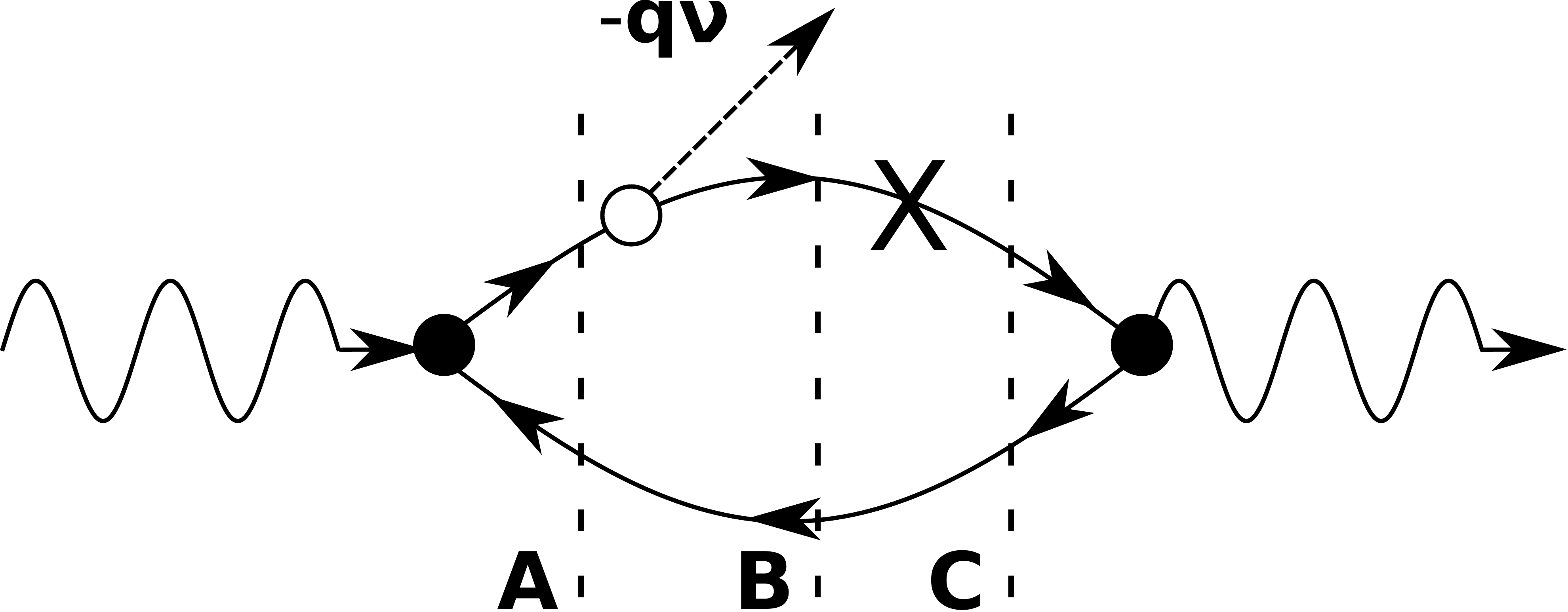}
\includegraphics[scale=.095]{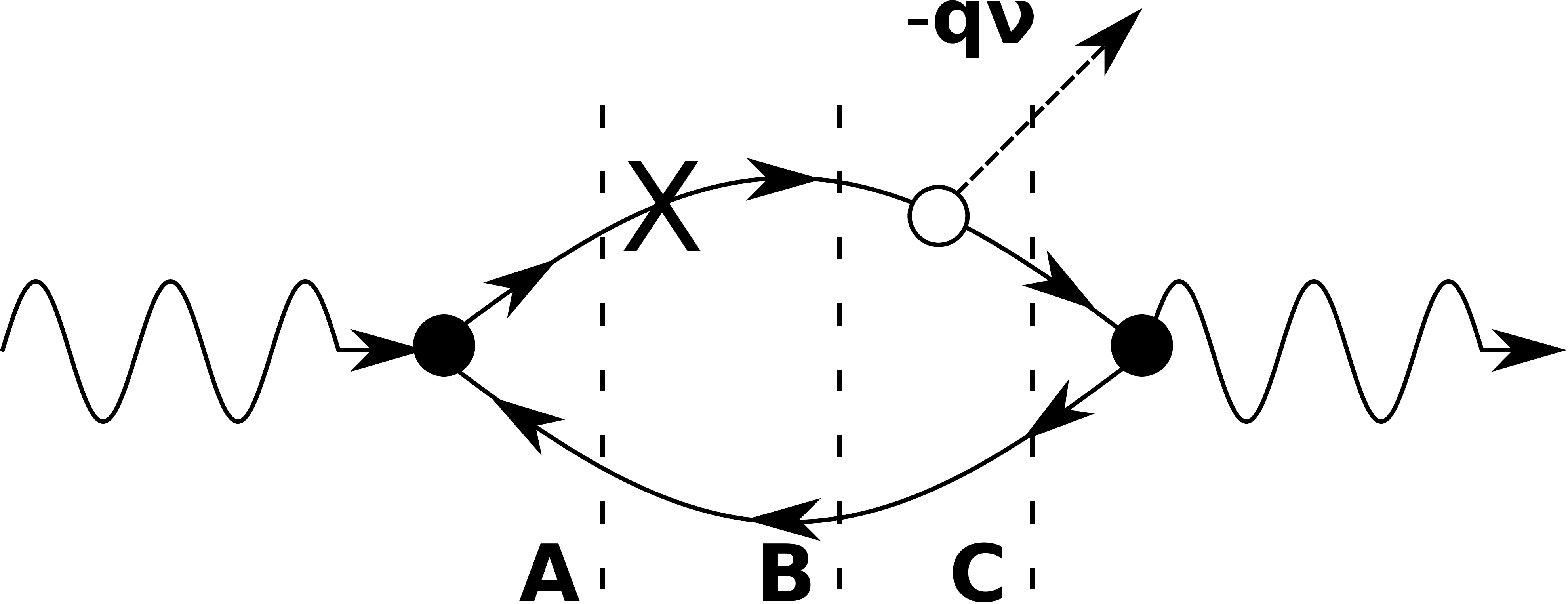}
\includegraphics[scale=.095]{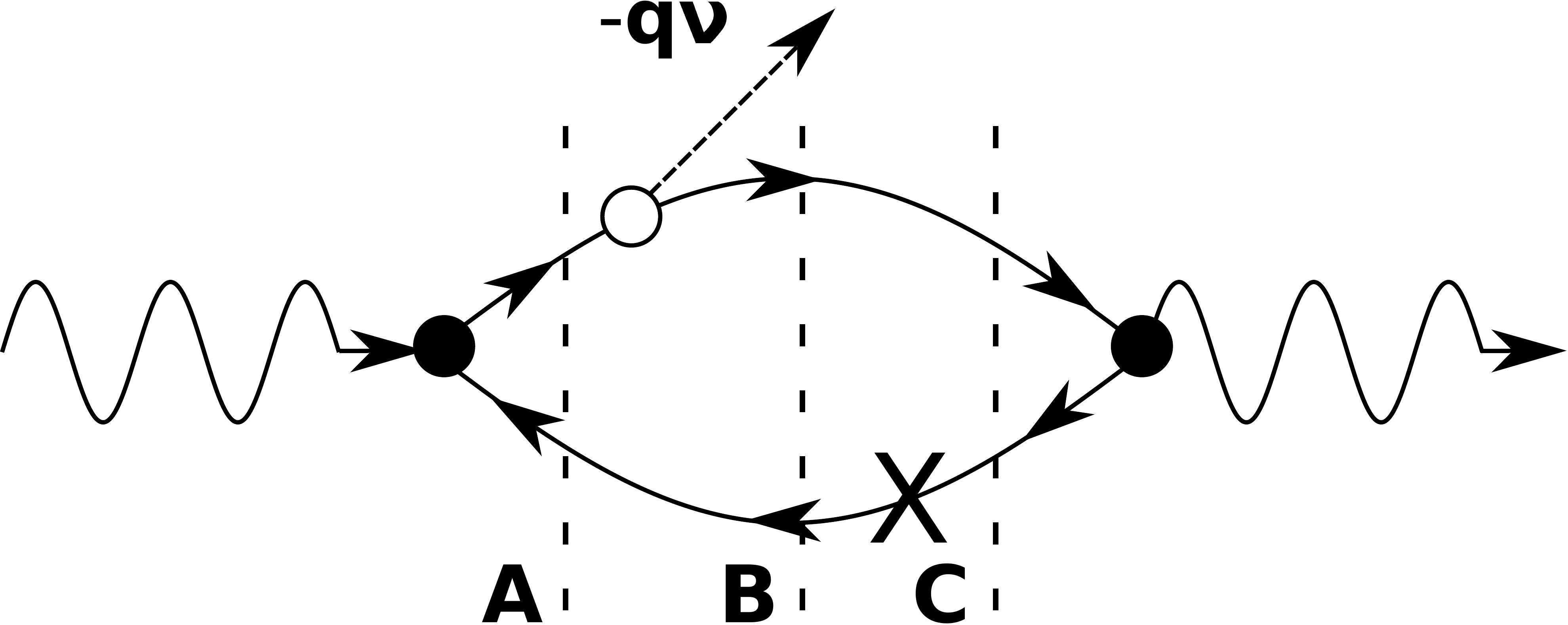}
\includegraphics[scale=.095]{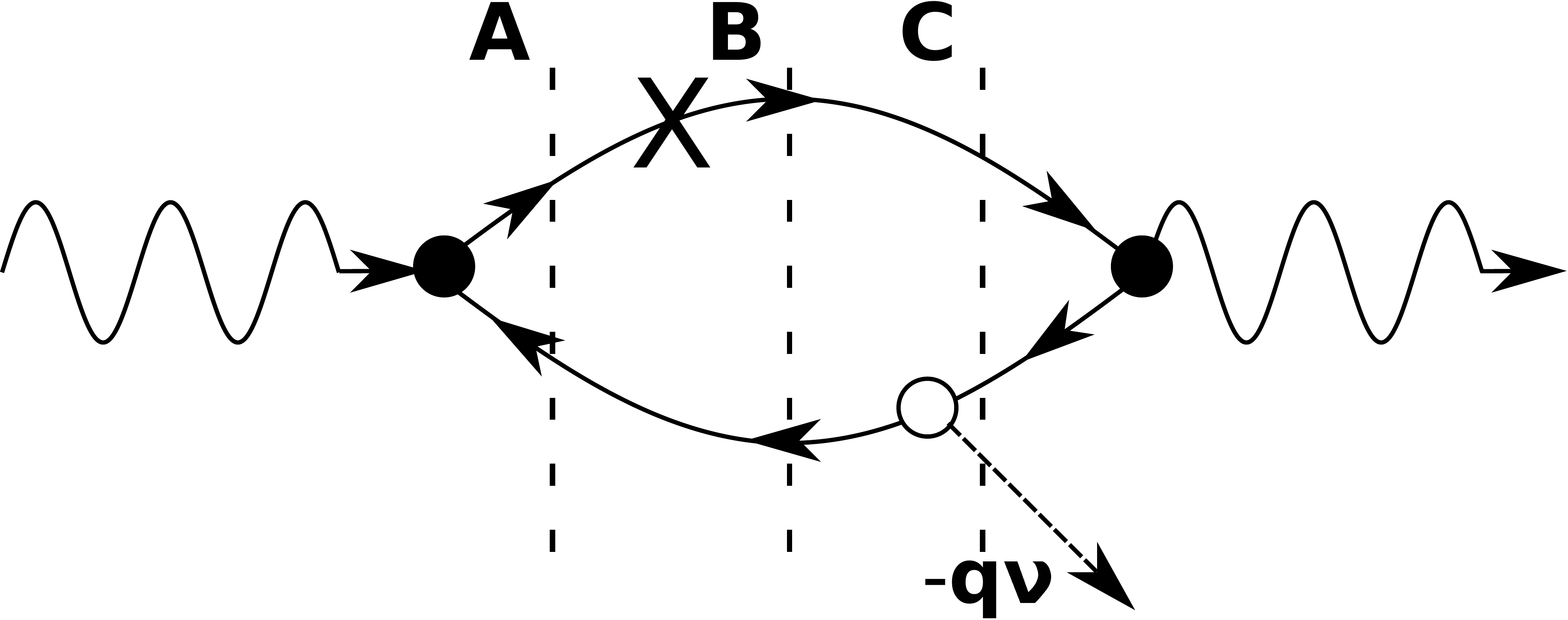}\\
\includegraphics[scale=.095]{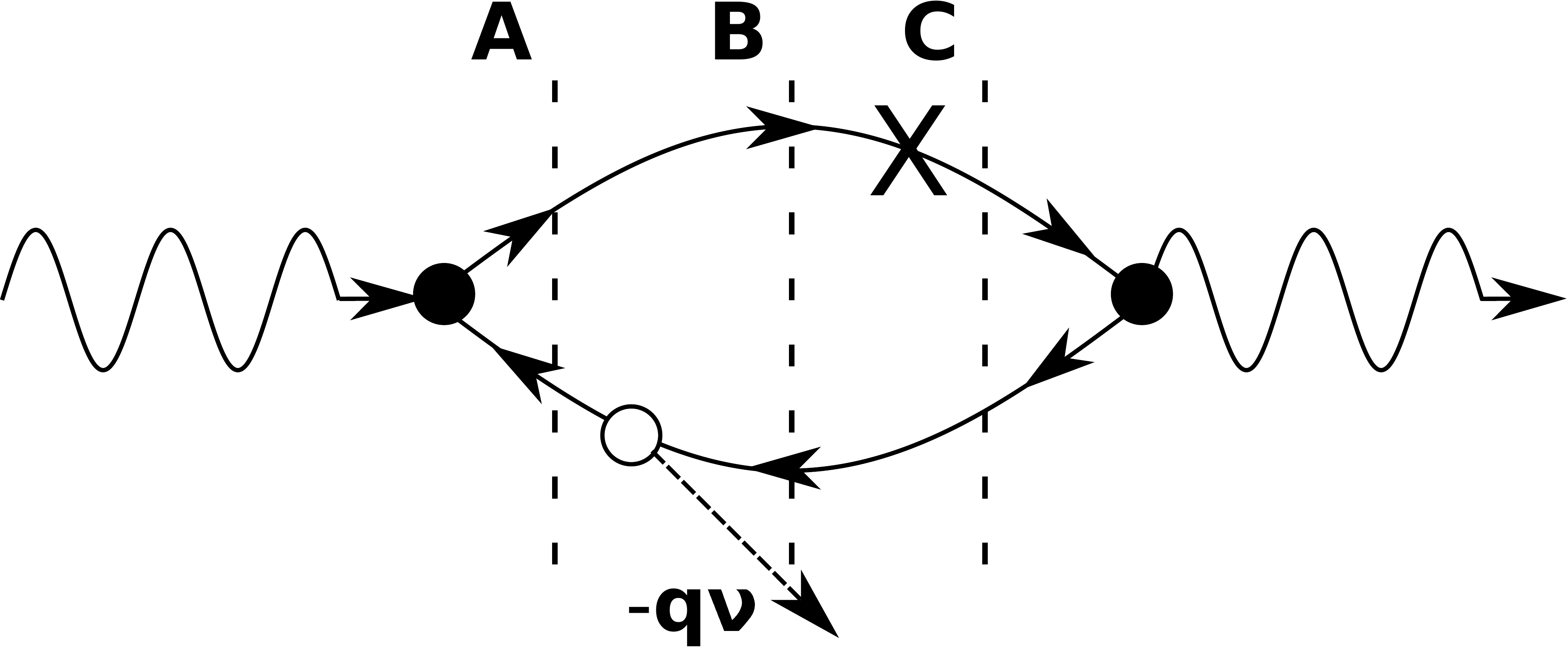}
\includegraphics[scale=.095]{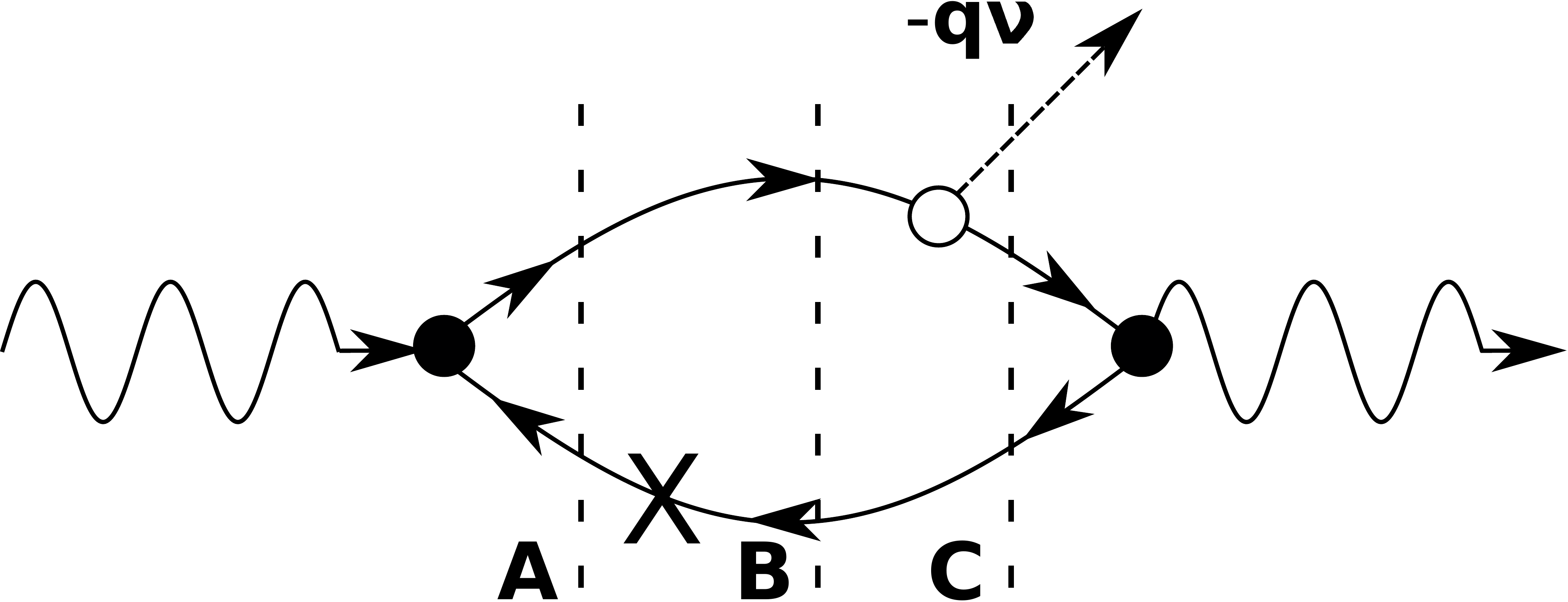}
\includegraphics[scale=.095]{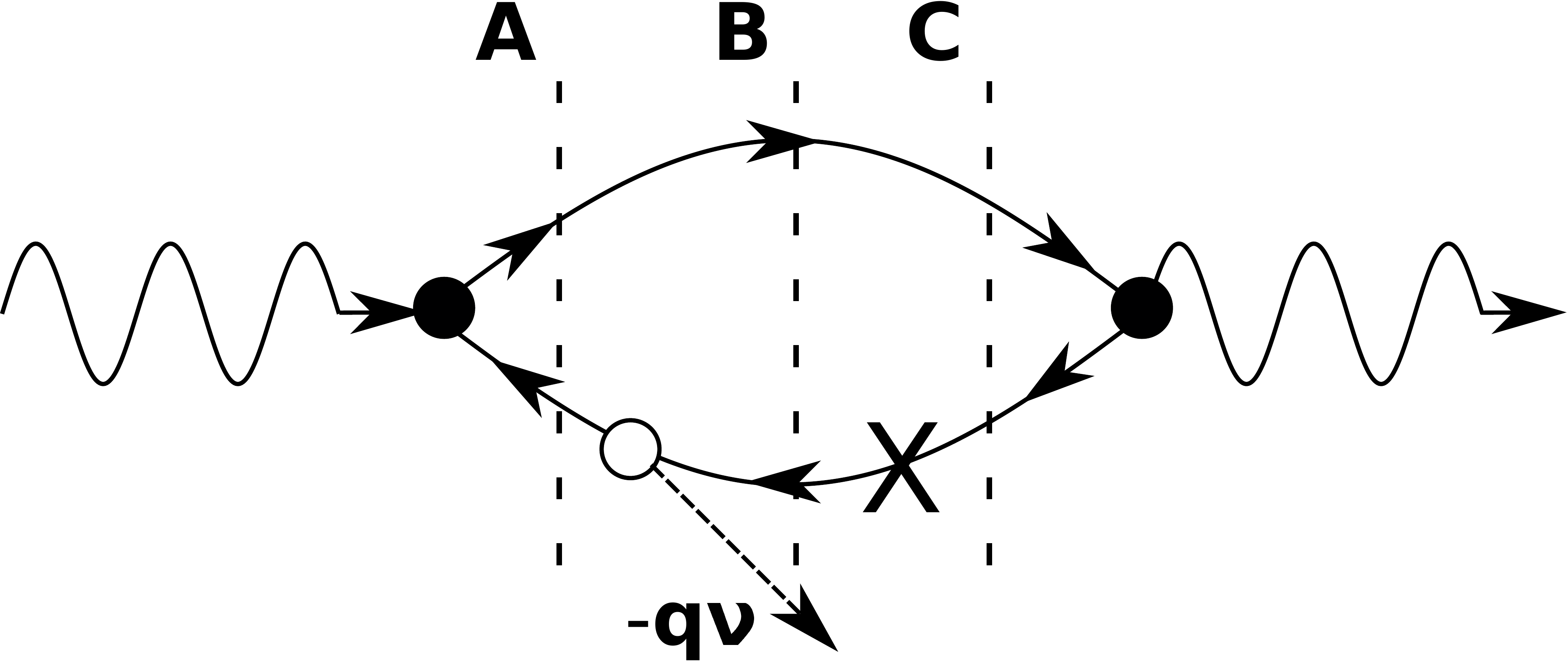}
\includegraphics[scale=.095]{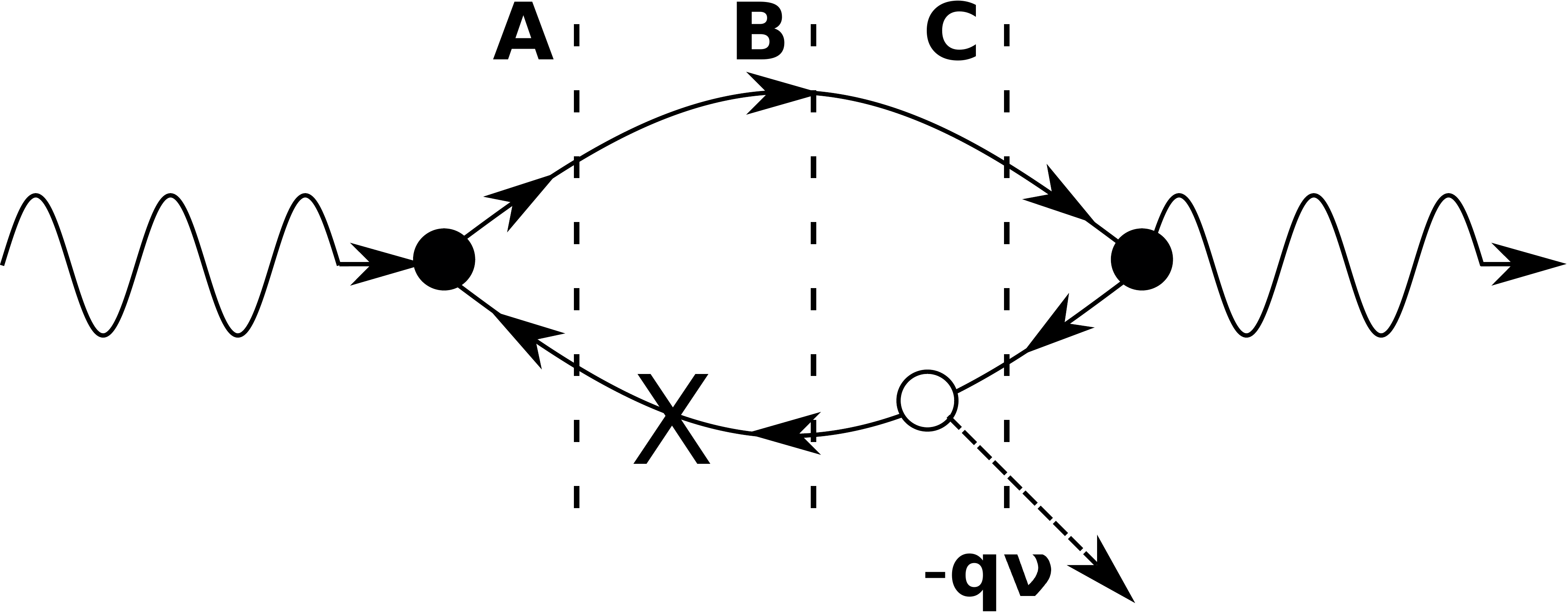}
\end{center}
\caption{\label{Feynman_def}Feynman diagrams of defect induced Raman processes
}
\end{figure}
\subsection{Raman intensities}
We compute the Raman cross sections within the established fourth order time-dependent perturbation theory \cite{martin_resonant_1975,zolyomi_resonance_2011,venezuela_theory_2011,popov_comparative_2016}. The scattering amplitudes ($K$) of the eight relevant Feynman diagrams \cite{venezuela_theory_2011} as presented in Fig. \ref{Feynman_2d} for two-phonon (pp) and in Fig. \ref{Feynman_def} defect induced (pd) diagrams can be written in the general form
\begin{equation}
K_{pp}^{\mu,\nu}=\sum_{A,B,C}\frac{M^{e-p}_{fC}M^{e-ph,\mu}_{CB}M^{e-ph,\nu}_{BA}M^{e-p}_{Ai}}{(E_i-E_C)(E_i-E_B)(E_i-E_A)},
\label{eq_pp}
\end{equation}
\begin{equation}
K_{pd}^{\mu}=\sum_{A,B,C}\frac{M^{e-p}_{fC}M^{e-ph,\mu}_{CB}M^{d}_{BA}M^{e-p}_{Ai}}{(E_i-E_C)(E_i-E_B)(E_i-E_A)},
\label{eq_pd}
\end{equation}

\noindent where $\mu,\nu$ are phonon branch indexes, $i,f$ denote the initial and final state of the system, respectively, $A,B,C$ are virtual intermediate states, $E_i,E_A,E_B,E_C,E_f$ denote the sum of the energies of all quasiparticles present in these states, and $M^{e-p}_{fC},M^{e-ph,\mu}_{CB},M^{d}_{BA}$ are the electron-photon, electron-phonon, and defect scattering matrix elements, respectively. The Raman cross section ($I$) can be calculated directly from these amplitudes by summing over all possible final states, and can be expressed as

\begin{equation}
I_{pp}(\omega)=\sum_{f,\mu\nu}\left|K^{\mu,\nu}_{pp}\right|^2\delta(\omega_\mu+\omega_\nu-\omega)(n(\omega_\mu)+1)(n(\omega_\nu)+1),
\end{equation}
\begin{equation}
I_{pd}(\omega)=\sum_{f,\mu}\left|K^{\mu}_{pd}\right|^2\delta(\omega_\mu-\omega)(n(\omega_\mu)+1),
\end{equation}

\noindent where $\delta(x)$ is the Dirac delta function ensuring the conservation of energy between the initial and final state, $\omega$ is the Raman shift and $n(\omega_\mu)$ is the Bose-Einstein distribution due to the induced emission of a phonon with frequency $\omega_\mu$. These matrix elements determine the amplitude of possible allowed and forbidden transitions. Their accurate description is essential in order to obtain correct peak intensities. In our model the matrix elements are calculated within the tight-binding approximation as presented below in the methods section.

Another important factor in calculating the Raman intensities is the resonance behaviour arising from the energy denominators. When the energy of a virtual state is close to the initial or final state energy, the corresponding energy denominator will become nearly zero, which increases the intensity dramatically. It is possible for two of the energy denominators to be simultaneously nearly zero, which is called double resonance, and can result in even larger intensities compared to the single resonant processes \cite{thomsen_double_2000,zolyomi_resonance_2011}. Furthermore, it is possible that all three denominators are nearly zero, which would be called triple or fully resonant process\cite{PhysRevB.78.125418}, albeit it was shown later to be less significant than quantum interference in real space \cite{venezuela_theory_2011}.
 However, if one of the denominators in Eqs. \ref{eq_pp} and \ref{eq_pd} becomes zero, the transmission amplitudes will diverge resulting in infinite peak intensities. In order to avoid these singularities numerically an imaginary component is introduced in the energy denominators, which in a physical sense relate directly to the finite lifetime of charge carriers. 

\subsection{DFT calculations}
\label{DFT}
\begin{figure}
\begin{center}
\includegraphics[scale=.33]{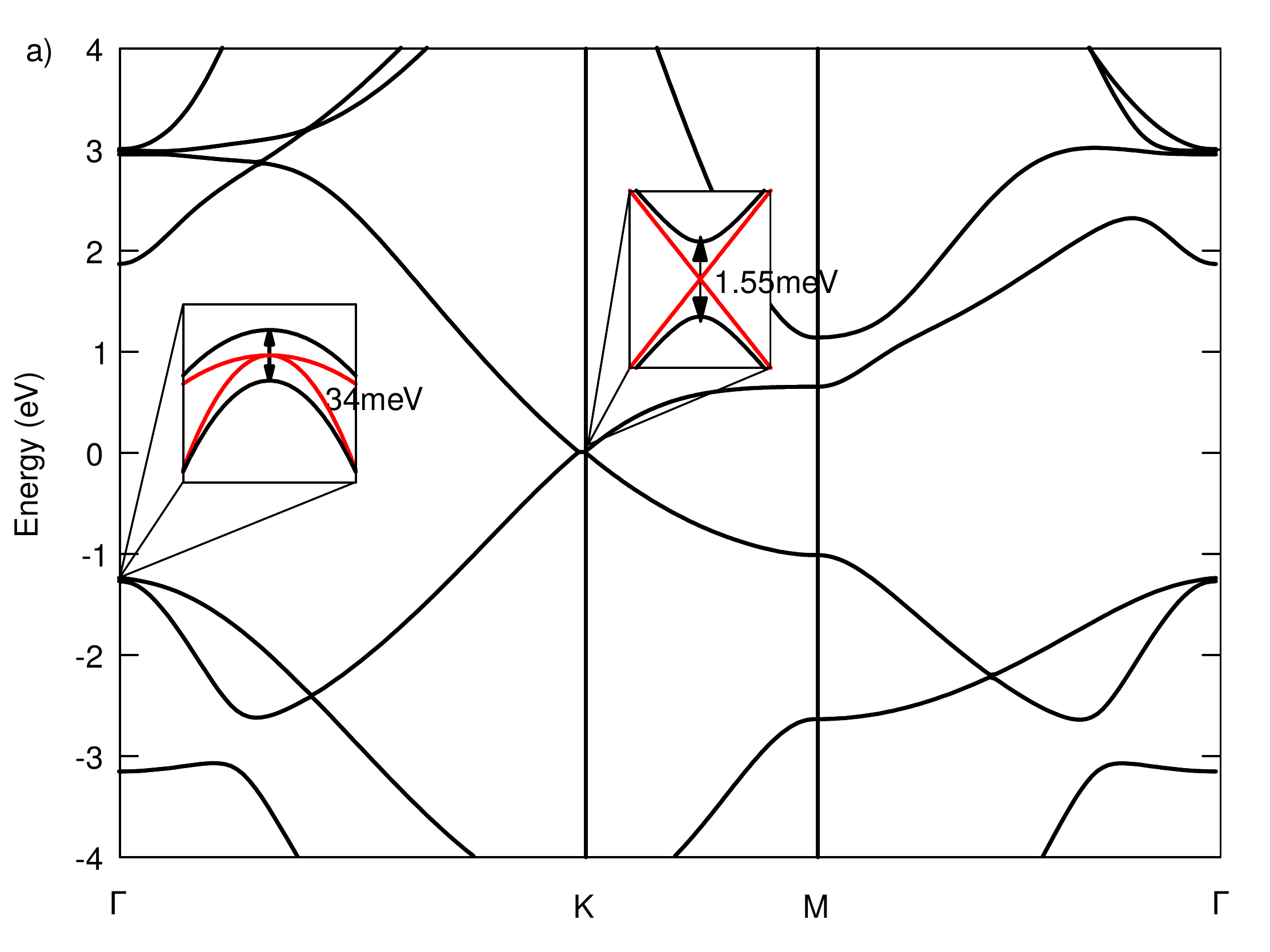}
\includegraphics[scale=.33]{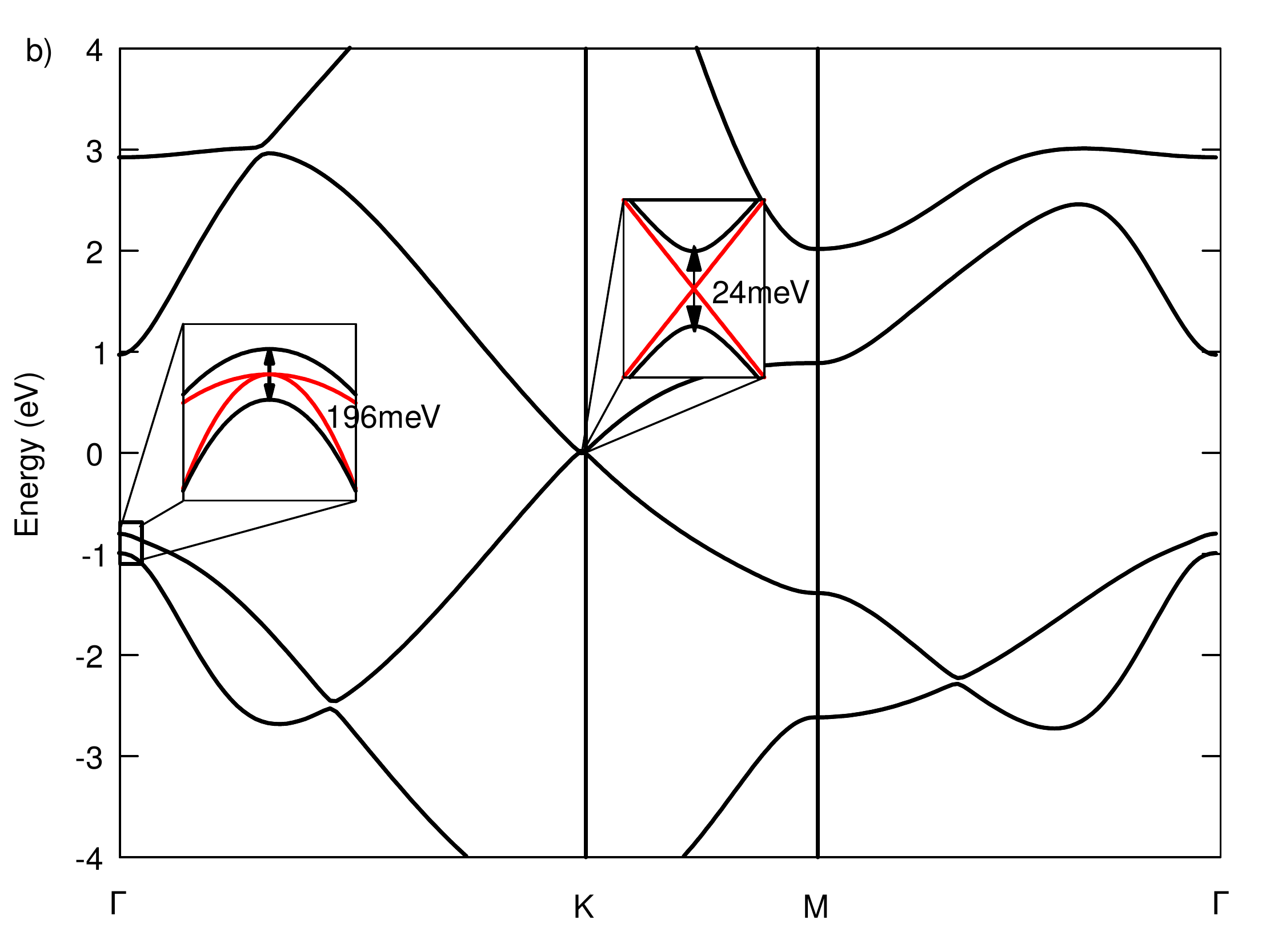}
\end{center}
\caption{Effect of spin-orbit coupling on the electronic band structure of silicene (left) and germanene (right). Insets indicate main differences between calculated band structures with (black) and without (red) spin-orbit coupling (values are taken from Ref. \cite{liu_low-energy_2011})}
\label{Si_Ge_DFT}
\end{figure}

In Fig. \ref{Si_Ge_DFT} we show the band structure of silicene and germanene with spin-orbit coupling (SOC) taken into account in order to investigate its effects and determine whether it needs to be included in the TB model and thus in the Raman calculations. In the case of silicene the band structure is rather unperturbed by the SOC \cite{liu_quantum_2011,liu_low-energy_2011,drummond_electrically_2012,singh_spinorbit_2017} apart  from the relatively small $1.55\,\rm meV$ and $34\,\rm meV$ splitting introduced at the K and $\Gamma$ points, respectively. These values are negligible next to the relevant Raman excitation energies ($1\,\rm eV$-$3\,\rm eV$), therefore we do not include SOC in our TB model for silicene. However, as the strength of SOC increases with the fourth power of the atomic number, in the case of germanene a much larger effect is expected. Several theoretical works \cite{liu_low-energy_2011,acun_germanene:_2015} indeed suggest that a $24\,\rm meV$ spin-orbit gap opens at the K point, however, compared to the relevant excitation energies ($1\,\rm eV$-$3\,\rm eV$) this is still negligible. On the other hand, SOC lifts the fourfold degeneracy of the highest valence bands at the $\Gamma$ point, splitting them into two Kramers doublets separated by a significant energy of $196\,\rm meV$. This latter effect can be explained by recognizing that in the first order SOC mixes spin states of different $p$ type orbitals, while transition between spin states on $p_z$ orbitals located on different atoms occurs in the second order through SOC coupling between $p_z$ and $p_x,p_y$ states followed by a hopping transition between the $p_x,p_y$ states and the neighbouring $p_z$ states. This effective coupling between $p_z$ orbitals located on neighbouring atoms can lift the degeneracy at the K point. Its magnitude, however, is small compared to the mixing of $p_x,p_y$ orbitals at the $\Gamma$ point.

\begin{figure}
\includegraphics[scale=.32]{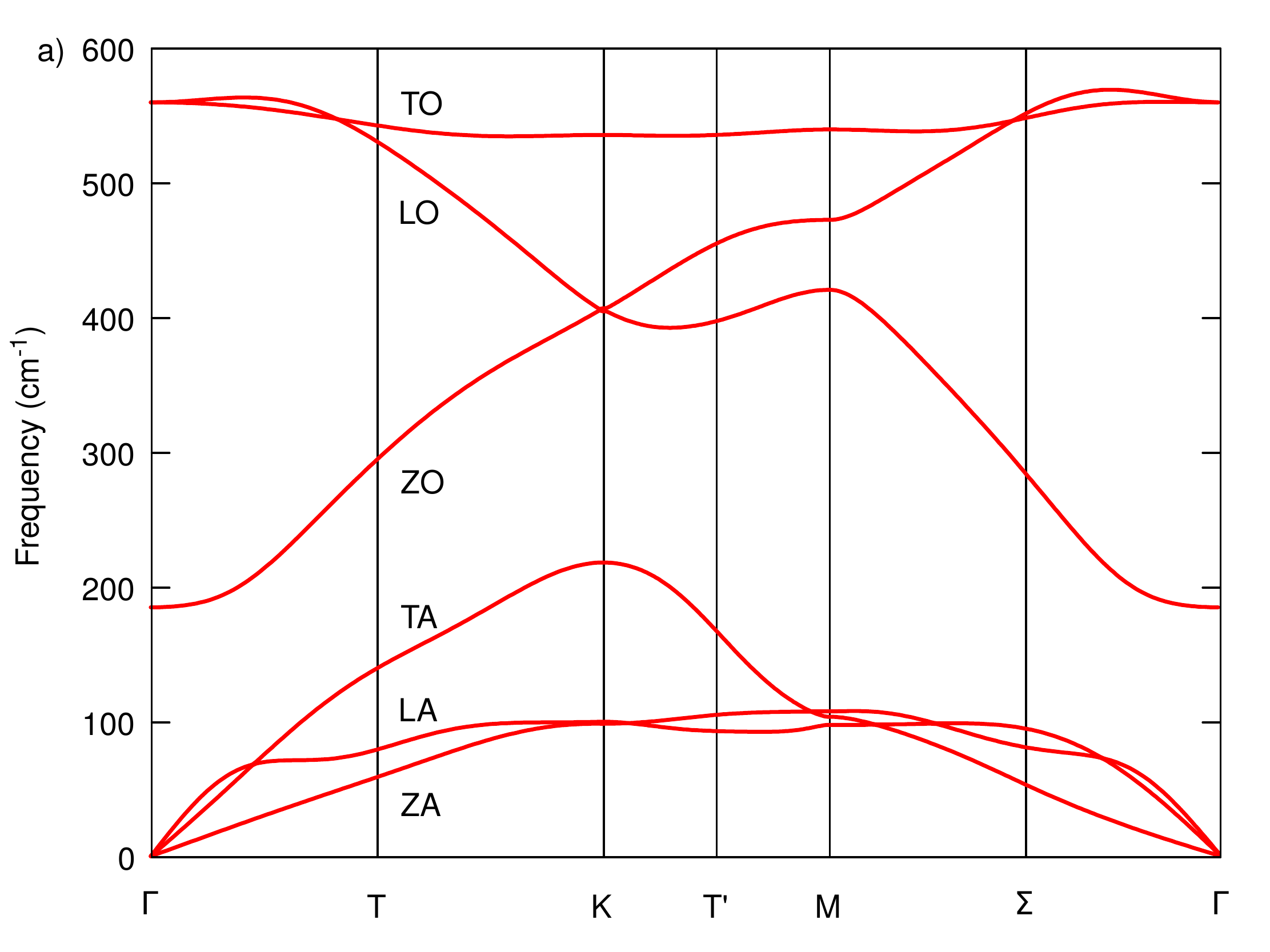}
\includegraphics[scale=.32]{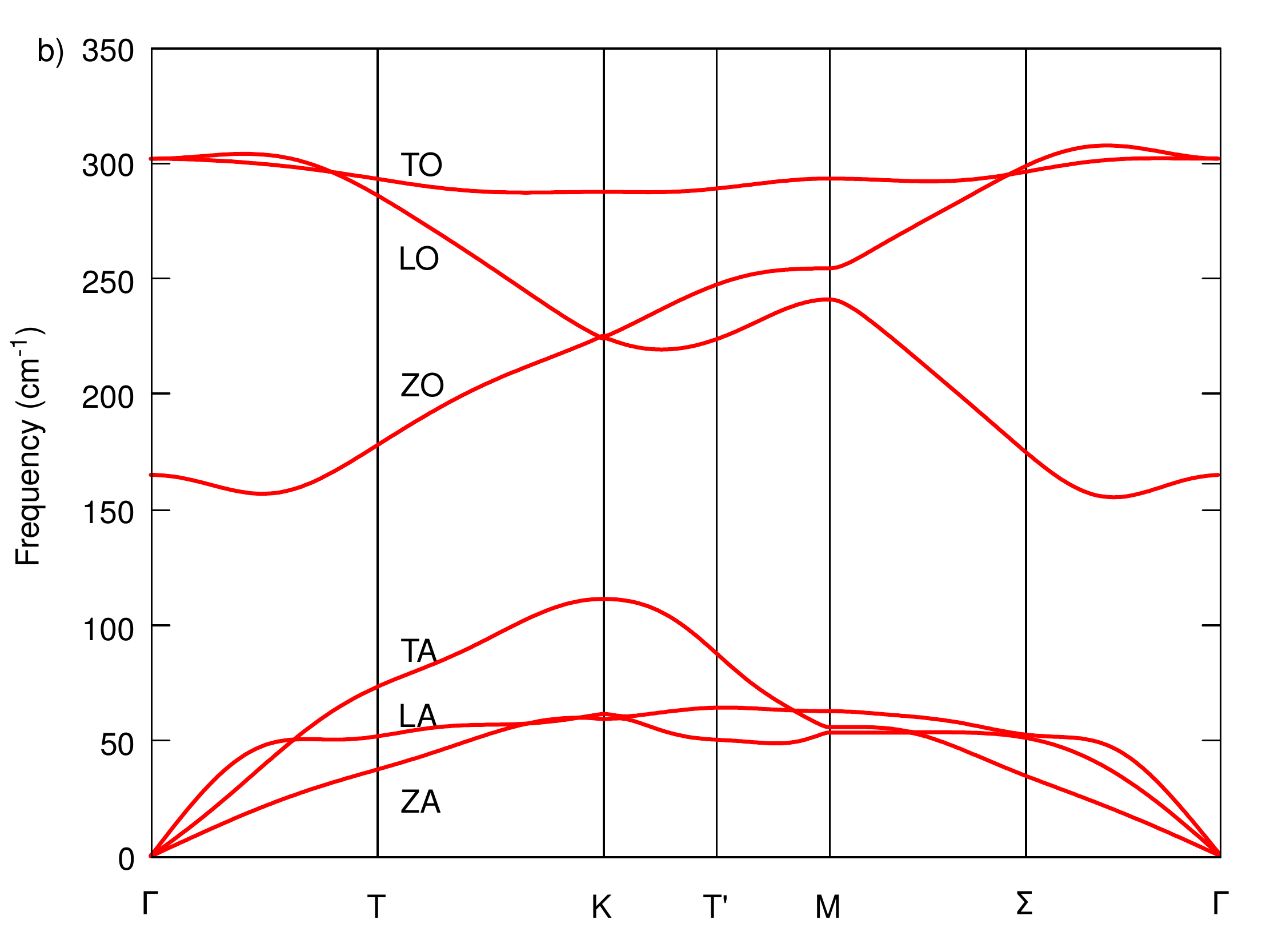}
\caption{Calculated phonon dispersion of silicene and germanene along high symmetry points with labels indicating symmetry of the modes.}
\label{Si_Ge_phonon}
\end{figure}
The phonon dispersion relation is obtained by calculating the Hessian matrix of a $7\times 7$ supercell within the frozen phonon approximation. Main features and phonon frequencies between the high symmetry points shown in Fig. \ref{Si_Ge_phonon} are in agreement with previous theoretical results \cite{drummond_electrically_2012,yang_phonon_2014,ge_comparative_2016}.

\subsection{Tight-binding model of silicene and germanene}

\begin{figure}
\begin{center}
\includegraphics[scale=.32]{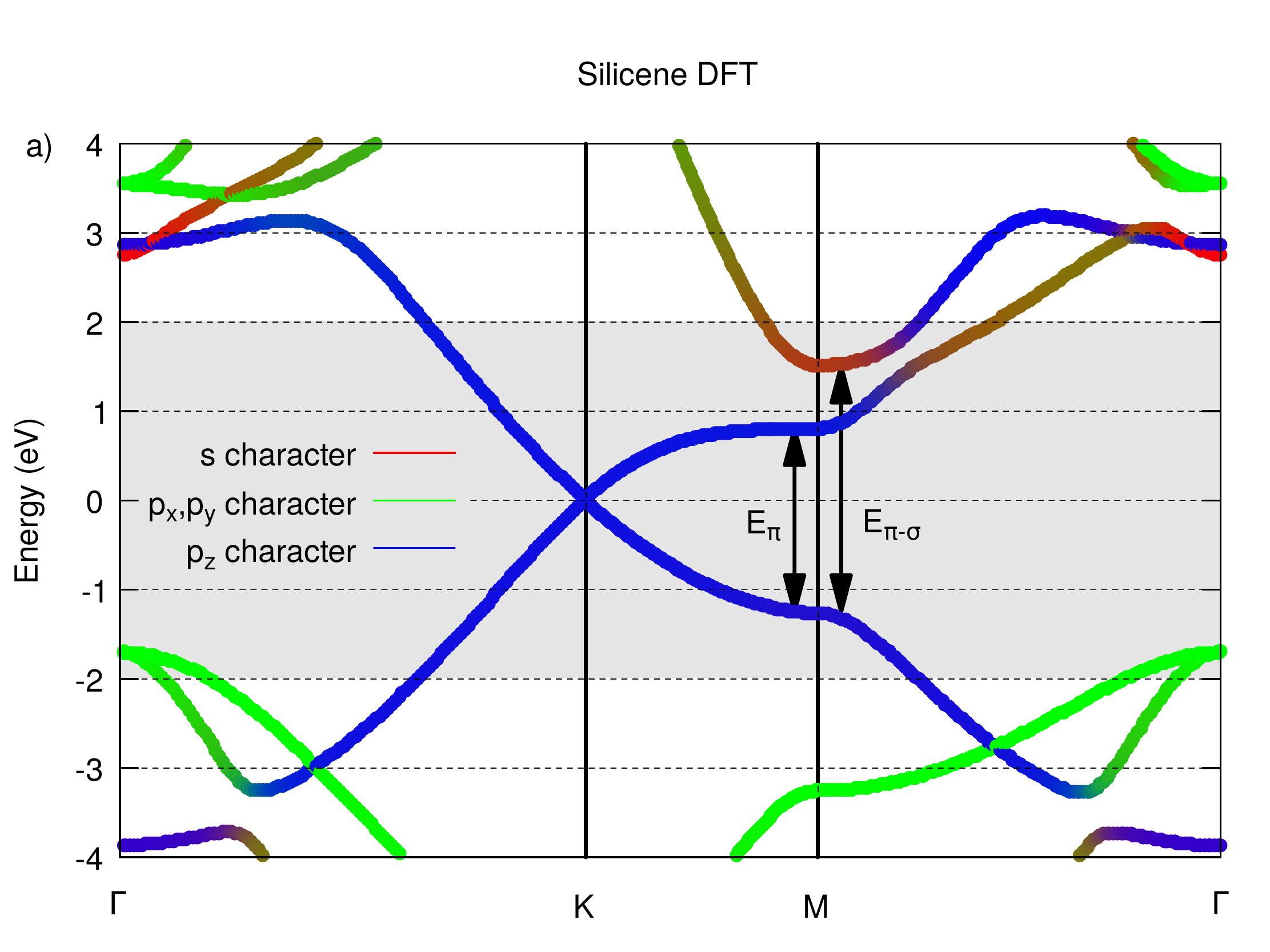}
\includegraphics[scale=.32]{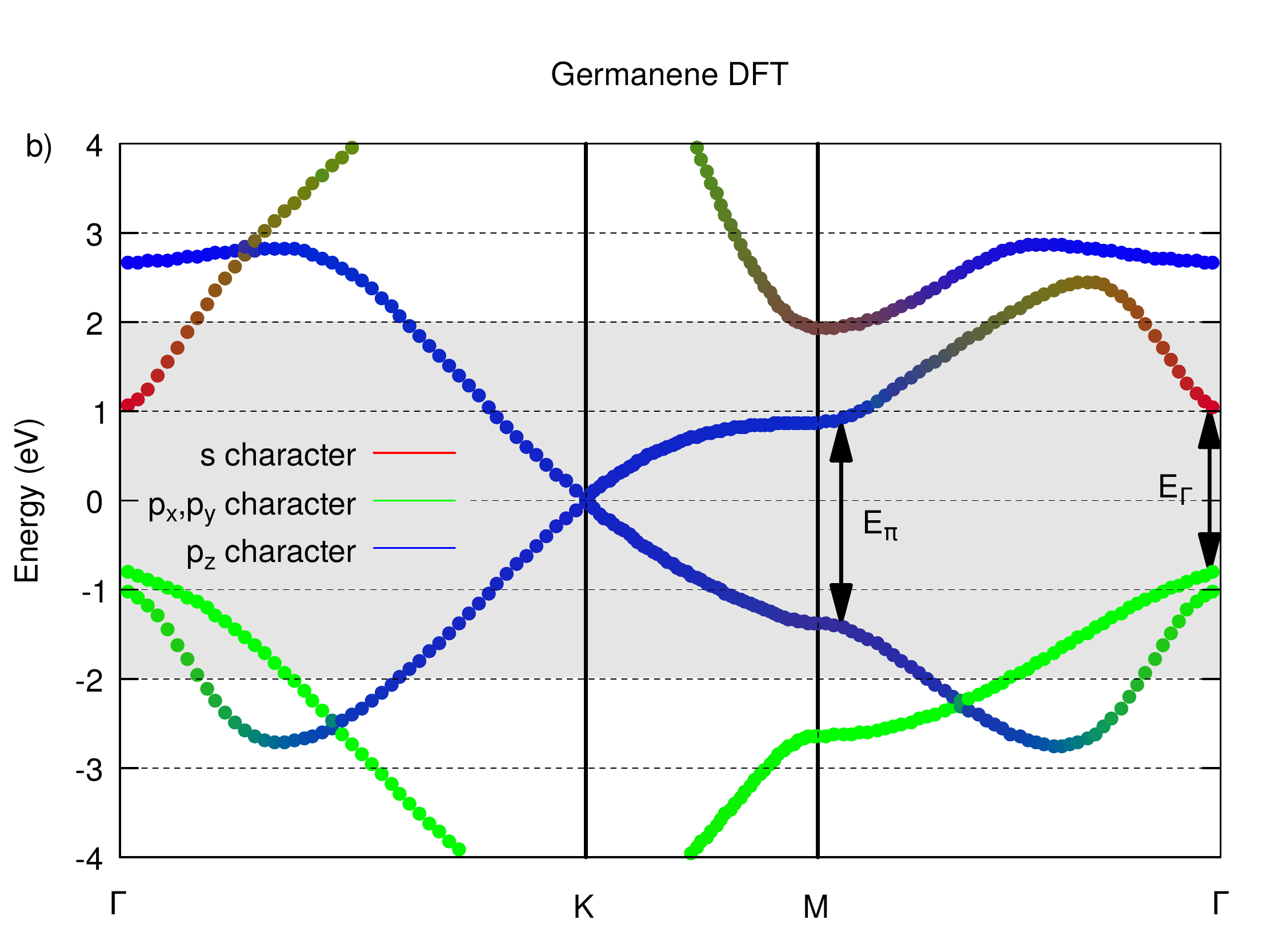}
\includegraphics[scale=.32]{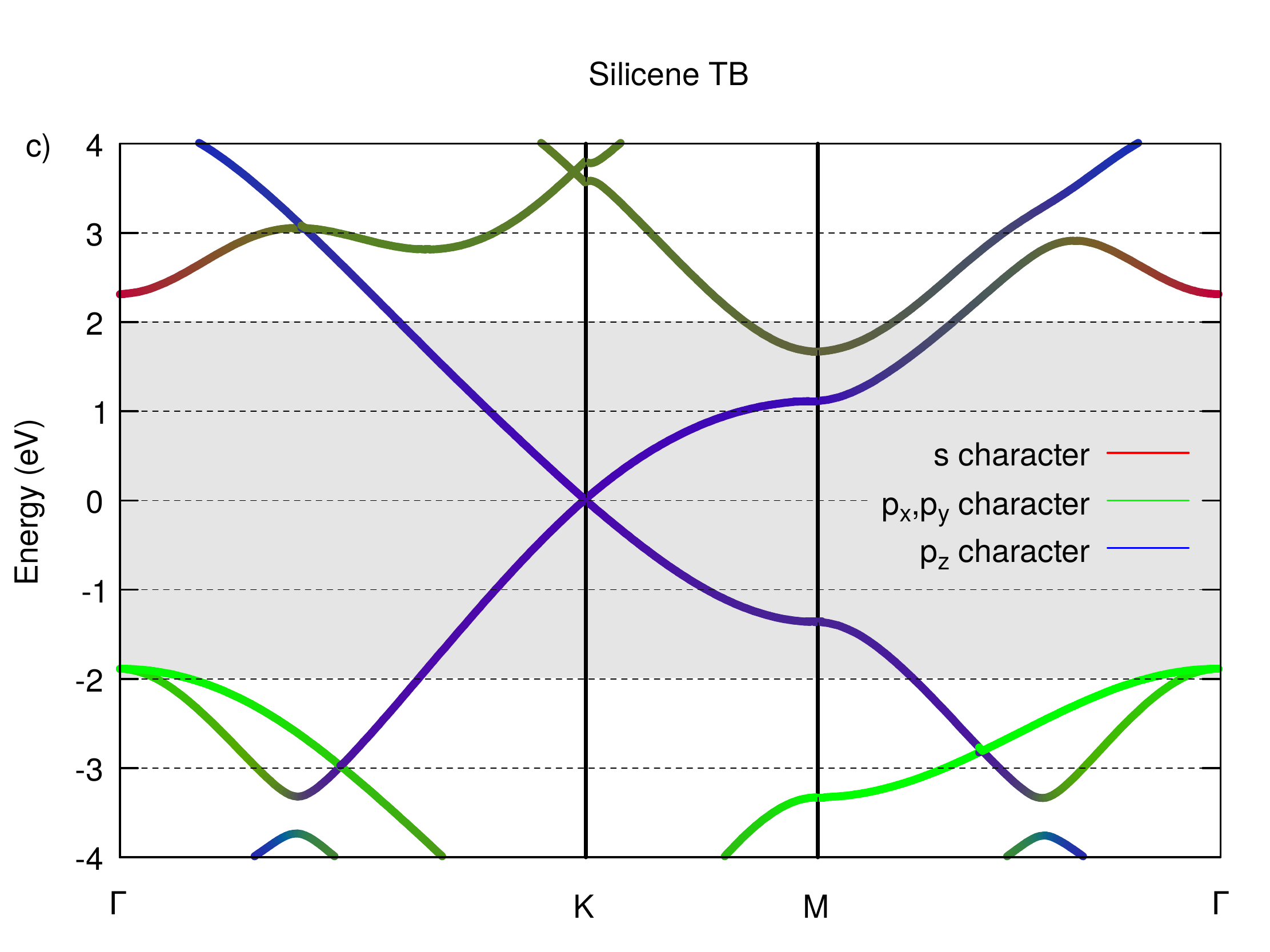}
\includegraphics[scale=.32]{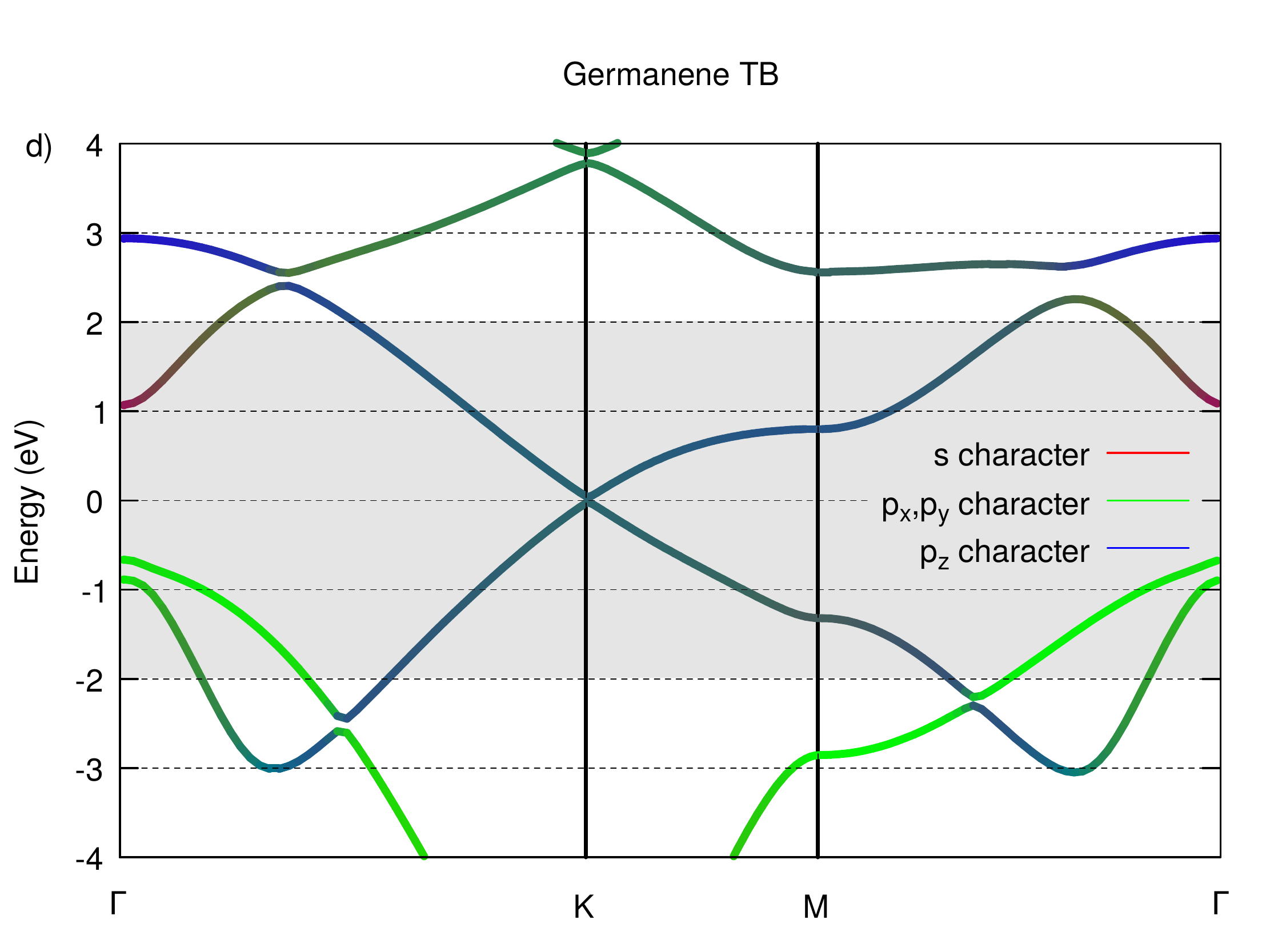}
\caption{Comparison of the characters of the DFT band structure and the fitted TB model for silicene (a,c) and germanene (b,d)}
\label{DFT-TB}
\end{center}
\end{figure}
To describe Raman scattering in silicene and germanene, we first construct a TB Hamiltonian with 4-orbital ($s$, $p_x$, $p_y$, $p_z$) basis on each atom, expressed as

\begin{equation}
|\psi_{n,\mathbf{k}}\rangle=\sum\limits_{i,j}e^{i\mathbf{k}\mathbf{R}_i}c_{n,i,j}(\mathbf{k})|\varphi_j(\mathbf{r}-\mathbf{R}_i)\rangle,
\end{equation}
or in the second quantized formalism as
\begin{equation}
|\psi_{n,\mathbf{k}}\rangle=a^\dagger_{n,\mathbf{k}}|0\rangle=\sum\limits_{i,j}e^{i\mathbf{k}\mathbf{R}_i}c_{n,i,j}(\mathbf{k})a^\dagger_{i,j}|0\rangle,
\end{equation}

\noindent where $\mathbf{R}_i$ is the atomic position of the $i$th atom, $|\varphi_j(\mathbf{r}-\mathbf{R}_i)\rangle$ is the $j$th basis centred on the $i$th atom, $a^\dagger_{i,j}$ is the creation operator of an electron on the $j$th basis centred on the $i$th atom, $a^\dagger_{n,\mathbf{k}}$ is the creation operator on the $n$th band with lattice momentum of $\mathbf{k}$, and $c_{n,i,j}(\mathbf{k})$ are the $i,j$th component of $n$th solution ($c_n(\mathbf{k})$) of the non-orthogonal Schr{\"o}dinger equation:
\begin{equation}
\hat{H}(\mathbf{k})c_n(\mathbf{k})=\epsilon_n(\mathbf{k})\hat{S}(\mathbf{k})c_n(\mathbf{k}),
\end{equation} 

\noindent where $\hat{H}(\mathbf{k})$ and $\hat{S}(\mathbf{k})$ are the tight-binding Hamiltonian and overlap matrices, respectively, and $\epsilon_n(\mathbf{k})$ is the $n$th eigenvalue. Overlap matrix elements between orbitals centred on different atoms are integrated numerically using hydrogen-like orbitals with effective nuclear charges \cite{clementi_atomic_1963}. The Hamiltonian is built by using the Slater-Koster method \cite{slater_simplified_1954}, taking up to third-nearest neighbour hopping interaction into account. Three on-site parameters are used to describe the different atomic energies of the orbitals, as the $p_x$ and $p_y$ on-site matrix elements are connected by symmetry.

Within the Slater-Koster method \cite{slater_simplified_1954} the Hamiltonian and the overlap matrix elements can be written as

\begin{equation}
\begin{split}
\hat{H}_{i',i}(\mathbf{k})&=\sum_{l}^{\mathrm{UC}}\sum_{L}^{\mathrm{3^{rd}NN}}\underbrace{\int\mathrm{d}^3\mathbf{r}\phi_{i'}^*(\mathbf{r}-\mathbf{r}_{L})\hat{H}
\phi_{i}(\mathbf{r}-\mathbf{r}_{l})}_{t_{i'i}(\mathbf{r}_{l}-\mathbf{r}_{L})}
e^{i\mathbf{k}(\mathbf{r}_{l}-\mathbf{r}_{L})}a^\dagger_{i,L}a_{i,l}+\cr 
&+\sum_{l}^{\mathrm{UC}}\underbrace{\int\mathrm{d}^3\mathbf{r}\phi_{i'}^*(\mathbf{r}-\mathbf{r}_{l})\hat{H}
\phi_{i}(\mathbf{r}-\mathbf{r}_{l})}_{\varepsilon_{i'i}}a^\dagger_{i',l}a_{i,l},
\end{split}
\end{equation} 

\begin{equation}
\hat{S}_{i',i}(\mathbf{k})=\sum_{l}^{\mathrm{UC}}\sum_{L}^{\mathrm{3^{rd}NN}}\underbrace{\int\mathrm{d}^3\mathbf{r}\phi_{i'}^*(\mathbf{r}-\mathbf{r}_{L})
\phi_{i}(\mathbf{r}-\mathbf{r}_{l})}_{s_{i'i}(\mathbf{r}_{l}-\mathbf{r}_{L})}
e^{i\mathbf{k}(\mathbf{r}_{l}-\mathbf{r}_{L})}a^\dagger_{i,L}a_{i,l},
\end{equation} 
\noindent where summation over $l$ goes over a unit cell, summation over $L$ goes over surrounding atoms in the crystal up to the third nearest neighbours, $\varepsilon_{i'i}$ are the on-site terms of each atomic orbital, while $t_{i'i}(\mathbf{r}_{l}-\mathbf{r}_{L})$ and $s_{i'i}(\mathbf{r}_{l}-\mathbf{r}_{L})$ are the hopping and overlap integrals, respectively, between atoms located at $\mathbf{r}_{l}$ and $\mathbf{r}_{L}$.

The hopping parameters are fitted to the DFT results within a $\pm2\,$eV range around the Fermi level. The rest of the band structure is ignored during the fitting process as we are interested in reproducing the optical transitions within the relevant regime ($1\,{\rm eV}-3\,{\rm eV}$). This assumption can be justified by the fact that the transitions between low and high energy bands would be suppressed by the energy denominators in Eqns \ref{eq_pp} and \ref{eq_pd}. During the fitting procedure we acquire several parameter sets using the least squares approach, which reproduces the first principles band structures fairly well. In the final step, we compare the wavefunction symmetries and $s$ and $p$ characters of the DFT data and the TB model. We used the least squares approach to determine which of the found symmetrically appropriate sets of parameters yields the best match with the DFT results in terms of the composition of wavefunctions in the fitted bands. We emphasize that fitting to the $s$ and $p$ characters is a stronger condition than considering the symmetries only, because it also ensures the proper mixing of the orbitals in our TB model. This is extremely important for the calculation of the Raman intensities as the magnitude of the matrix elements, and therefore the transition probabilities, are mainly determined by the symmetries and characters of the wavefunctions.

As stated in the previous section, in the tight binding model of germanene we take SOC into account between $p$ orbitals. We implement this using the atomic matrix elements of the $\hat{\mathbf{L}}\hat{\mathbf{S}}$ operator on the $|p_x\uparrow\rangle$,$|p_x\downarrow\rangle$,$|p_y\uparrow\rangle$,$|p_y\downarrow\rangle$,$|p_z\uparrow\rangle$,$|p_z\downarrow\rangle$ basis,

\begin{equation}
\hat{H}_\mathrm{SOC}=
\frac{\lambda_{\mathrm{SOC}}}{2}
\left( \begin{array}{cccccc}
0& 0& -i& 0& 0& 1\\
0& 0& 0& i& -1& 0\\
i& 0& 0& 0& 0& -i\\
0& -i& 0& 0& -i& 0\\
0& -1& 0& i& 0& 0\\
1& 0& i& 0& 0& 0\\
\end{array}
\right)
\end{equation}

The SOC parameter $\lambda_\mathrm{SOC}$ is chosen to reproduce the numerical value of the HSE06 gap at the $\Gamma$ point, that is, $\lambda_\mathrm{SOC}=196$meV.

The fitted hopping integrals and calculated overlaps for silicene and germanene are shown in Tables \ref{si-tb} and \ref{ge-tb}, respectively. The characters presented in Fig. \ref{DFT-TB} show that the Dirac-like band of both silicene and germanene mostly contains $s$ and $p_z$ character, whilst the $p_x,p_y$ character is suppressed. The mixing of $s$ and $p_z$ dominated bands is expected from the DFT calculations as the buckled structure yields the mixing of $sp^2$ hybrid orbitals with the $p_z$ orbital. Moreover, in the case of silicene the flat conduction band around the M point also possesses $s$ and $p_z$ character, and optical transitions between the Dirac-like valence band and this band are allowed. 
The excitation energy of $E_{\pi}^{Si}=2.12\,\rm eV$ between the Dirac-like bands at the M point is referred as the $\pi$ plasmon energy in the graphene literature or $\pi$-like plasmon energy for silicene\cite{PhysRevB.95.085419}. While bearing in mind that it is not a clear $\pi$-type plasmon, we will adopt the shortest notation by calling it $\pi$ plasmon. The $E_{\pi-\sigma}^{Si}=2.77\,\rm eV$ excitation energy between the valence band and the higher conduction band is a more conventional $\pi-\sigma$ plasmon, therefore it can be safely referred as the $\pi-\sigma$ plasmon energy. In the case of germanene we find that although the $\pi$ plasmon energy of $E_{\pi}^{Ge}=2.12\,\rm eV$ is lower compared to that of silicene, the $\pi-\sigma$ plasmon energy of $E_{\pi-\sigma}^{Ge}=3.31\,\rm eV$ is significantly higher. As this energy is larger than the relevant energy range, we neglect this band during the fitting procedure. On the other hand, at the $\Gamma$ point of germanene in-plane polarized optical transitions of $1.61\,\rm eV$ and $1.82\,\rm eV$ can be found between the mostly $s$ and $p_z$ conduction band and the SOC-perturbed $p_x$ and $p_y$ valence bands. Therefore, two resonances can be expected in the Raman spectra of germanene, and the splitting between them gives the on-site spin-orbit coupling strength. This effect is suppressed in silicene as the SOC strength is around one order of magnitude smaller compared to germanene, and the corresponding excitation energy of $4.45\,\rm eV$ is too large for the regime accessible in typical Raman measurements.

\begin{table}
\begin{center}
\begin{tabular}{|c|cccc|}
\hline
&$s$&$p_x$&$p_y$&$p_z$\\
On-site&-2.451&-0.198&-0.198&-0.055\\
\hline
\hline
Hopping&$t_{ss\sigma}$&$t_{sp\sigma}$&$t_{pp\sigma}$&$t_{pp\pi}$\\
1$^\mathrm{st}$ neighbour&-1.675&2.868&3.207&-1.372\\
2$^\mathrm{nd}$ neighbour&-0.793&0.605&0.721&-0.019\\
3$^\mathrm{rd}$ neighbour&-1.165&0.859&0.982&-0.104\\
\hline
\hline
Overlap&$s_{ss\sigma}$&$s_{sp\sigma}$&$s_{pp\sigma}$&$s_{pp\pi}$\\
1$^\mathrm{st}$ neighbour&0.031&0.032&-0.036&0.022\\
2$^\mathrm{nd}$ neighbour&0.003&0.005&-0.003&0.001\\
3$^\mathrm{rd}$ neighbour&0.002&0.004&-0.002&0.001\\
\hline
\end{tabular}
\caption{Fitted tight-binding (eV) and calculated overlap parameters of silicene}
\label{si-tb}
\end{center}
\end{table}
\begin{table}
\begin{center}
\begin{tabular}{|c|cccc|}
\hline
&$s$&$p_x$&$p_y$&$p_z$\\
On-site&-8.189&0.327&0.327&0.619\\
\hline
\hline
Hopping&$t_{ss\sigma}$&$t_{sp\sigma}$&$t_{pp\sigma}$&$t_{pp\pi}$\\
1$^\mathrm{st}$ neighbour&-2.040&3.080&2.933&-1.089\\
2$^\mathrm{nd}$ neighbour&0.317&0.254&0.604&-0.180\\
3$^\mathrm{rd}$ neighbour&0.117&0.339&0.218&-0.137\\
\hline
\hline
Overlap&$s_{ss\sigma}$&$s_{sp\sigma}$&$s_{pp\sigma}$&$s_{pp\pi}$\\
1$^\mathrm{st}$ neighbour&0.043&0.047&-0.047&0.029\\
2$^\mathrm{nd}$ neighbour&0.009&0.008&-0.005&0.002\\
3$^\mathrm{rd}$ neighbour&0.006&0.007&0.002&0.002\\
\hline
\end{tabular}
\caption{Fitted tight-binding (eV) and calculated overlap parameters of germanene}
\label{ge-tb}
\end{center}
\end{table}

\subsection{Defect scattering}
We model defect induced Raman scattering through a variety of possible point defects that may occur in silicene or germanene. From an experimental point of view, four categories of point defects can be distinguished: substitutional atoms, Stone-Wales defects, adatoms, and vacancies. Within the framework of our tight-binding model, there are two ways in which we can describe scattering through such defects: the perturbative  approach, and the scattering potential approach.

In the perturbative approach we introduce small perturbations into the tight-binding parameters. In this method, two types of scattering matrix elements can be defined: on-site and hopping scatterers. In either case, the corresponding tight-binding parameter is changed, resulting in a small perturbation of the system. The on-site scattering Hamiltonians perturb the on-site energy of a given orbital, while the hopping scattering Hamiltonians change a Slater-Koster hopping parameter. In our model we use 3 on-site parameters on every atom and 4 non-equivalent hoppings between atoms. Since the main contribution to the electron-photon and electron-phonon matrix elements arises from nearest-neighbour interaction, we only take into account defect scattering induced by changes in the on-site terms or in the nearest-neighbour hopping integrals. The defect scattering Hamiltonians for a nearest-neighbour hopping perturbation and for an on-site perturbation located on $\mathbf{R_0}$, respectively, can be written as

\begin{equation}
\begin{split}
&M^{\mathrm{t}}_{BA}=\langle \psi_{n,\mathbf{k-q}}|\hat{H}_\mathrm{t}|\psi_{m,\mathbf{k}}\rangle=\sum\limits_{i'}^{1^{st}NN}\sum\limits_{j,j'}c^*_{n,i',j'}(\mathbf{k-q}) c_{m,0,j}(\mathbf{k})\times\cr
&\times\delta t_{jj'}(\mathbf{R}_i') e^{i\mathbf{k}\mathbf{R}_0}e^{-i(\mathbf{k-q})\mathbf{R}_{i'}} a^\dagger_{i',j'}a_{0,j}.
\end{split}
 \label{2nd_def_hopping}
\end{equation}

\begin{equation}
M^{\varepsilon}_{BA}= \langle \psi_{n,\mathbf{k}}|\hat{H}_\mathrm{\varepsilon}|\psi_{m,\mathbf{k}}\rangle=\sum\limits_{j}c^*_{n,0,j}(\mathbf{k}) c_{m,0,j}(\mathbf{k})\delta\varepsilon_j a^\dagger_{0,j}a_{0,j}.
\label{2nd_def_on-site}
\end{equation}

The above method can be used to describe a variety of real defects in the crystal. Substitutional impurities mostly change the atomic ionization energies, Stone-Wales defects perturb the in-plane $\sigma$ bonds \cite{sahin_stone-wales_2013,zhang_tight-binding_2016}, adatoms mostly change $\pi$ orbitals \cite{zhang_tight-binding_2016}, whilst vacancies remove a site which eliminates an on-site term from the Hamiltonian along with the corresponding nearest-neighbour hopping terms \cite{zhang_tight-binding_2016}. Therefore, the $t_{ss\sigma}$ and $t_{pp\pi}$ defects combined can be used to describe the presence of adatoms, $t_{sp\sigma}$ and $t_{pp\sigma}$ defects together can provide a qualitative model of Stone-Wales defects, and a change in the on-site matrix elements can model substitutional impurities. Vacancies could in principle be described through a combination of on-site and hopping scattering, but since this defect produces strong, local effects, it is not suitable to describe them through the perturbative approach.

Therefore, for vacancies, we use the scattering potential approach instead. In this method, rather than perturbing the parameters in the model, we introduce a scattering electron-ion potential that can be used to model the presence of a vacancy in the crystal. The scattering potential has the same form as the atomic potential used in the electron-phonon coupling in Eq. \ref{ep-coupling}, which allows us to express scattering by a vacancy located at the position $\mathbf{R}_v$ as

\begin{equation}
\begin{split}
&M^{vac}_{BA}=\langle \psi_{n,\mathbf{k-q}}|\hat{H}_\mathrm{vac}|\psi_{m,\mathbf{k}}\rangle=\sum\limits_{i,i',j,j'}c^*_{n,i',j'}(\mathbf{k-q}) c_{m,i,j}(\mathbf{k})\times\cr
&\times\left\langle\varphi_{j'}(\mathbf{r}-\mathbf{R}_{i'})\left|V_\mathrm{e-ion}(\mathbf{r}-\mathbf{R}_v)\right|\varphi_j(\mathbf{r}-\mathbf{R}_i)\right\rangle e^{i\mathbf{k}\mathbf{R}_i}e^{-i(\mathbf{k-q})\mathbf{R}_{i'}}a^\dagger_{i',j'}a_{i,j}.
\end{split}
\end{equation}

Note, that the scattering potential approach is also useful for the description of substitutional defects, if said defects introduce strong, local changes to bonding in the crystal which cannot be accurately described by the perturbative approach.

\subsection{Lifetime of charge carriers}
\label{lifetime_sec}
\begin{figure}
\includegraphics[scale=.62]{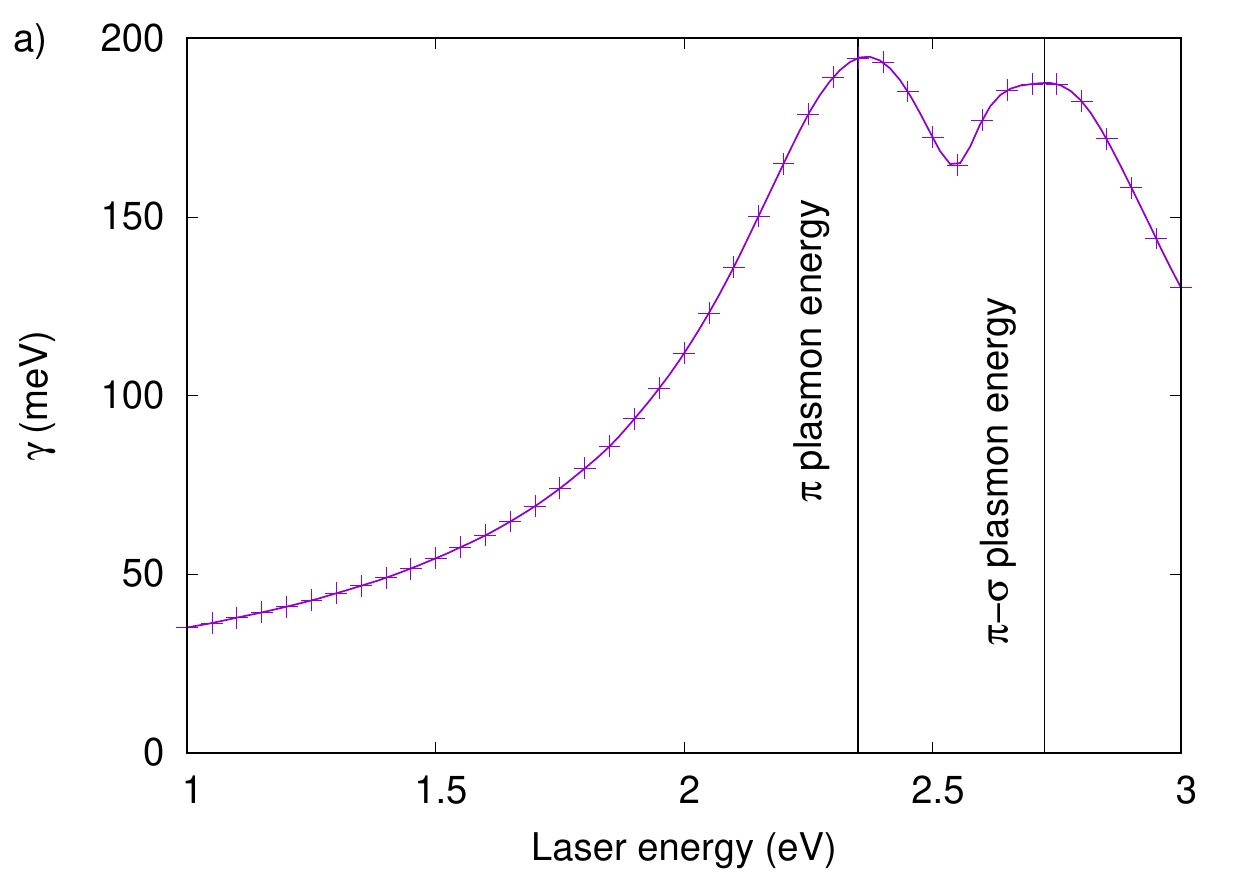}
\includegraphics[scale=.62]{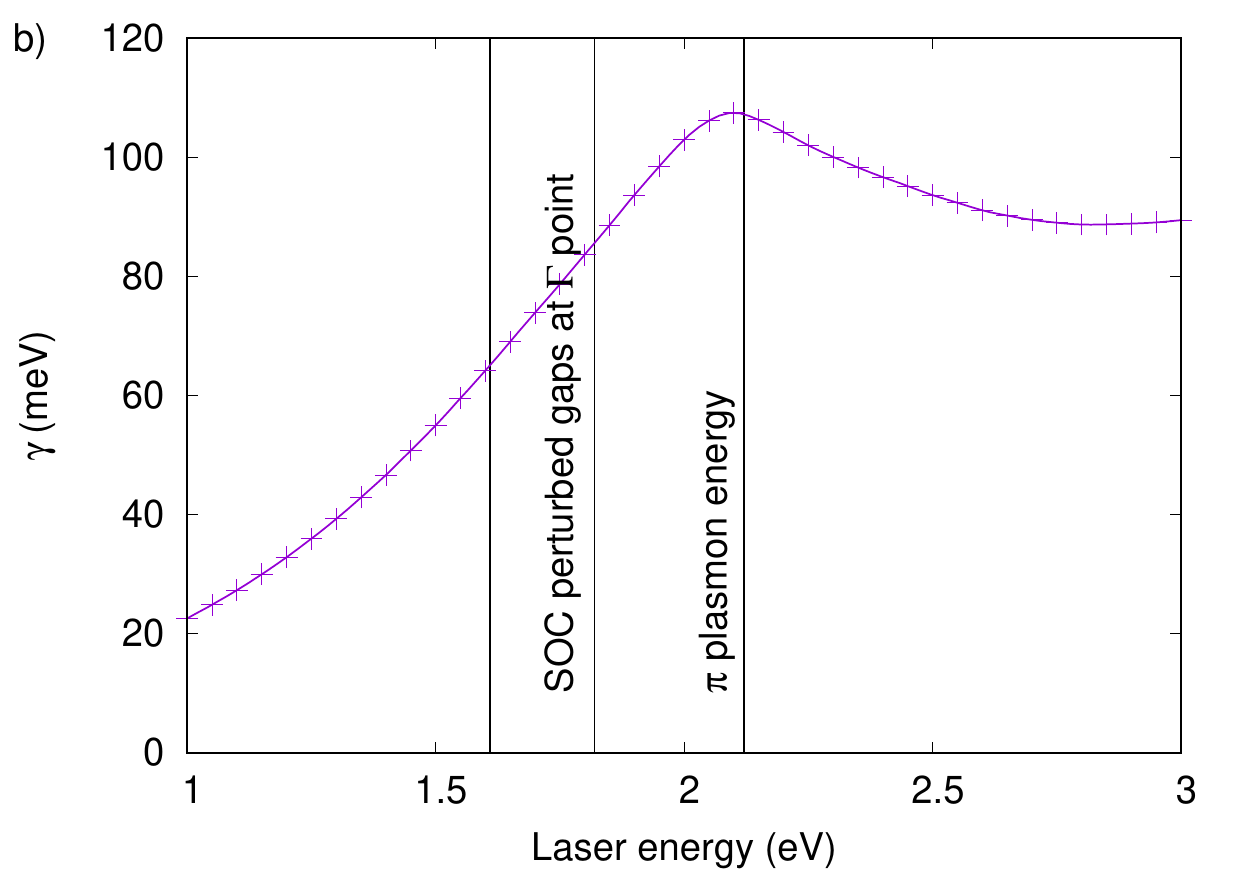}
\caption{Laser energy dependence of the inverse electronic lifetime $\gamma(\epsilon_L)$ in silicene (a) and germanene (b)}
\label{lifetime}
\end{figure}
The imaginary part of the energy denominators are calculated by taking into account the scattering of charge carriers using Fermi's golden rule. In leading order \cite{venezuela_theory_2011} the electron-phonon interaction determines the lifetime of excitations, while contributions from electron-photon interaction and even defect scattering can be neglected, the former due to the negligible momentum transfer during optical transitions, and the latter under the assumption of low defect concentration.
In this approximation the inverse electronic lifetime can be calculated as

\begin{equation}
\gamma_{m\mathbf{k}}=\frac{2\pi}{N_\mathbf{q}}\sum\limits_{\nu,\mathbf{q},m}\left|\langle\psi_{n,\mathbf{k-q}}|\hat{H}_\mathrm{e-ph,\nu}|\psi_{ m,\mathbf{k}}\rangle\right|^2\delta(\epsilon_{m\mathbf{k}}-\epsilon_{n\mathbf{k-q}}-\hbar\omega_{\nu}(\mathbf{q})),
\end{equation}

\noindent where $N_\mathbf{q}$ is the number of $\mathbf{q}$ points used in the Brillouin zone integration. In order to simplify the calculations we remove the $\mathbf{k}$-dependence from $\gamma$ by taking an average over the electron-hole pairs which can be excited at a given laser energy \cite{venezuela_theory_2011,popov_comparative_2016} as

\begin{equation}
\gamma(\epsilon_L)=\frac{1}{N_\mathbf{k}}\sum\limits_{n,m,\mathbf{k}}(\gamma_{n\mathbf{k}}+\gamma_{m\mathbf{k}})\delta(\epsilon_L-(\epsilon_{n\mathbf{k}}-\epsilon_{m\mathbf{k}})),
\end{equation}

\noindent where $N_\mathbf{k}$ is the number of $\mathbf{k}$ points taken into account in the summation, $\epsilon_L$ is the exciting laser energy, and the Dirac $\delta(x)$ is approximated with a Gaussian function with $0.05\,\rm eV$ standard deviation. To calculate $\gamma(\epsilon_L)$ we use a $360\times360\times1\,\Gamma$-centred Monkhorst-Pack grid in the Brillouin zone for the electronic $\mathbf{k}$ points and a $180\times180\times1\,\Gamma$-centred Monkhorst-Pack set for the phonon $\mathbf{k}$ points.

The energy dependence of $\gamma(\epsilon_L)$ for silicene and germanene is shown in Fig. \ref{lifetime}. The inverse lifetime of silicene shows similar dependence in the low energy ($<2\,\rm eV$) region as previous works suggest \cite{popov_comparative_2016}. A resonance can be found at the $\pi$ plasmon energy (depicted by the dashed line) as expected from the band structure. A second resonance is visible at the $\pi-\sigma$ plasmon energy, as expected due to the large density of states arising from the flat conduction band near the M point.

In the case of germanene, as illustrated in Fig. \ref{lifetime}, only one resonance can be found near the excitation energy of the $\pi$-plasmon. The absence of the resonance of the $\pi-\sigma$ plasmon is due to the fact that its excitation energy is approximately $1\,\rm eV$ higher than that of silicene, thus it drops out of the relevant range. However as the $\Gamma$ point gap is smaller compared to silicene, a twin-resonance is expected due to the SOC. The apparent absence of this feature can be explained by taking into account two factors: the relatively lower electron-phonon coupling strength between states of $p_x,p_y$ character compared to the dominantly $s,p_z$ bands, and the small difference between the excitation energies at the $\Gamma$ point ($1.61\,\rm eV$, $1.82\,\rm eV$) and at the M point ($2.12\,\rm eV$). The former results in smaller transition matrix elements, thereby a reduced contribution to the lifetime, whilst the latter indicates that the overlap between resonances makes it difficult to separate them. The latter argument is supported by the highly asymmetric shape of the resonance compared to the inverse lifetime of silicene.

\section{Two-phonon Raman spectra}

\begin{figure*}
\includegraphics[scale=.6]{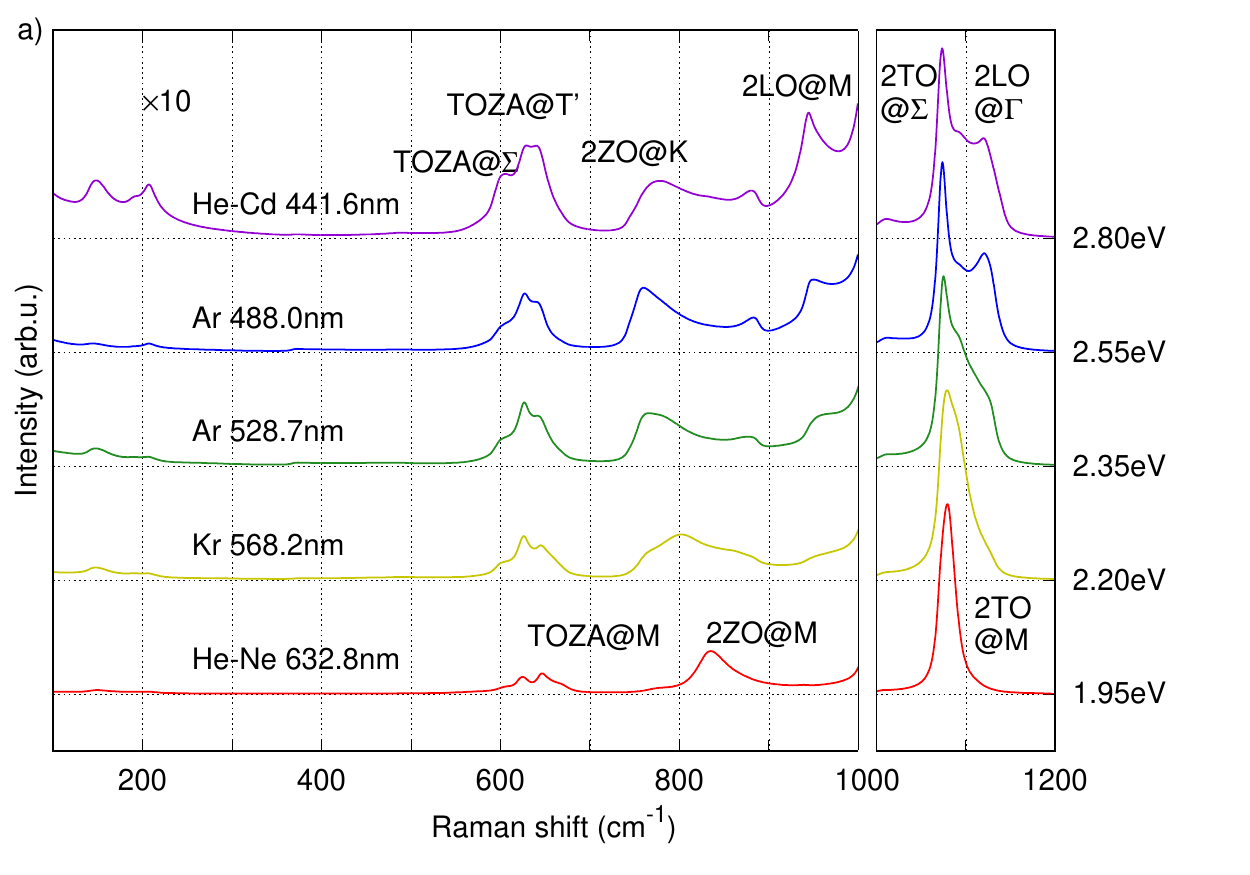}
\includegraphics[scale=.6]{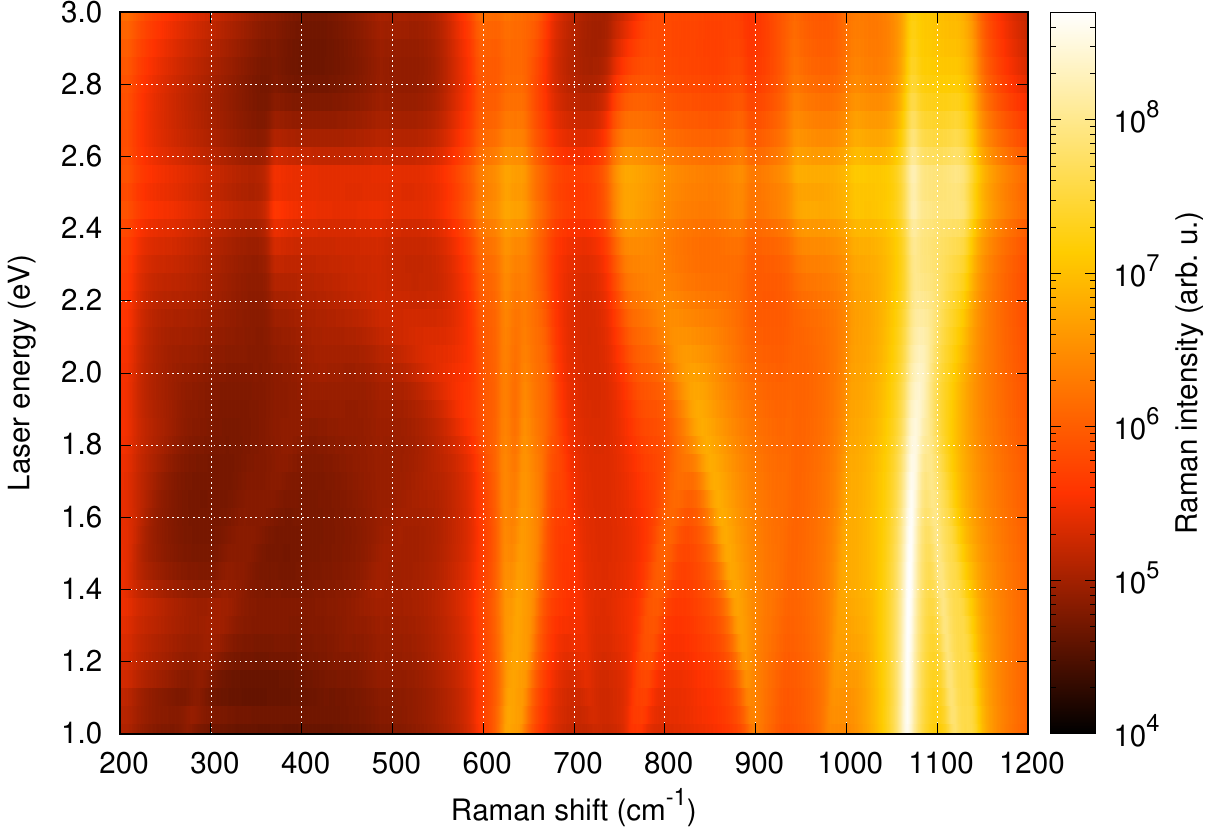}$\quad$\\
\includegraphics[scale=.6]{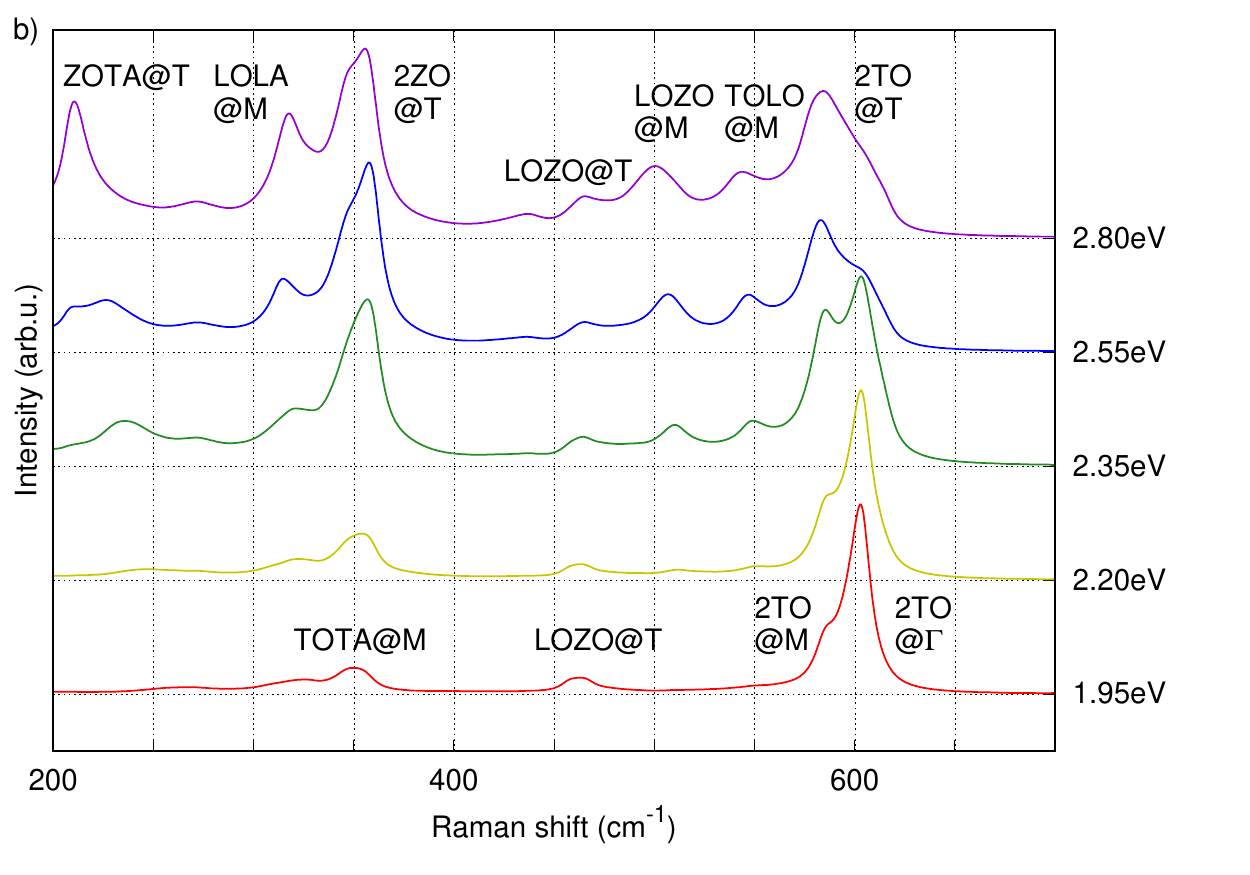}
\includegraphics[scale=.6]{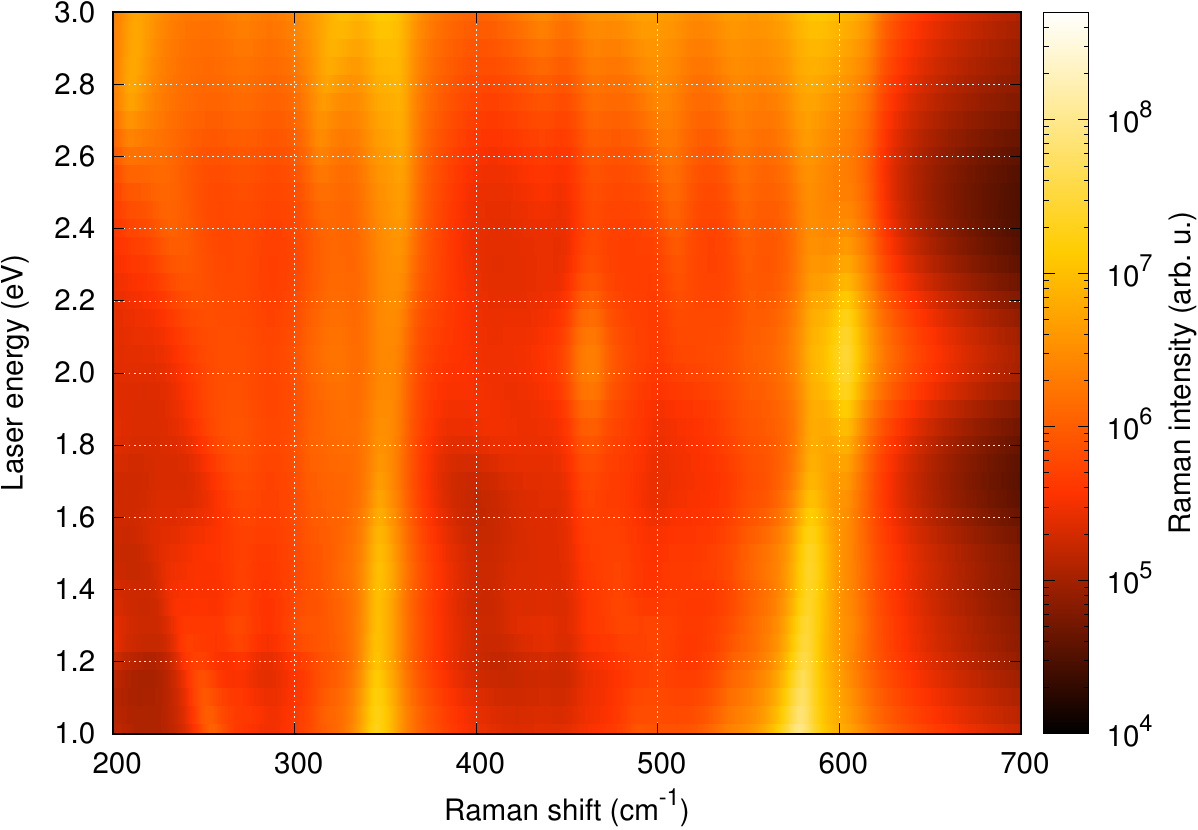}$\quad$\\
\caption{Two-phonon Raman spectra of silicene (a) and germanene (b): Characteristic spectra at a few widely used excitation laser energies (left) and the full resonance profile (right)}
\label{2phon_Si_Ge}
\end{figure*}
Apart from the plasmon excitations, resonance condition occurs only in a small area of the Brillouin zone, therefore, a dense $\mathbf{k}$ point grid is needed to calculate the Raman spectra. In our model we achieve convergence with integration over a $360\times 360\times 1$ Monkhorst-Pack grid in the electronic $\mathbf{k}$ space (virtual states) and a $180\times 180\times 1$ Monkhorst-Pack grid in the phonon $\mathbf{k}$ space (final states).

Recently, a comparative study of two phonon Raman processes in graphene and silicene was published \cite{popov_comparative_2016}, based on a non-orthogonal tight-binding model \cite{popov_comparative_2004} to calculate the electronic and vibrational properties. This approach utilized tight-binding parameters based on previous studies on silicon \cite{frauenheim_density-functional-based_1995} and carbon \cite{porezag_construction_1995} dimers which can describe the low energy electronic properties accurately, but the model neglects contributions from the second lowest conduction band at the M point, and there are missing features in the energy range of visible light ($1.8\,{\rm eV}$-$3\,\rm eV$). In comparison, using our model, which is parametrized from first principles density functional theory and takes the second conduction band at the M point into account, we obtain similar but different results.

In Fig. \ref{2phon_Si_Ge}a we present the normalized two-phonon Raman spectra of silicene at commonly used laser excitation energies. For better visibility of the low energy ($<1000\,{\rm cm}^{-1}$) peaks, their region is enhanced by a factor of 10. At lower excitation energies (1eV-2eV), electron-hole excitations can occur on the Dirac cone near the K points, resulting in phonons that originate from the vicinity of the K point when charge carriers are scattered between neighbouring Dirac cones (inter-valley processes) and from the $\Gamma$ point when scattering occurs within the same Dirac cone (intra-valley processes). As shown in the right panel of Fig. \ref{2phon_Si_Ge}a, in this region the spectrum is dominated by two peaks around 1100cm$^{-1}$: the 2TO peak with phonons originating from the K point ($\approx 1080$cm$^{-1}$) and the 2LO peak with phonons originating from the $\Gamma$ point ($\approx 1120$cm$^{-1}$) (these bands are referred as 2D and 2D' respectively in the literature of resonant Raman scattering in graphene). By increasing the excitation energy, location of electron-hole pairs on $\mathbf{k}$-space shifts towards the M point, which pushes the phonon wave vector towards the M point as well, and consequently merges the 2TO and 2LO peaks as can be seen on the lowest energy spectrum in Fig. \ref{2phon_Si_Ge}a. At the first plasmon energy electron-hole pairs can be excited from a wide area of the Brillouin zone, resulting in activation of multiple peaks from various regions.

These results are consistent with the findings of Ref. \cite{popov_comparative_2016}, however, at larger excitation energies our spectra differ from those in Ref. \cite{popov_comparative_2016}. The difference is due to the $\pi-\sigma$ plasmon, which only appears in our calculations as the model in Ref. \cite{popov_comparative_2016} neglects the higher energy conduction band and does not take the M point into account. Similarly to the excitations involving the Dirac cones at low excitation energies, the $\pi-\sigma$ plasmon introduces charge carrier scattering between states belonging to the same M point (intra-valley) and between different M points (inter-valley). Intra-valley scattering at the M point yields phonons originating from the $\Gamma$ point, while inter-valley M-point scattering results in phonons originating from the vicinity of the K point, which can be seen in the spectrum with the largest excitation energy in Fig. \ref{2phon_Si_Ge}a.

\begin{figure*}
\includegraphics[scale=.33]{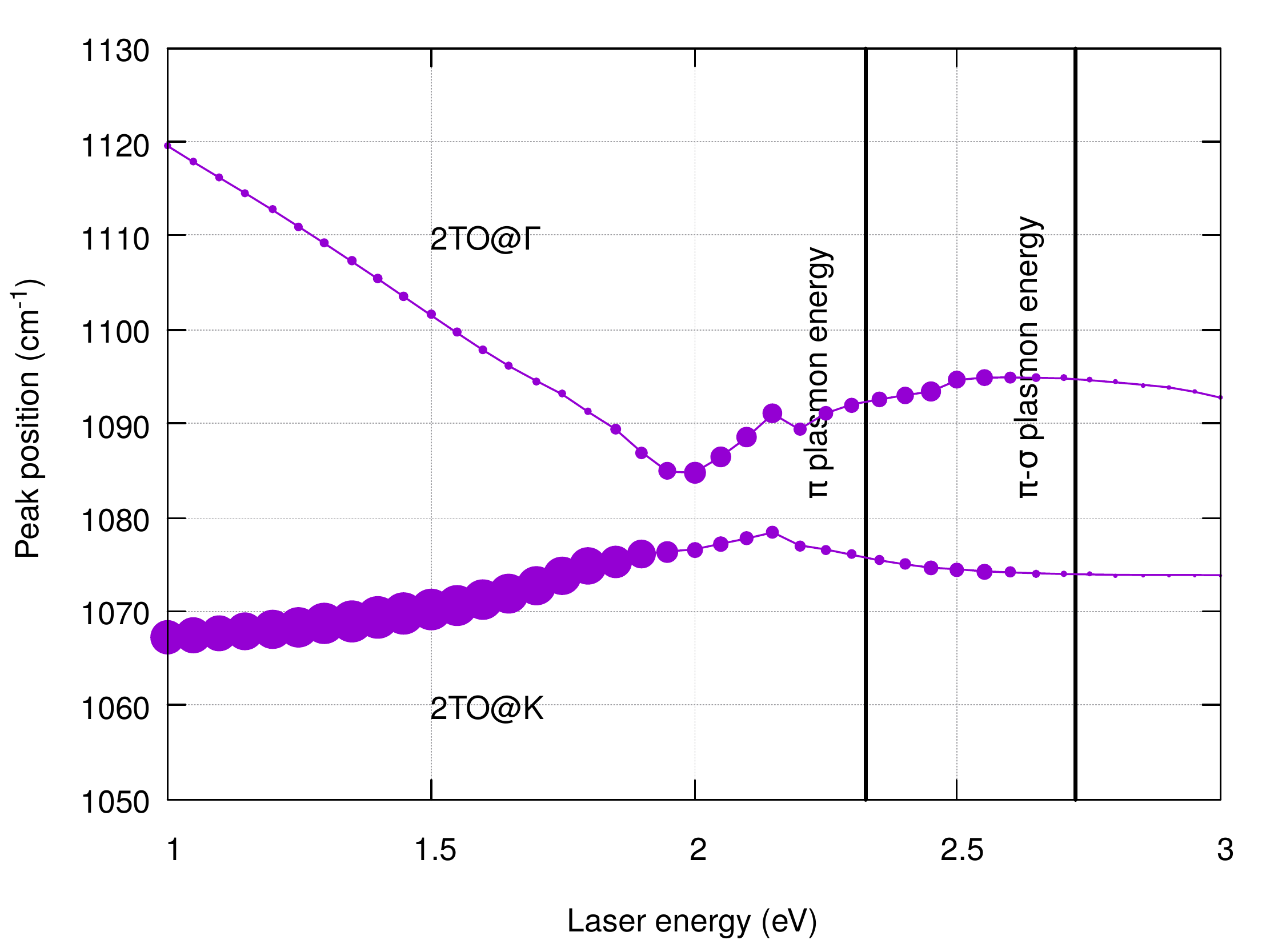}
\includegraphics[scale=.33]{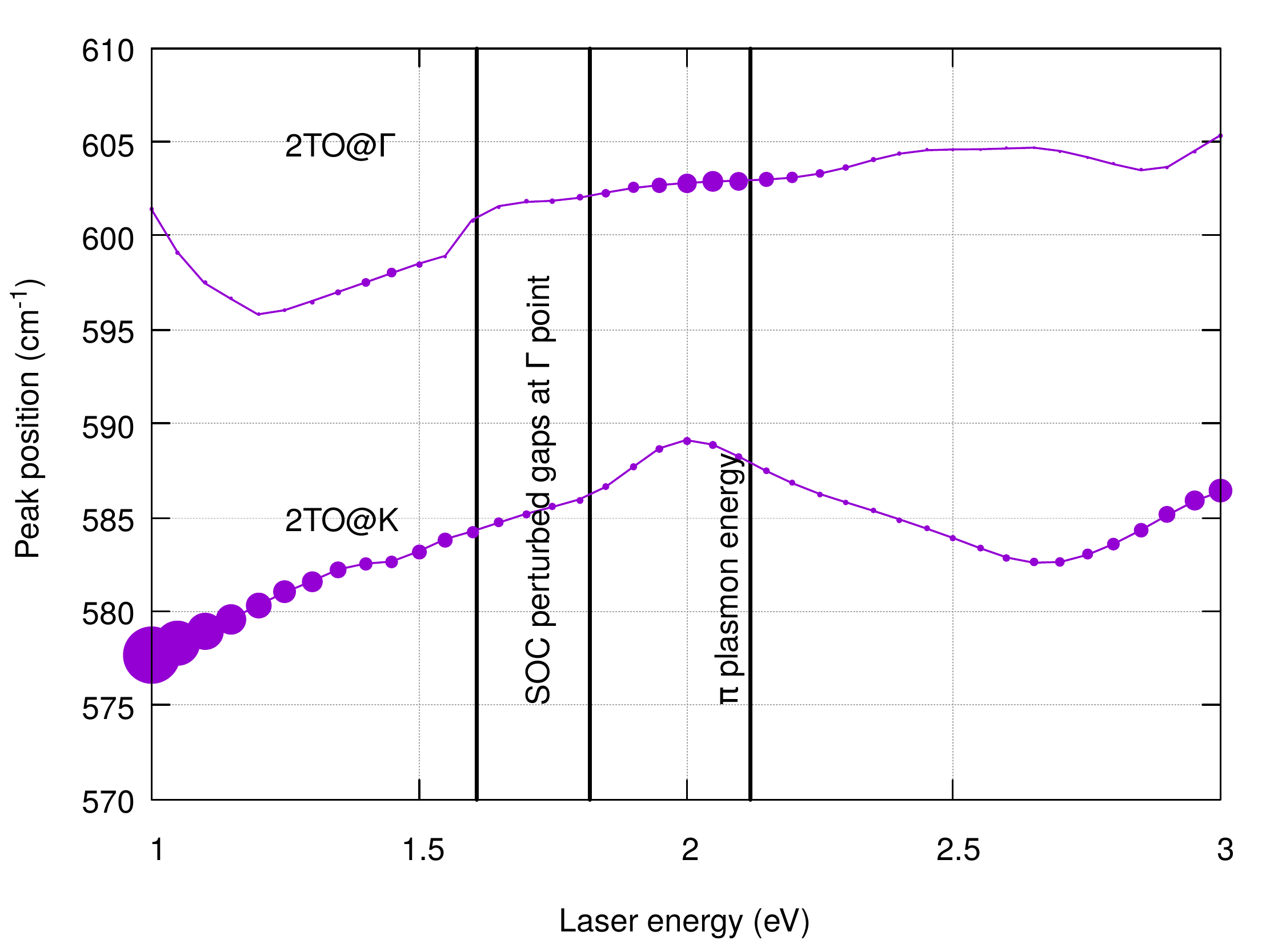}$\quad$\\
\caption{Dispersion of the main in-plane peaks of silicene (left) and germanene (right), point size indicates the intensity of peaks}
\label{2phon_Si_Ge_2}
\end{figure*}

The intensity and dispersion of the two main peaks are shown in Fig. \ref{2phon_Si_Ge_2}. Below the excitation energy of $2$ eV both peaks exhibit a linear dispersion similar to that of graphene. Around the $\pi$ plasmon energy the peaks converge on one another, making them difficult to separate between $2-2.3$ eV. Above this energy range but below the $\pi-\sigma$ plasmon energy both peaks exhibit and increase in Raman shift within the range of $2.4-2.6$ eV. Both peaks become approximately non-dispersive past the $\pi-\sigma$ plasmon energy, as expected from Fig. \ref{Si_Ge_phonon}.

Due to resonance effects numerous peaks are also activated in the low energy region. Peaks visible at the lowest excitation energy can be identified as 2ZO bands resulted from both inter- and intra-valley scattering. The position and origin of these peaks agrees with the findings in Ref. \cite{popov_comparative_2016}. However, our model predicts quite different relative intensities for these peaks. One reason for the difference is that the $\pi-\sigma$ plasmon, which is missing from the model in Ref. \cite{popov_comparative_2016}, enhances these peaks and also introduces new peaks assigned as combination peaks of the ZO and in-plane bands as depicted in the largest excitation energy spectrum in Fig. \ref{2phon_Si_Ge}. The second reason why we obtain different results is that the sublattice buckling in the silicene crystal is approximately 50\% larger in Ref. \cite{popov_comparative_2016} compared to our DFT results. First principles methods have been proven to yield accurate structural parameters for germanene, matching experiments \cite{madhushankar_electronic_2017}, therefore we expect that the LDA structures used in the present work provide qualitatively accurate predictions for the Raman spectra of germanene as well as silicene, which is expected to be described to similar accuracy by the LDA. In the Raman spectra of graphene the ZO phonon bands are suppressed, which implies that reduced sublattice buckling can be responsible for the relatively smaller intensity of ZO combination peaks in our calculations. Moreover, the sublattice buckling value calculated for germanene is closer to the value used in the previous work \cite{popov_comparative_2016}, thus a closer agreement can be expected in the relative intensity ratios.

We calculate the Raman spectra of silicene and germanene for excitation energies between $1\,\rm eV$ and $3\,\rm eV$ with an energy resolution of $0.05\,\rm eV$. This is plotted on the right hand side in Fig. \ref{2phon_Si_Ge}a-b, where the colours indicate the Raman intensity on logarithmic scale. The dispersion of the peaks can be clearly seen for high and low intensity peaks as well. Resonance effects in the overall and peak intensities are captured around both plasmonic excitations, and their effect on the spectra can be distinguished. Resonance with the $\pi$ plasmon occurs when electron-hole excitations take place around the flat band at the M point. At this energy the results of intra- and inter-valley processes merge, resulting in larger linewidth and integrated intensity. At the $\pi-\sigma$ plasmon energy, other scattering processes between M points are activated, resulting in larger overall intensity. Moreover, as the flat conduction band responsible to the $\pi-\sigma$ plasmon is mostly of $s$ character, it enhances the contribution of the out-of-plane modes due to non-zero coupling between $s$ electrons and ZO modes.

During double resonant processes the peak positions usually shift at different laser energies due to the different origin of phonons dictated by the double resonant condition. In the case of graphene this shift is on the order of $100\,\mathrm{cm}^{-1}/\mathrm{eV}$ for the 2D band \cite{venezuela_theory_2011}, while in the calculated excitation profile in Fig. \ref{2phon_Si_Ge} much lower shift can be found. This is the result of the smaller dispersion of the TO phonon band shown in Fig. \ref{Si_Ge_phonon} and the overall smaller vibrational frequencies compared to graphene. The positions of the lower intensity peaks such as the 2ZO or LOZO peak exhibit larger shift in accordance with their calculated phonon dispersions in Fig. \ref{Si_Ge_phonon}. 

We show the calculated two-phonon Raman spectra of germanene in Fig. \ref{2phon_Si_Ge}b for the same excitation energies as shown for silicene. Similarly the low energy spectra are dominated by the 2TO peak, however, the largest contribution originates from the $\Gamma$ point, as both intra-valley processes and scattering between valence states located near $\Gamma$ gives contribution to scattering with near $\Gamma$ point phonons. Therefore relative contribution of inter-valley scatterings to the spectra is smaller although their presence can be captured in multiple peaks depicted in Fig. \ref{2phon_Si_Ge}b. Low intensity ZO peaks are also present, even at the lowest excitation energy in Fig. \ref{2phon_Si_Ge}b peaks composed of inter- and intra-valley scattered ZO phonons are visible. Their relative intensity ratio compared to the intensity of 2TO phonon peaks, is much closer to the ratio calculated for silicene in Ref. \cite{popov_comparative_2016}. This also confirms that the difference in buckling is responsible for the qualitative differences between our and the previous model.

Above the plasmon energy, other combination peaks become visible in the $200\,{\rm cm}^{-1}$-$500\,\rm cm^{-1}$ region. These peaks originate with no exception from the vicinity of the K point, as above the plasmon energy electron-hole excitations can occur around the M point and scattering between neighbouring M points result in emission of near K point phonons. Moreover, intensity of the ZO peaks show dramatic increase compared to the resonant spectra of silicene or graphene. The general enhancement of the out-of-plane peaks with larger buckling can be understood by taking into account the hybridization of the $\pi$ electrons. By increasing the buckling, the $\pi$ orbitals which have purely $p_z$ character will hybridize with the $s$ orbitals. The qualitative difference between $s$ and $p_z$ orbitals is that interaction between $p_z$ orbitals and out-of-plane phonons is prohibited, whilst $s$ orbitals do not distinguish between in-plane and out-of-plane modes, therefore, by increasing the $s$ character, the intensity of ZO peaks will increase.

The resonance profile of the two-phonon spectra of germanene shown in Fig. \ref{2phon_Si_Ge}b show similar trends as seen in the case of silicene. Similarly to the results of the lifetime calculation, resonance effect can only be found around the plasmon excitation energy, whilst the splitting of the bandgap by the SOC at $\Gamma$ cannot be captured.

Finally we plot the dispersion of the main two in-plane peaks in Fig. \ref{2phon_Si_Ge_2}. The dispersion of the 2TO peak originating from K point contains several linear dispersion regions, however, within the excitation energy range of 2--2.7 eV, unlike the previously investigated cases, the peak position shifts downwards with increasing laser energy. Similarly the 2TO peak originating from $\Gamma$ point exhibits linear dispersion, however the amplitude of this dispersion is much smaller compared to the previous cases. Unlike in the case of silicene the peak originating from the $\Gamma$ point shows a resonance in the intensity, whilst the peak originating from the K point does not exhibit an increase in intensity.
\section{Defect induced Raman spectra}

\begin{figure*}
\begin{center}
\includegraphics[scale=.55]{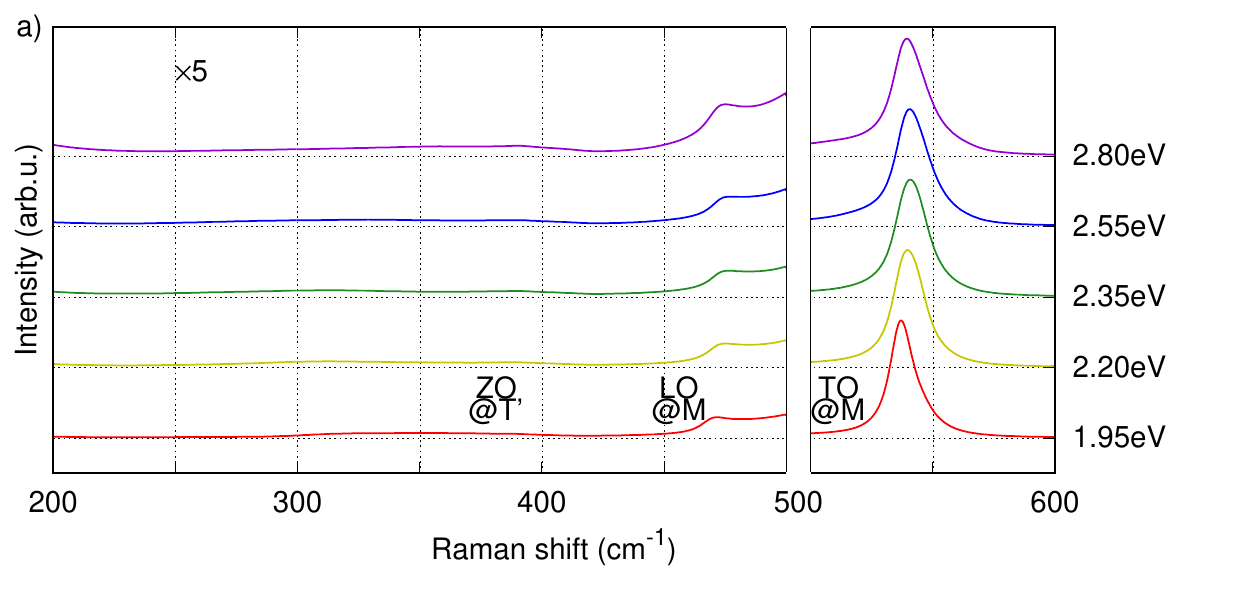}
\includegraphics[scale=.55]{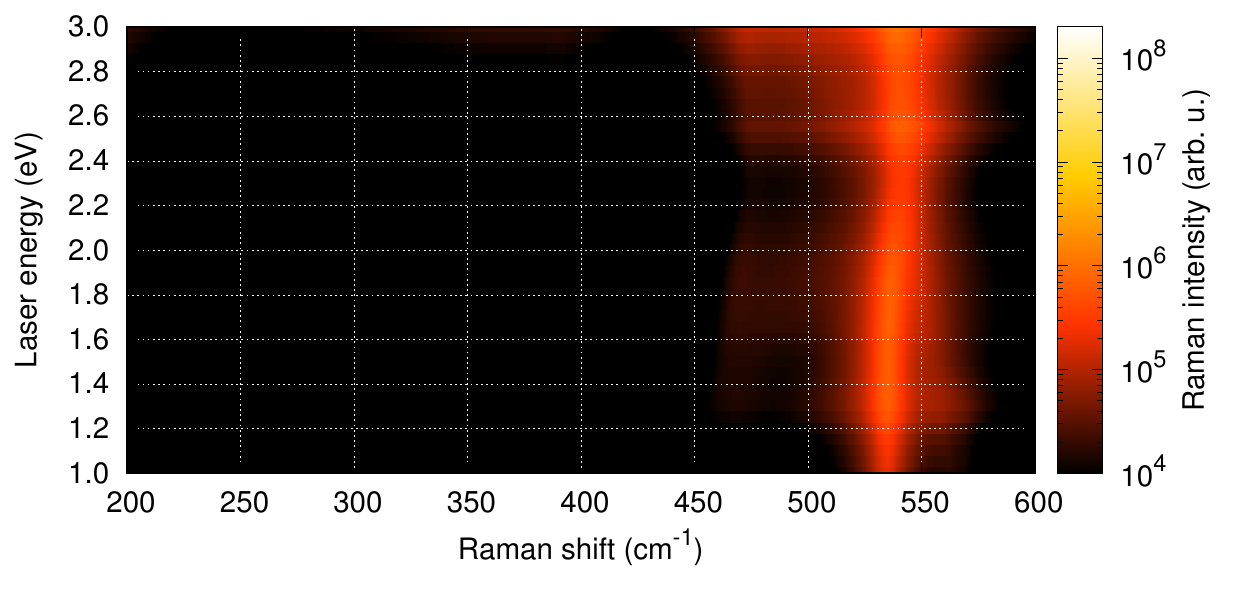}\\
\includegraphics[scale=.55]{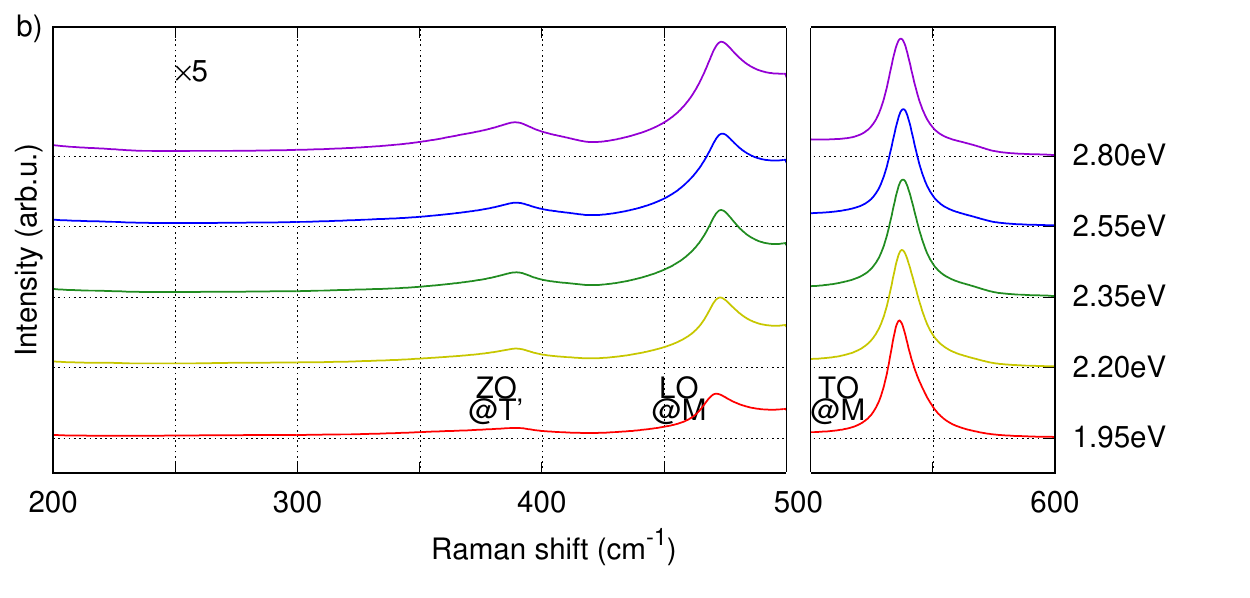}
\includegraphics[scale=.55]{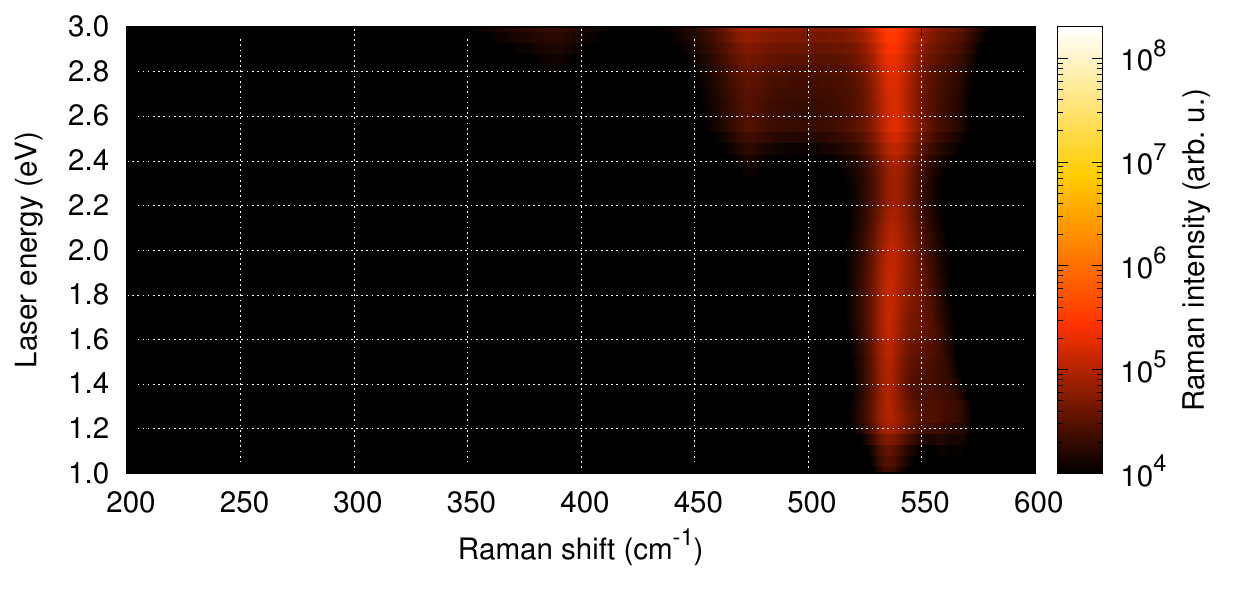}\\
\includegraphics[scale=.55]{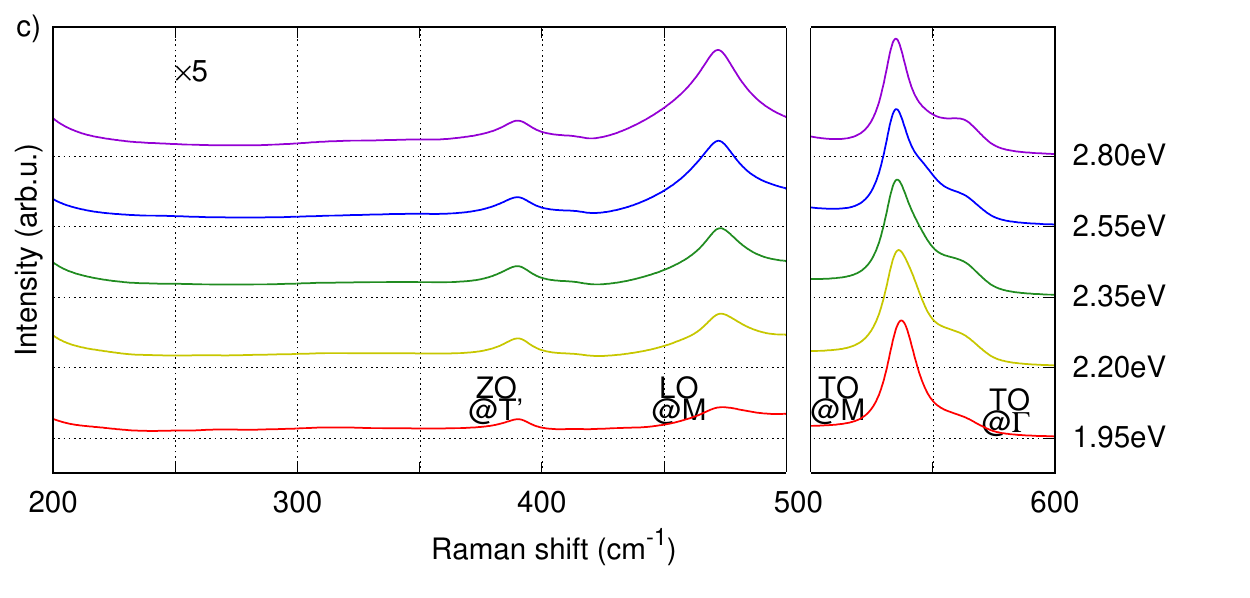}
\includegraphics[scale=.55]{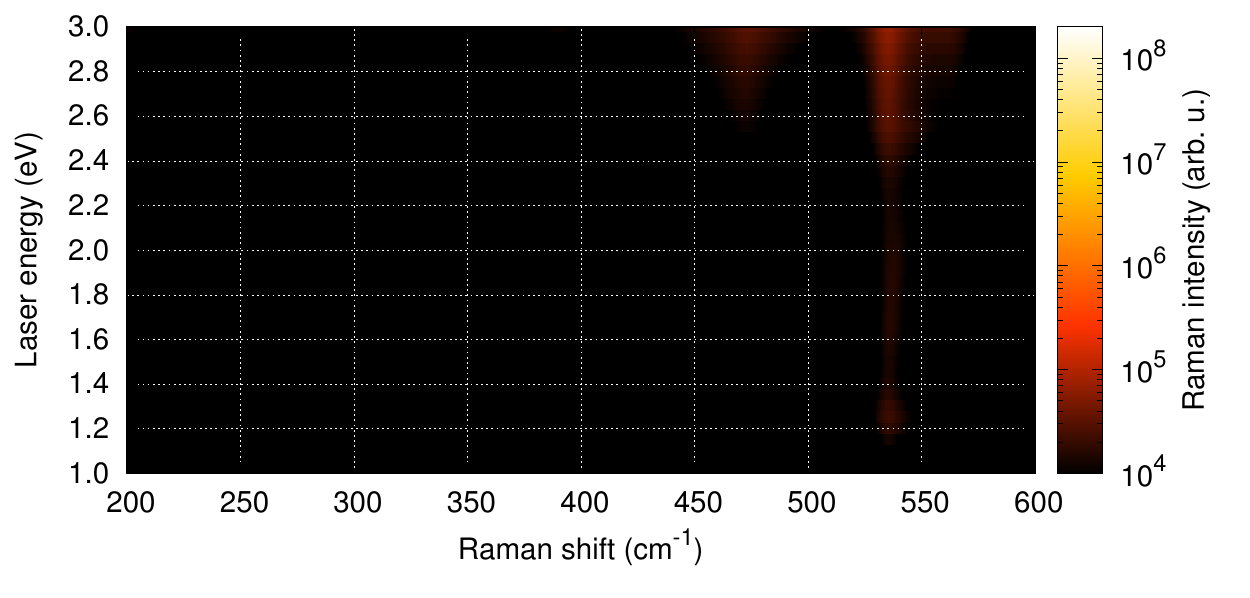}\\
\includegraphics[scale=.55]{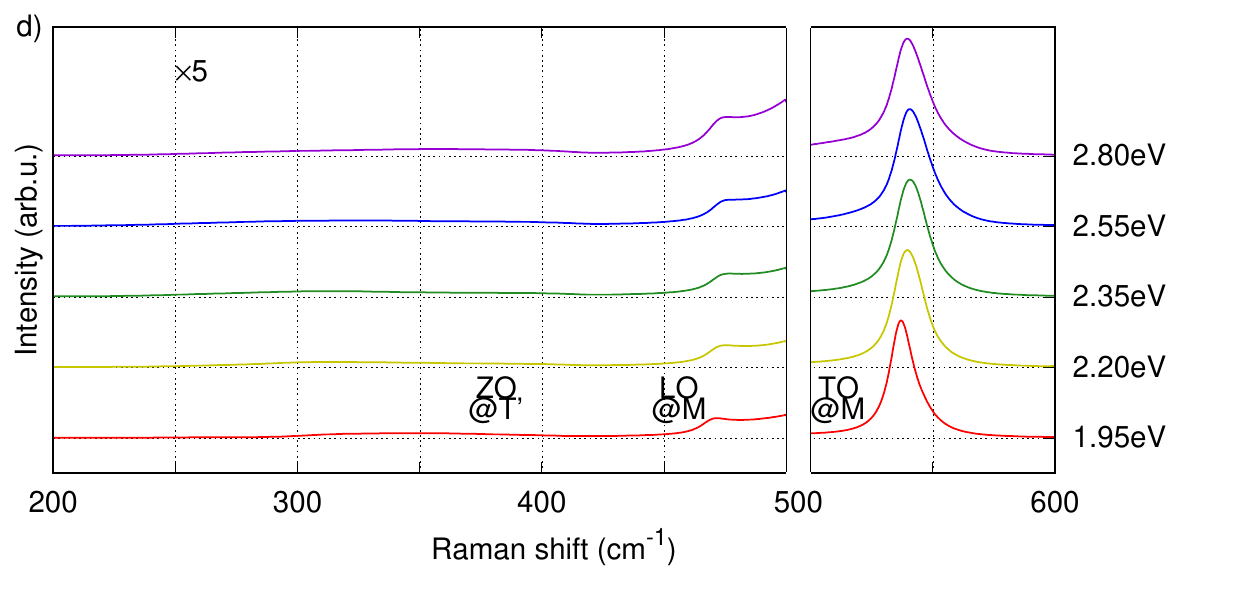}
\includegraphics[scale=.55]{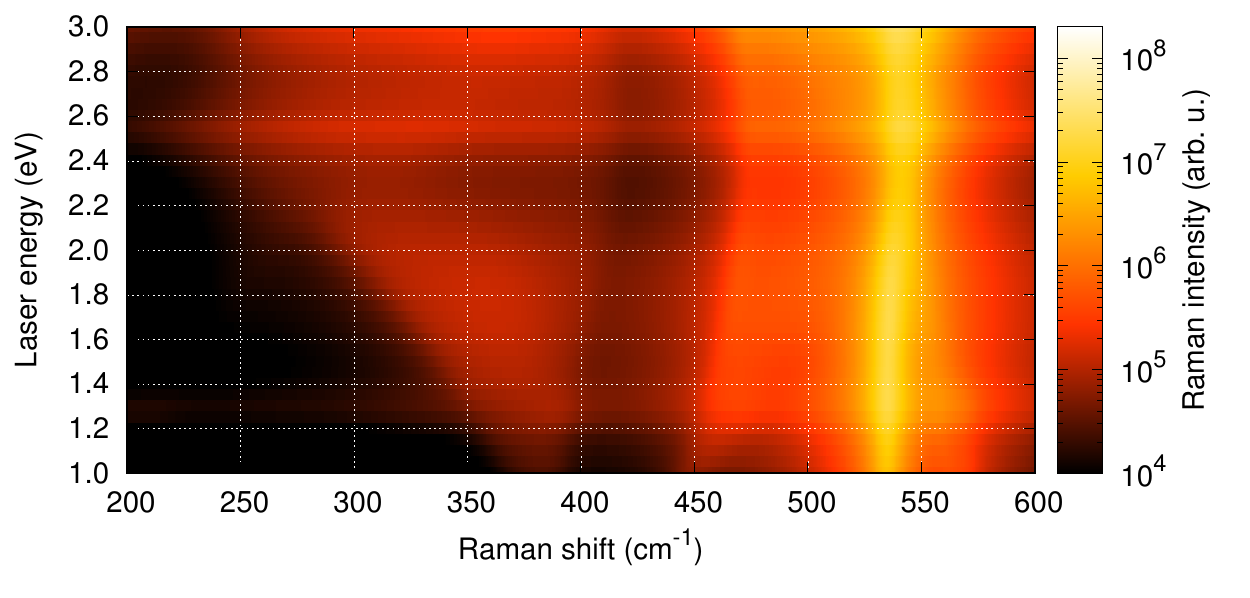}\\
\includegraphics[scale=.55]{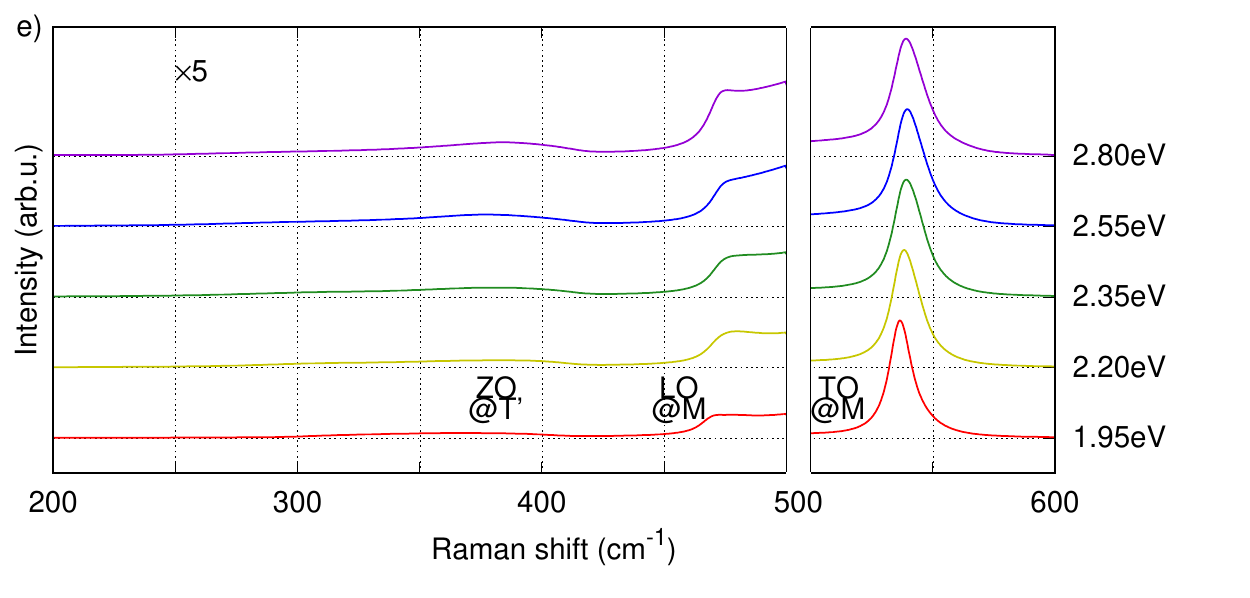}
\includegraphics[scale=.55]{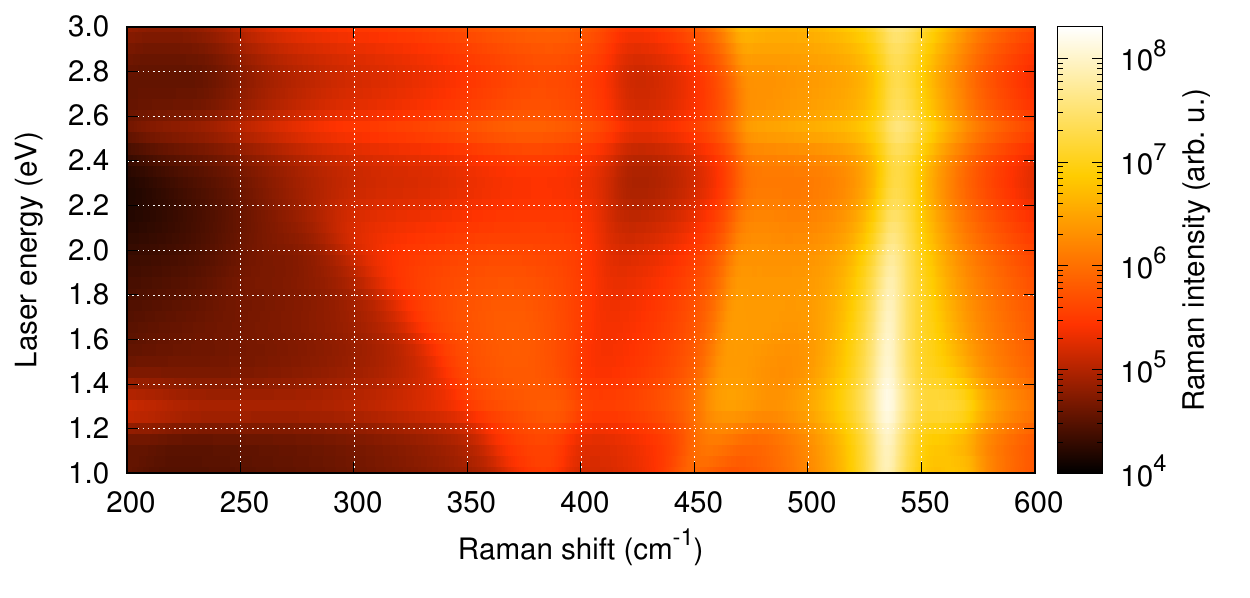}\\
\includegraphics[scale=.55]{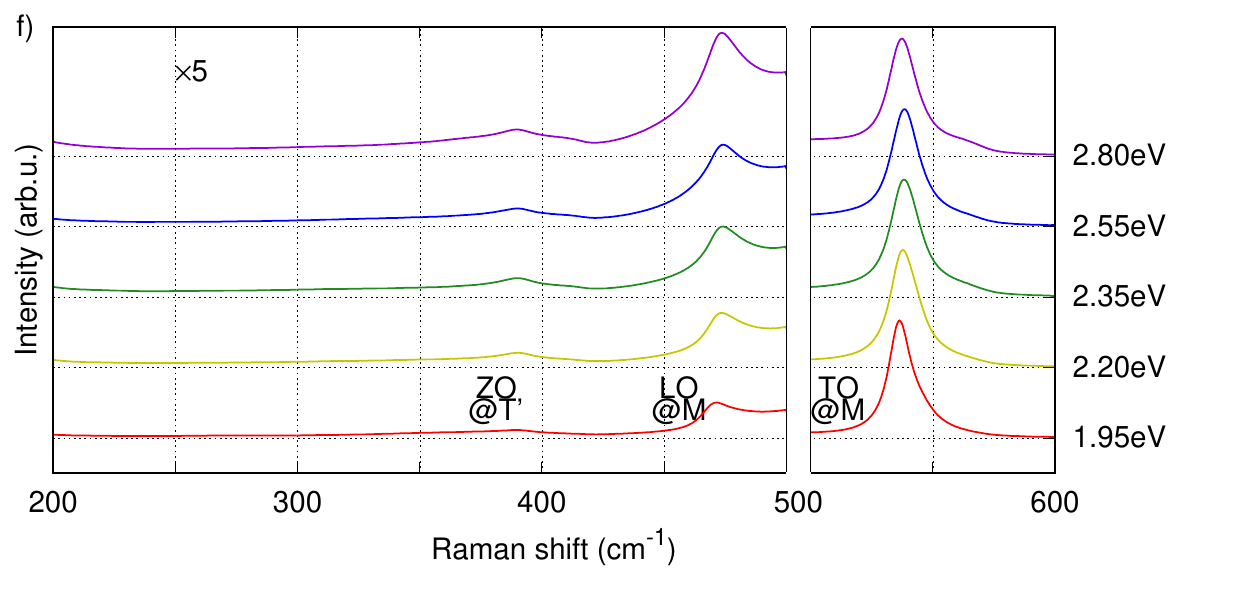}
\includegraphics[scale=.55]{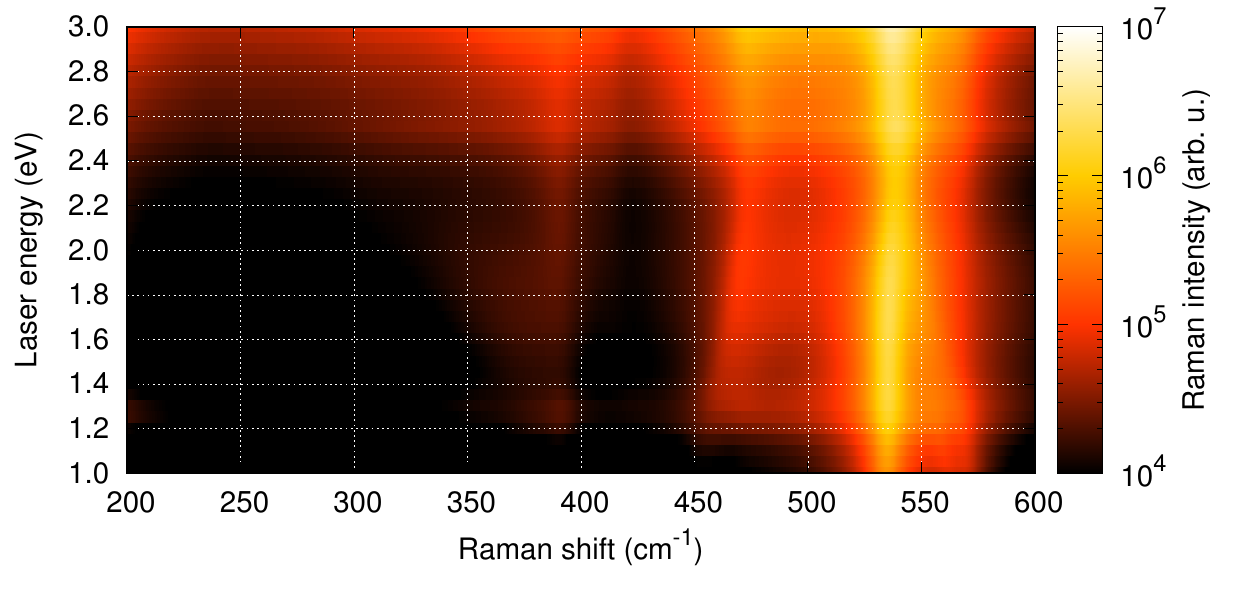}\\
\end{center}
\caption{Defect induced Raman spectra of silicene at selected laser excitation energies (left side) and excitation profiles on a logarithmic scale (right side) for different hopping scattering matrix elements: a) $t_{ss\sigma}$, b) $t_{sp\sigma}$, c) $t_{pp\sigma}$, d) $t_{pp\pi}$, e) On-site, f) Vacancy}
\label{def_hop_Si}
\end{figure*}

The calculated defect induced Raman spectra for silicene are shown in Fig. \ref{def_hop_Si}. We find a peak from ZO phonons, however, its intensity is small compared to the LO and TO peaks. In particular, Raman scattering induced by $t_{ss\sigma}$ and $t_{pp\pi}$ defects contain a wide background, rather than a distinguishable peak. In contrast, defects which perturb the $\sigma$ bonds ($t_{sp\sigma}$ and $t_{pp\sigma}$) activate this peak as shown in the left side of Fig. \ref{def_hop_Si}b-c. Moreover the $t_{pp\sigma}$ defect activates the TO peak with phonons originating from the $\Gamma$ point, this peak is only visible at this defect, therefore its presence is a clear indication of perturbations in the $t_{pp\sigma}$ matrix element.

In order to calculate comparable intensity for a given defect concentration, the magnitudes of the hopping and on-site defects are chosen to be equal. As shown in the right hand side of Fig. \ref{def_hop_Si}, the largest contribution results from $t_{pp\pi}$ defect, in accordance with the $p_z$ dominated electronic bands around the Fermi level. These bands, however, have a significant amount of $s$ character due to hybridization of $s$ and $p$ orbitals caused by the sublattice buckling, thus $t_{ss\sigma}$ and $t_{sp\sigma}$ defects can introduce similar (lower) intensities as plotted in Fig.  \ref{def_hop_Si}a-b. The lowest overall Raman intensity results from $t_{pp\sigma}$ defects, as these hopping matrix elements describe bands well below the Fermi level.

Raman spectra of on-site defects are calculated by perturbing all on-site matrix elements on a given atom equally. Results shown in Fig. \ref{def_hop_Si}e resemble the $t_{pp\pi}$ induced defects in many ways: the spectra contain the wide background arising from ZO phonon band, and the absolute intensity is also remarkably high. Due to the $p_z$ dominated electronic bands, the largest Raman intensity is achieved by perturbing these atomic states, therefore by changing the on-site energy of each state equally results in scattering dominated by the $p_z$ orbital.

\begin{figure*}
\begin{center}
\includegraphics[scale=.55]{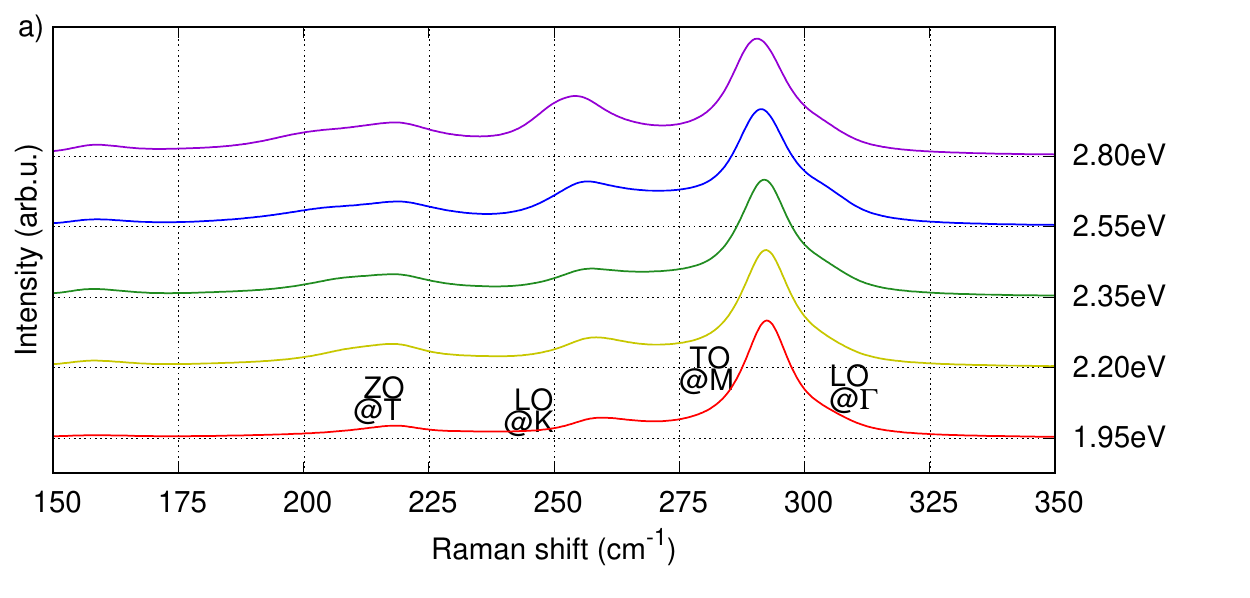}
\includegraphics[scale=.55]{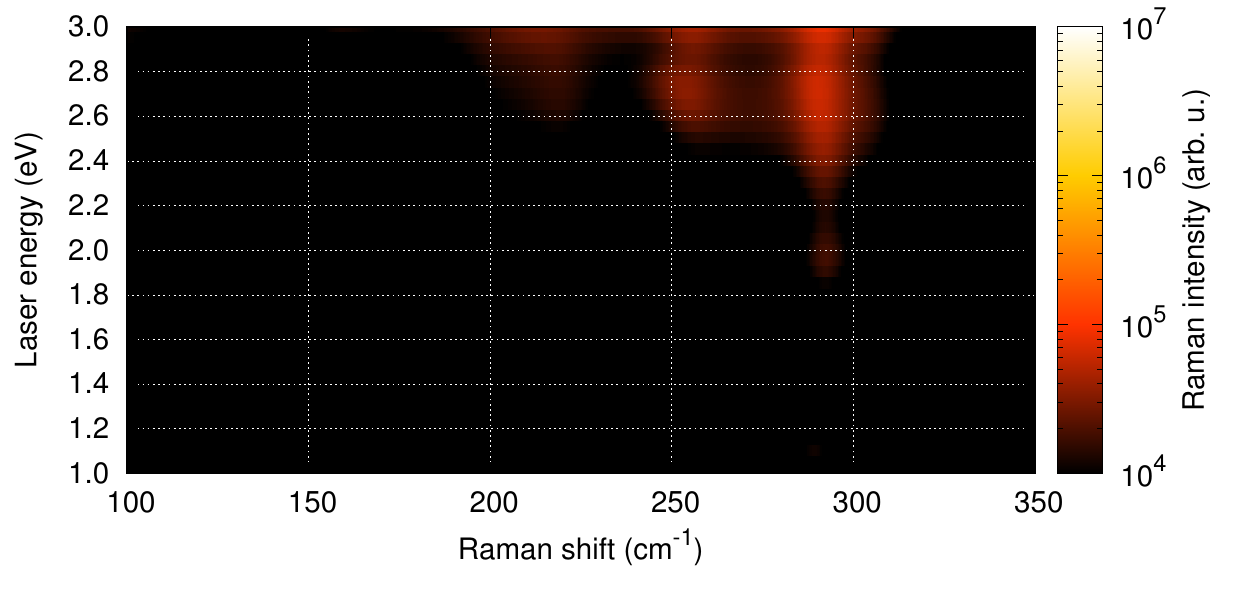}\\
\includegraphics[scale=.55]{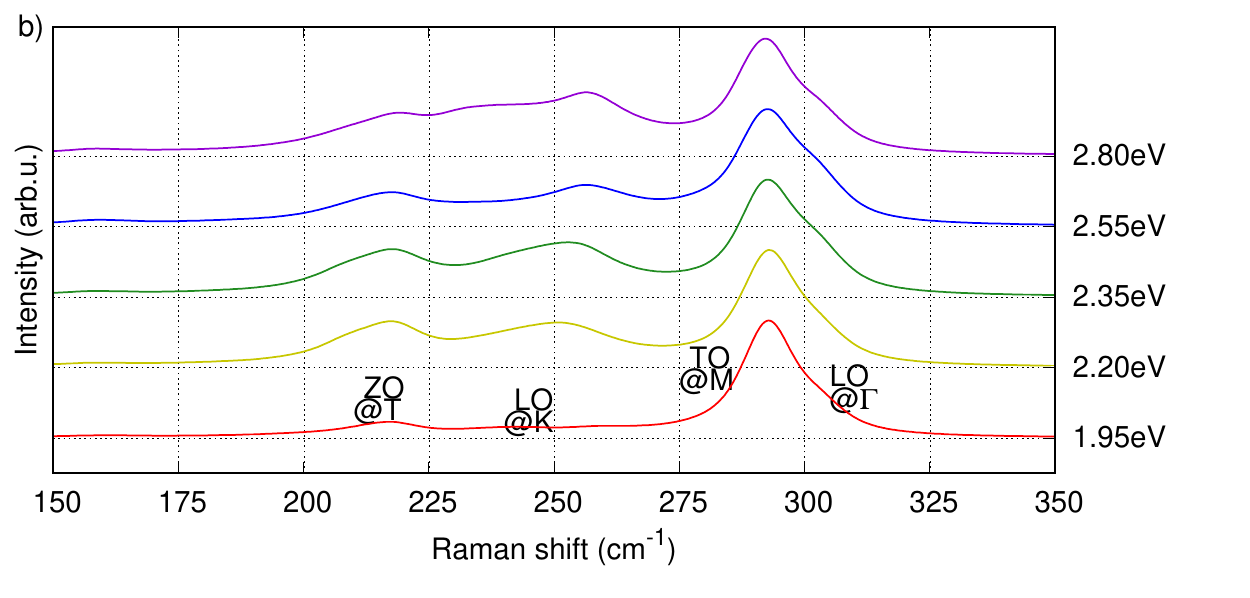}
\includegraphics[scale=.55]{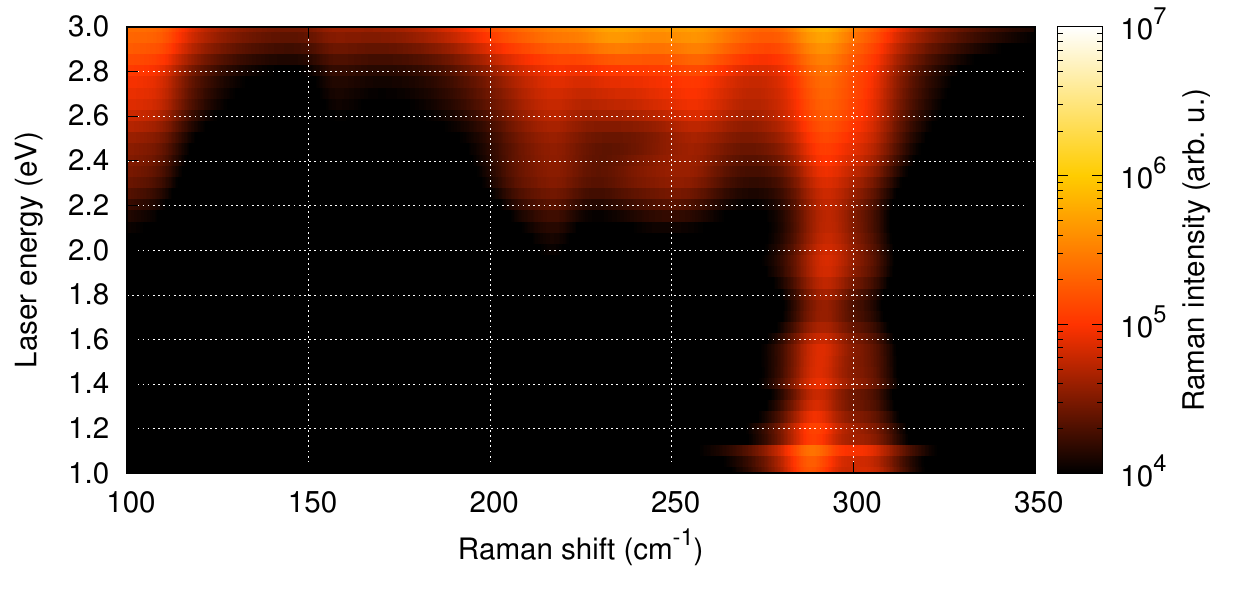}\\
\includegraphics[scale=.55]{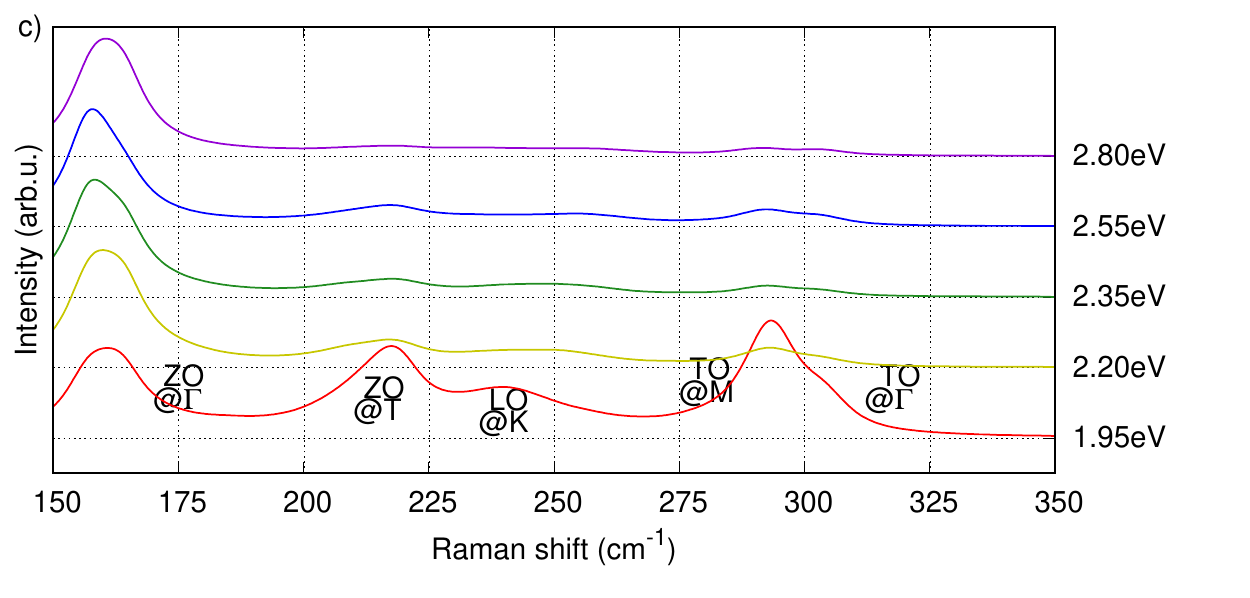}
\includegraphics[scale=.55]{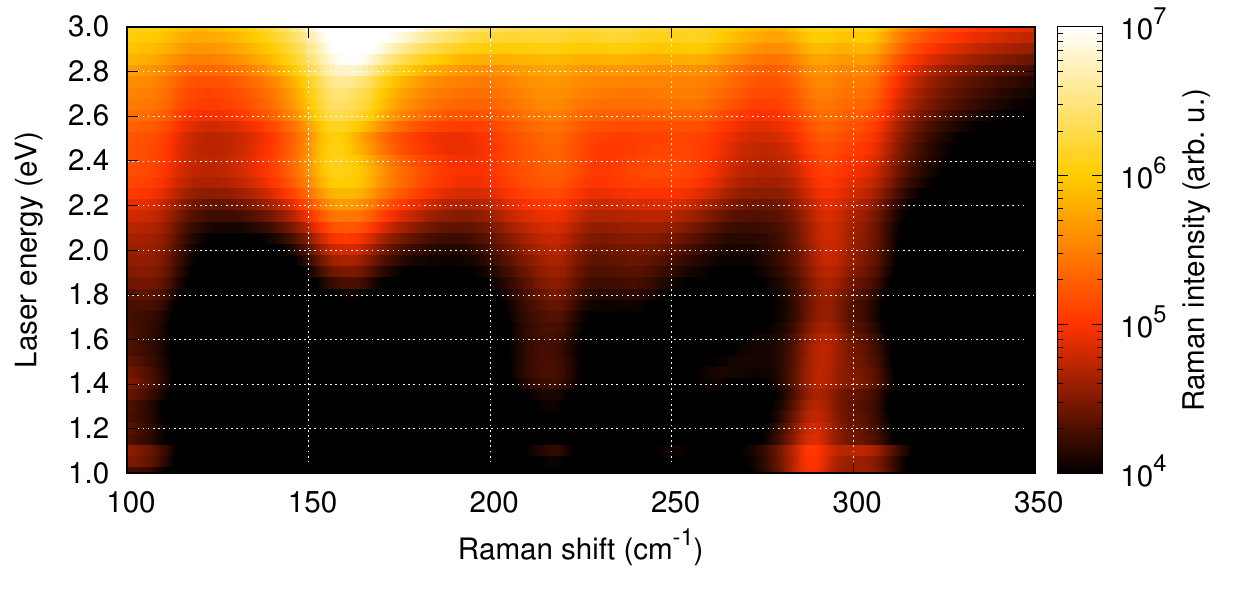}\\
\includegraphics[scale=.55]{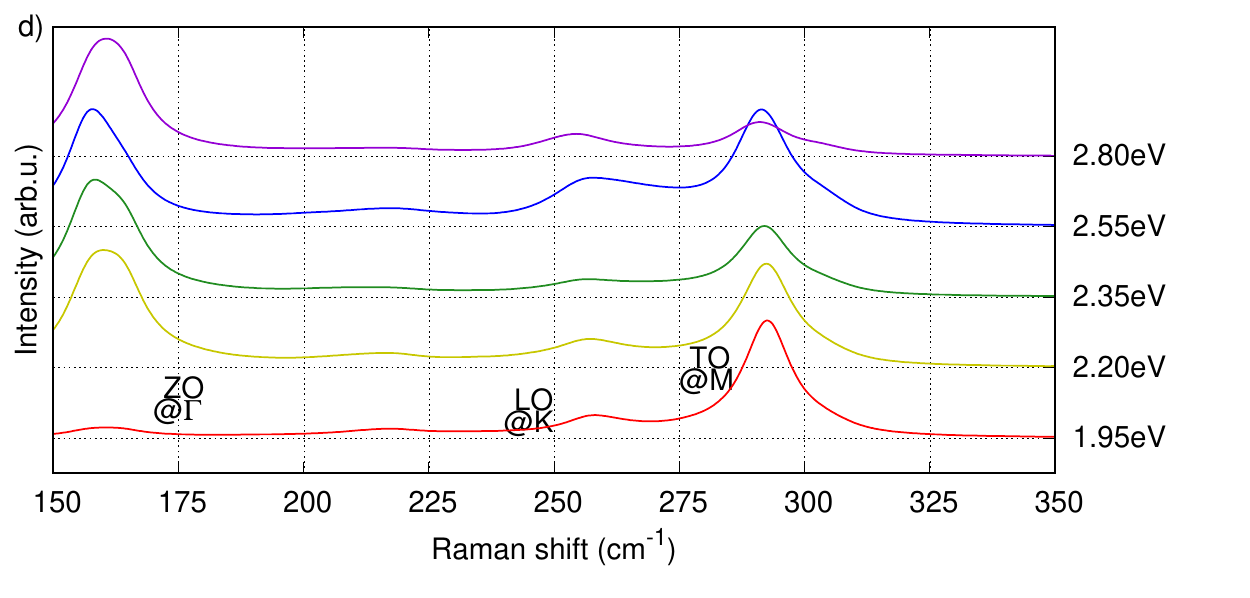}
\includegraphics[scale=.55]{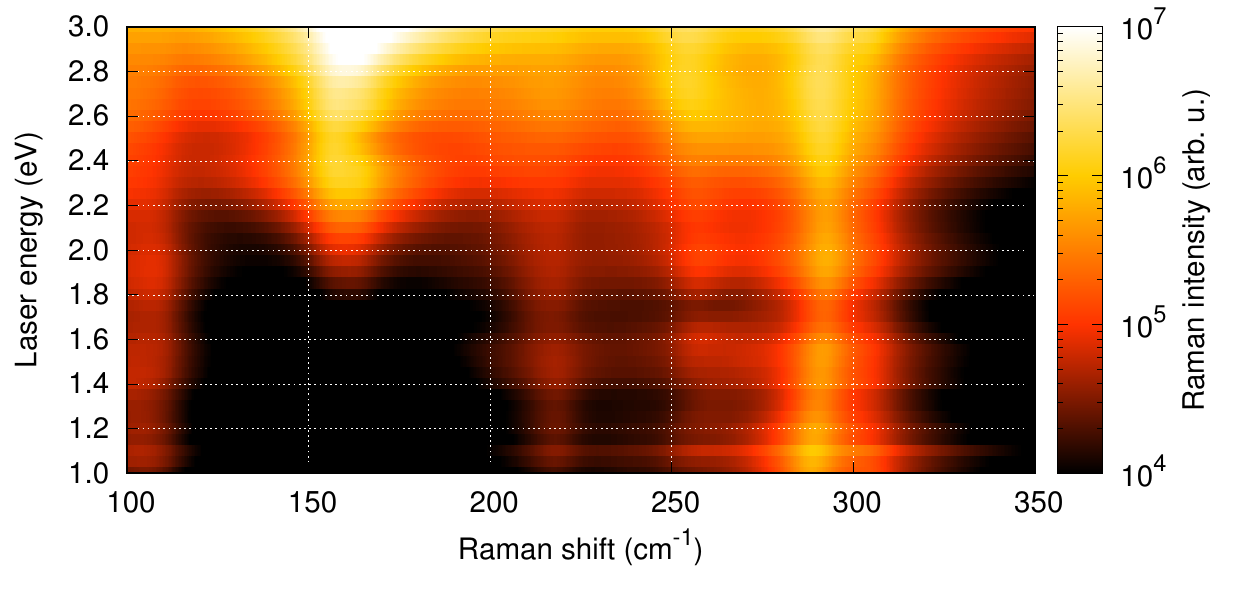}\\
\includegraphics[scale=.55]{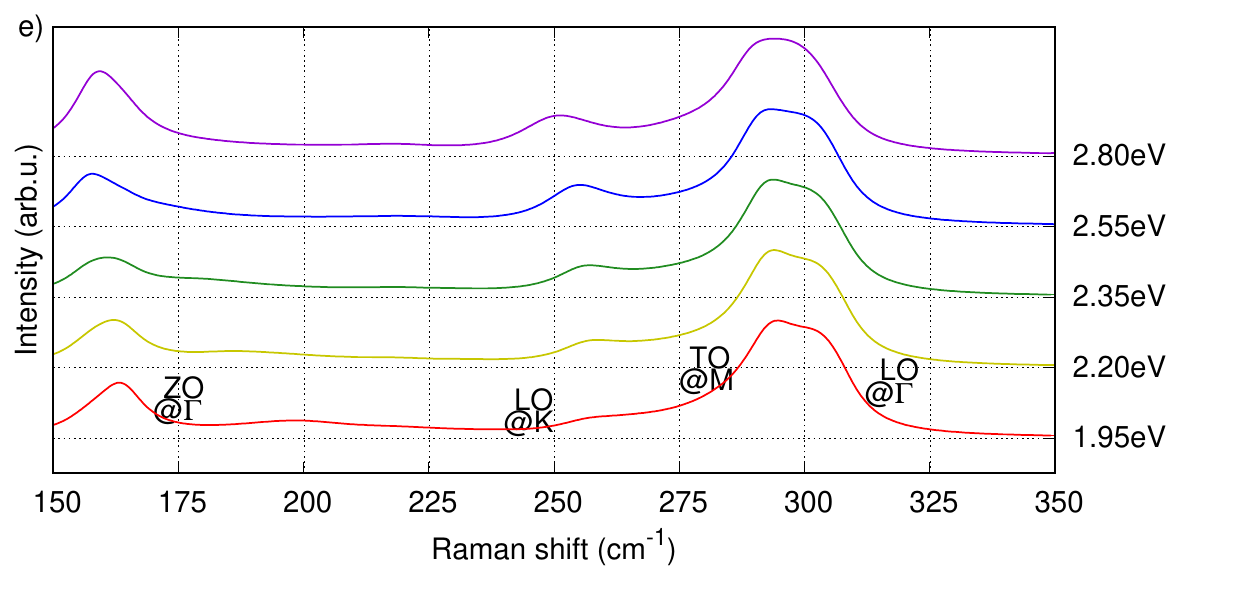}
\includegraphics[scale=.55]{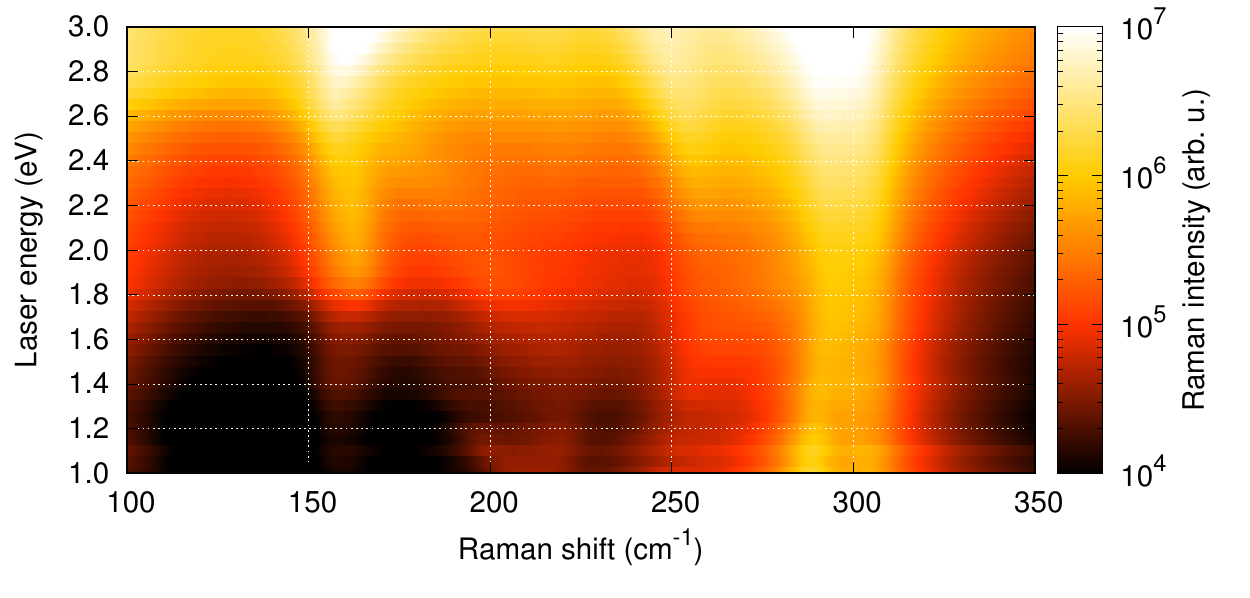}\\
\includegraphics[scale=.55]{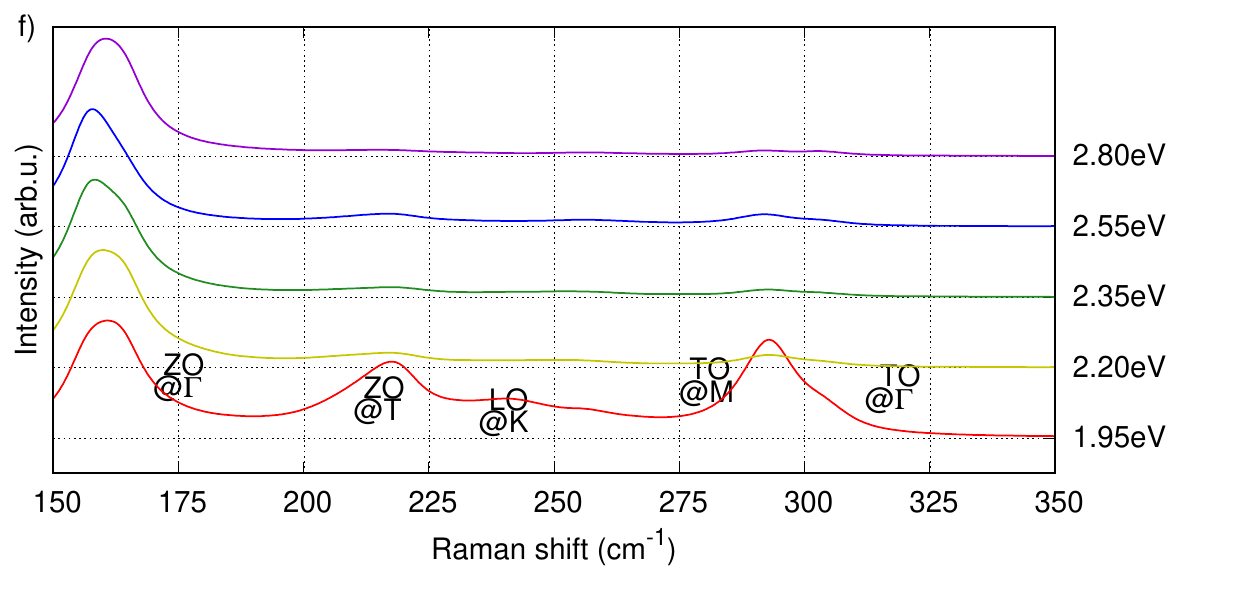}
\includegraphics[scale=.55]{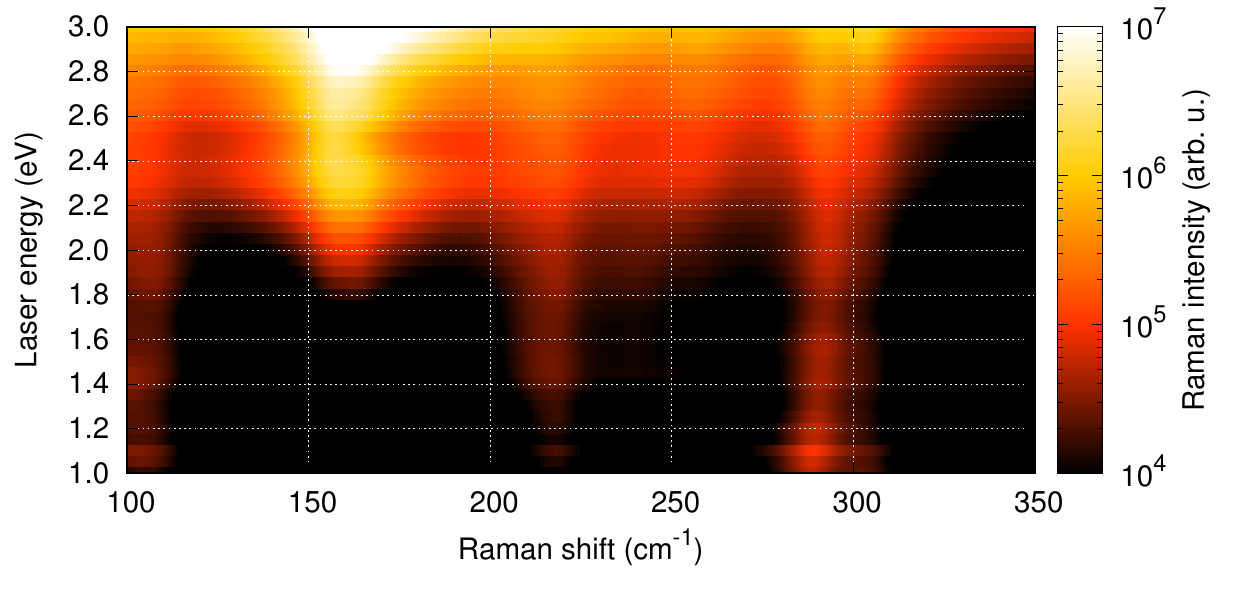}\\
\caption{Defect induced Raman spectra of germanene at selected laser excitation energies (left side) and excitation profiles on a logarithmic scale (right side) for different hopping scattering matrix elements: a) $t_{ss\sigma}$, b) $t_{sp\sigma}$, c) $t_{pp\sigma}$, d) $t_{pp\pi}$, e) On-site, f) Vacancy}
\label{def_hop_Ge}
\end{center}
\end{figure*}

Vacancy induced spectra combine the effects of multiple hopping defects and the on-site defect: they are still dominated by the TO peak, with a somewhat distinguishable ZO peak, and at larger laser energies the TO peak originating from the $\Gamma$ point is visible as a shoulder. Particullarly, spectra of samples with large number of vacancies can be found in the literature\cite{jung_situ_2018}, thus comparison with experimental results can be made. In Ref. \citenum{jung_situ_2018} several features are identified as result of two phonon or defect induced peaks. In the measured spectra the intense peak of defect-free germanene is accompanied by a wide shoulder, which can be identified as the TO@M peak in Fig. \ref{def_hop_Ge}f. Moreover the ZO@T peak can also be seen around $200\,\textrm{cm}^{-1}$ as well as the previously described 2TO@$\Gamma$ peak around $600\,\textrm{cm}^{-1}$.

Resonance profiles of these defects can provide guidelines to distinguish between them. Defects perturbing bonds including $t_{sp\sigma}$, $t_{pp\sigma}$ and the vacancy show enhanced intensity above the $\pi-\sigma$ plasmon energy. On the contrary defects perturbing $p_z$ orbitals exhibit a resonance near this energy, but no enhancement effect can be seen.

Defect induced spectra of germanene with the aforementioned defect scattering matrix elements are presented in Fig. \ref{def_hop_Ge}. In general, the defect induced spectra of germanene show more features compared to silicene, similarly to the two phonon spectra. Due to the multiple activated peaks in the spectra, distinguishing different scatterers is easier. Results of hopping scatterers presented in Fig. \ref{def_hop_Ge} show that $t_{ss\sigma}$ and $t_{sp\sigma}$ defects do not activate the ZO peak originating from the $\Gamma$ point, but several peaks can be seen originating from the M point. These peaks are only present above the plasmon excitation energy, as presented in the right hand side of Fig. \ref{def_hop_Ge}a-b. Although at the same point significant enhancement of TO peak can be caught, the absolute intensity of these defects does not reach the intensity of the other defects. In the case of $t_{pp\sigma}$ and $t_{pp\pi}$, enhancement of the ZO peak originating from the $\Gamma$ point is shown in Fig. \ref{def_hop_Ge}c-d. As the spectra on the left hand side are normalized to the largest peak intensity, apparently the TO peak loses its intensity in the case of the $t_{pp\sigma}$ defect. From the examination of the excitation profile on the right hand side, it is evident that even though the TO peak is enhanced, the enhancement factor is relatively small.

On-site defect induced spectra in Fig. \ref{def_hop_Ge}e show similar features to $t_{pp\pi}$, however, the LO peak intensity is remarkably higher. Spectra of vacancy induced scattering shown in Fig. \ref{def_hop_Ge}f resemble the $t_{pp\sigma}$ defect, although the ZO intensity is even larger.

\begin{figure*}
\begin{center}
\includegraphics[scale=.6]{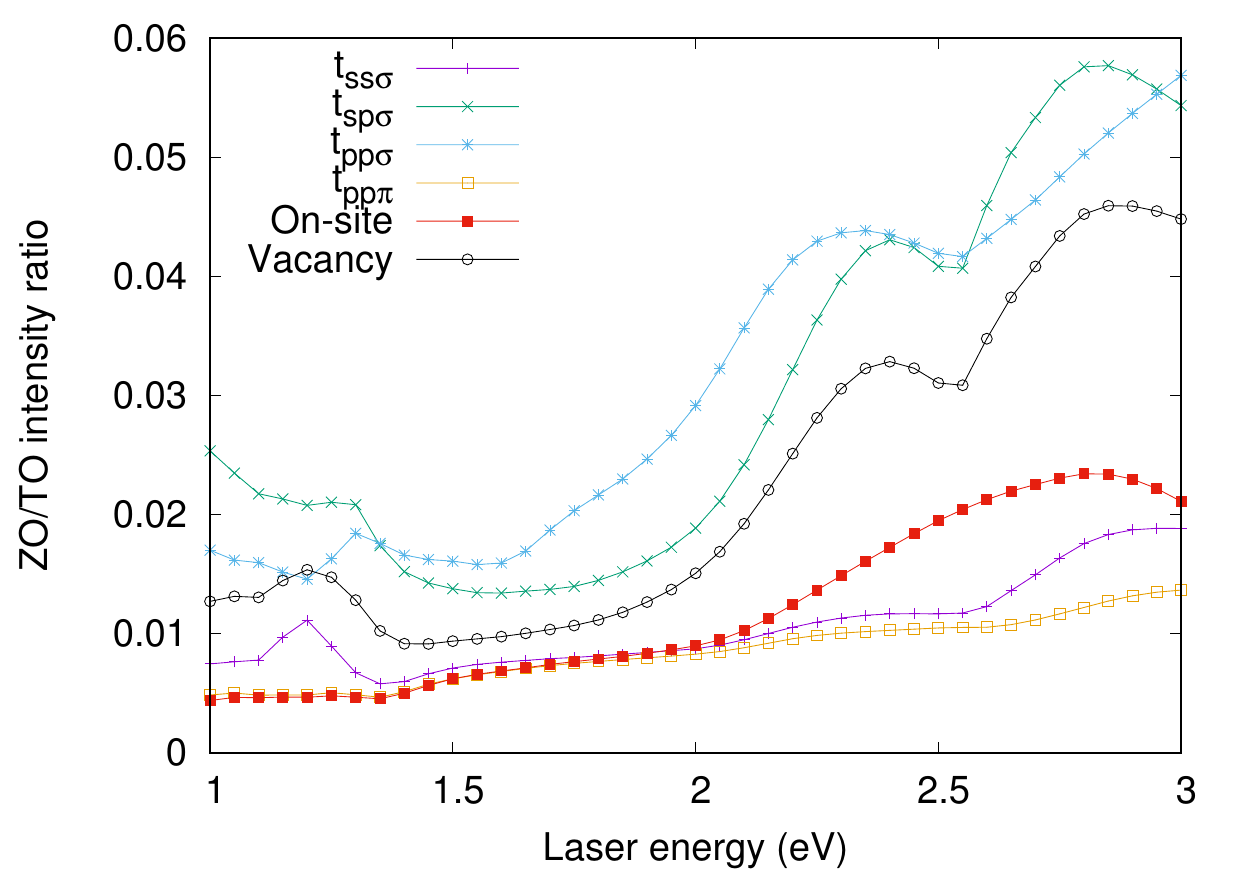}
\includegraphics[scale=.6]{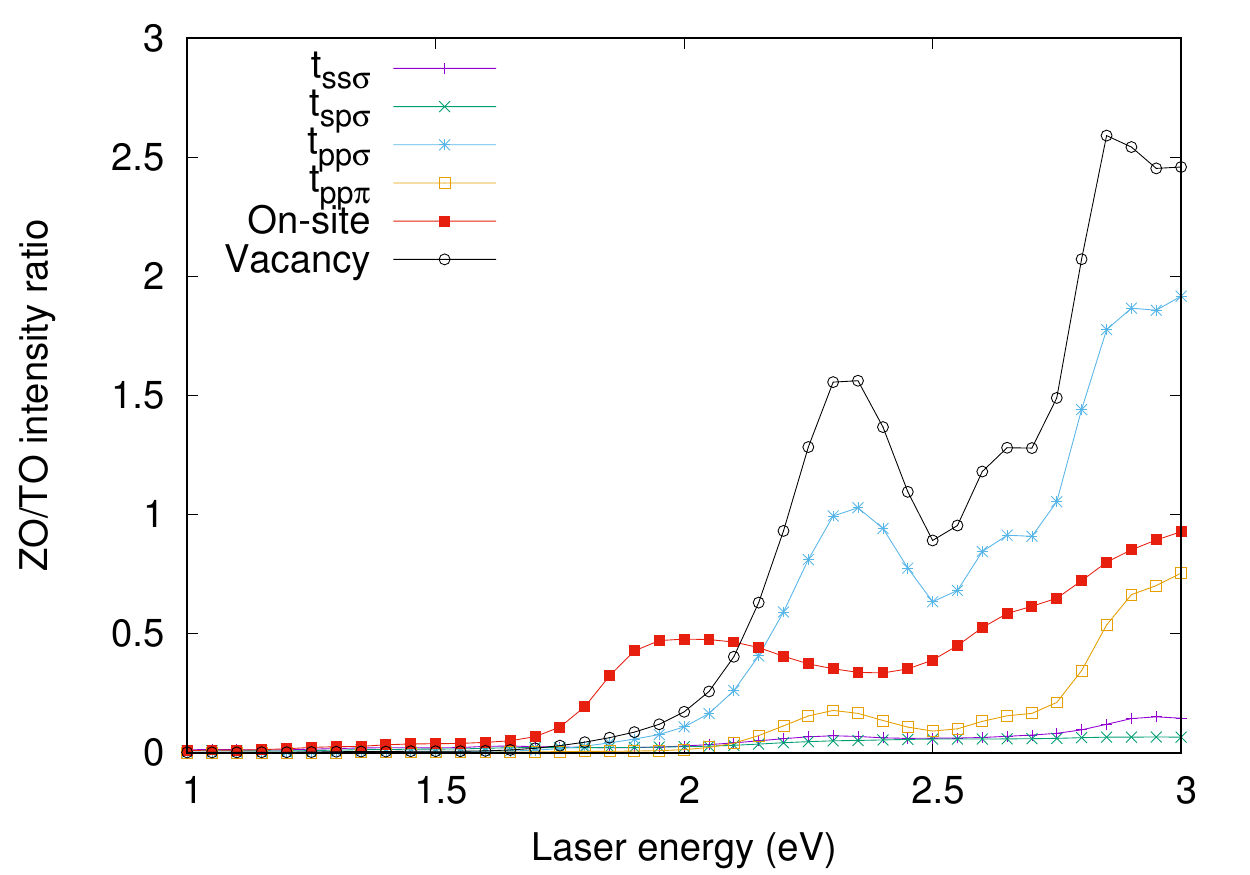}
\caption{Comparison of the ZO/TO intensity ratio of different defect induced processes in silicene (left) and germanene (right).}
\label{def_comp}
\end{center}
\end{figure*}

The ZO/TO intensity ratio for all considered defects is presented in Fig. \ref{def_comp}. Although the intensity of the ZO peak on silicene is small, differences between defect scatterers is visible in Fig. \ref{def_comp}. Generally the ZO peak intensity can be correlated to the effect on the $\sigma$ bonds, as larger ZO intensity is achieved for $t_{sp\sigma}$, $t_{pp\sigma}$, and vacancies. Resonances near the plasmon excitation energies are also visible, implying greater enhancement of the ZO intensity compared to the intensity of the TO peak. Similar resonance effect can be seen in the case of germanene at the plasmon energy, although resonance at the $\Gamma$ gap value is not present. Similarly to silicene, the largest relative intensity can be achieved with $t_{pp\sigma}$ defects and vacancies, which perturb mostly the in-plane bonds. Similarly large intensity ratio can be found at the $t_{pp\pi}$ hopping defect, however, the ZO peak intensity does not reach the intensity of the TO peak, whilst the former defects induce larger ZO peak intensity above the plasmon energy. Finally, $t_{ss\sigma}$, $t_{sp\sigma}$ and the on-site defect introduce a ZO peak around one order of magnitude smaller than the ZO peak.

\section{Conclusions}

We used a tight-binding model parametrized from first principles density functional theory to describe resonant Raman scattering in silicene and germanene. We found that spin-orbit coupling has a significant effect on the resonance profile of germanene, whereas spin-orbit is negligible in silicene. We showed that the $\pi-\sigma$ plasmon transition at the M point between the top valence band and the second lowest conduction band introduces an additional resonance in silicene. By analyzing the two phonon spectra we derived a relation between sublattice buckling and the relative intensity ratio of the intensity of out-of-plane modes. We calculated the Raman response of defect induced single phonon scattering for substitutional atoms (on-site), Stone-Wales defects ($t_{sp\sigma}$,$t_{pp\sigma}$), adatoms ($t_{ss\sigma}$,$t_{pp\pi}$) and vacancies. We demonstrated that the relative intensity ratio of out-of-plane and in-plane vibrations can be exploited to identify the presence of these defects from the Raman spectrum.

\section{Methods}
\subsection{Details of DFT calculations}
To compute the electronic band structures and phonon dispersions of silicene and germanene, we rely on first principles density functional theory, using the plane-wave-based VASP code \cite{kresse_textitab_1993,kresse_efficient_1996}. We use the local density approximation (LDA) to relax the structures and compute phonon frequencies, as it is well established that the LDA gives a quantitatively reliable description of these properties in solids. In contrast, we compute the electronic band structures using the HSE06 hybrid functional \cite{heyd_hybrid_2003,heyd_erratum:_2006}, as hybrid functionals yield more accurate electronic band structures than the LDA \cite{muscat_prediction_2001,heyd_energy_2005,deak_accurate_2010,lucero_improved_2012}. Note, that the optimized structural parameters, shown in Table \ref{struct}, are in good agreement between LDA and HSE06, further reinforcing the expectation that LDA is accurate enough to describe the atomic structure in these materials. The structural optimization using  the LDA and HSE06 functional is performed on a $30\times 30\times 1$ and  $18\times 18\times 1$ $\Gamma$-centred Monkhorst-Pack grid \cite{monkhorst_special_1976}, respectively, until all atomic forces decrease below $0.003\,\rm eV/$\AA. The plane-wave cutoff energy is set to $700\,\rm eV$ in all calculations.

Vibrational properties are calculated using the LDA functional using a $6\times 6\times 1$ $\Gamma$-centred Monkhorst-Pack grid. The atomic displacements are set to 0.01$\,$\AA.
\begin{table}
\begin{tabular}{|c|c|c|c|c|}
\hline
&$a_\mathrm{Si}$&$\Delta_\mathrm{Si}$&$a_\mathrm{Ge}$&$\Delta_\mathrm{Ge}$\\
\hline
LDA&3.825$\,$\AA &0.439$\,$\AA &3.968$\,$\AA &0.647$\,$\AA\\
\hline
HSE06&3.841$\,$\AA &0.434$\,$\AA &3.989$\,$\AA &0.645$\,$\AA\\
\hline
\end{tabular}
\caption{Structural parameters of silicene and germanene}
\label{struct}
\end{table}
\subsection{Matrix elements}
\label{matel}
The electron-photon matrix elements are numerically evaluated on the atomic basis set using the interaction Hamiltonian

\begin{equation}
\hat{H}_\mathrm{e-p}=\frac{e\hbar}{im}\sum\limits_{n,m,\mathbf{k},\mathbf{q},\lambda}\frac{\partial}{\partial \mathbf{r}_\lambda}\underbrace{\sqrt{\frac{\hbar}{2\omega\epsilon_0}}e^{-i\mathbf{qr}}\left\lbrace\mathbf{e}_\lambda b_\lambda( -\mathbf{q})+\mathbf{e}^*_\lambda b^\dagger_\lambda( \mathbf{q})\right\rbrace}_{\mathbf{A}(\mathbf{q},\mathbf{r})}a^\dagger_{n,\mathbf{k}-\mathbf{q}}a_{m,\mathbf{k}},
\end{equation}

\noindent where $\mathbf{e}_\lambda$ is the $\lambda$th component of polarization vector $\mathbf{e}$, $b^\dagger_\lambda( \mathbf{q})$ and $b_\lambda( \mathbf{q})$ are the bosonic creation and annihilation operators of a photon with momentum $\mathbf{q}$ and frequency $\omega$ in the dielectric environment described by $\epsilon_0$ dielectric constant and $\mathbf{A}(\mathbf{q},\mathbf{r})$ is the vector potential of the photon. Since the vector potential in the long wavelength limit ($\mathbf{q}\approx 0$) is independent of the coordinates ($\mathbf{A}(\mathbf{q},\mathbf{r})\approx\mathbf{A}(0,0)$), we compute the electron-photon matrix element by calculating the transition matrix elements of the $\nabla$ operator between the atomic orbitals. In usual experimental setups only backscattering photons are measured, therefore we calculate the in-plane components of the matrix element between the atomic orbitals. The integrals are evaluated between atomic sites up to third-nearest neighbours in accordance with our tight-binding model. Similarly to previous theoretical works \cite{gruneis_inhomogeneous_2003,jiang_electron-phonon_2005} we find that the largest matrix elements are between nearest-neighbours; this is due to the overlap between the orbitals decaying exponentially with increasing distance, and numerous on-site transition matrix elements being forbidden by symmetry. The electron-photon matrix element can be expressed as

\begin{equation}
\begin{split}
&M^{e-p}_{Ai}=\langle \psi_{n,\mathbf{k}}|\hat{H}_\mathrm{e-p}|\psi_{m,\mathbf{k}}\rangle=\cr
&=\frac{e\hbar}{im}\sqrt{\frac{\hbar}{2\omega\epsilon_0}}\sum\limits_{i,j,i',j',\lambda}c^*_{n,i',j'}(\mathbf{k}) c_{m,i,j}(\mathbf{k})\left\langle\varphi_{j'}(\mathbf{r}-\mathbf{R}_{i'})\left|\frac{\partial}{\partial \mathbf{r}_\lambda} \right|\varphi_j(\mathbf{r}-\mathbf{R}_i)\right\rangle \times\cr
&\times e^{i\mathbf{k}(\mathbf{R}_i-\mathbf{R}_{i'})}\left\lbrace\mathbf{e}_\lambda b_\lambda( \mathbf{0})+\mathbf{e}^*_\lambda b^\dagger_\lambda( \mathbf{0})\right\rbrace a^\dagger_{i',j'}a_{i,j},
\end{split}
\end{equation}

\noindent which shows that in the approximation of a constant vector potential, the matrix element only allows electron-hole excitations where the two quasiparticles have the same $\mathbf{k}$. Furthermore, the conservation rule of lattice momentum $\mathbf{k}$ during electron-photon interaction means that in order to absorb or emit a photon, the first $A$ and last $C$ virtual state involved in Eqns (\ref{eq_pp}) and (\ref{eq_pd}) should include an electron and a hole with the same $\mathbf{k}$. 

Next we calculate the electron-phonon matrix element $M^{e-ph,\mu}_{CB}$ describing the emission of a phonon from band $\mu$ with momentum $\mathbf{q}$, which we obtain  numerically from the interaction Hamiltonian,

\begin{equation}
\begin{split}
\hat{H}_\mathrm{e-ph,\mu}&= \sum\limits_{l,n,m,\mathbf{k},\mathbf{q}}\sqrt{\frac{\hbar}{2M_l\omega_\mu}}\frac{\partial V_\mathrm{e-ion}(\mathbf{r}-\mathbf{R}_l)}{\partial \mathbf{R}_{l}}\mathbf{Q}_{\mu,l}(\mathbf{q})e^{-i\mathbf{q}\mathbf{R}_l}\times\cr
&\times\lbrace d_\mu(\mathbf{-q})+d^\dagger_\mu(\mathbf{q})\rbrace a^\dagger_{n,\mathbf{k}-\mathbf{q}}a_{m,\mathbf{k}}
\end{split},
\end{equation}

\noindent where $d^\dagger_\mu(\mathbf{-q})$ and $d_\mu(\mathbf{q})$ are the creation and annihilation operators of a phonon with momentum $\mathbf{q}$ on the $\mu$ phonon band, $V_\mathrm{e-ion}$ is the electron-ion potential, $M_l$ is the mass of the nuclei positioned at $\mathbf{R}_l$, $\omega_\mu$ is the frequency of the phonon with $\mathbf{Q}_{\mu,l}(\mathbf{q})$ normal mode. The matrix elements of the derivative of the electron-ion potential on our basis are calculated by \cite{jiang_electron-phonon_2005}. In our tight-binding formalism the matrix element between different electronic bands can be written as

\begin{equation}
\begin{split}
&M^{e-ph,\mu}_{CB}=\langle \psi_{n,\mathbf{k-q}}|\hat{H}_\mathrm{e-ph,\mu}|\psi_{m,\mathbf{k}}\rangle=\cr
&=\sum\limits_{i,i',j,j',l}\sqrt{\frac{\hbar}{2M_l\omega_\mu}}c^*_{n,i',j'}(\mathbf{k-q}) c_{m,i,j}(\mathbf{k})\mathbf{Q}_{\mu,l}(\mathbf{q})\times\cr
&\times\left\langle\varphi_{j'}(\mathbf{r}-\mathbf{R}_{i'})\left|\frac{\partial V_\mathrm{e-ion}(\mathbf{r}-\mathbf{R}_l)}{\partial \mathbf{R}_{l}}\right|\varphi_j(\mathbf{r}-\mathbf{R}_i)\right\rangle e^{i\mathbf{k}\mathbf{R}_i}e^{-i(\mathbf{k-q})\mathbf{R}_{i'}}e^{-i\mathbf{q}\mathbf{R}_l}\times\cr
&\times\lbrace d_\mu(\mathbf{q})+d^\dagger_\mu(\mathbf{-q})\rbrace a^\dagger_{i',j'}a_{i,j}.
\end{split}
\label{ep-coupling}
\end{equation}


\begin{acknowledgement}
Support from the Hungarian National Research, Development and Innovation Office (NKFIH, Grant No. K-108676 and K-115608) is acknowledged. We acknowledge [NIIF] for awarding us access to resource based in Hungary at Debrecen. This research was supported by the National Research Development and Innovation Office of Hungary within the Quantum Technology National Excellence Program  (Project No. 2017-1.2.1-NKP-2017-00001). This work was completed in the ELTE Excellence Program (783-3/2018/FEKUTSRAT) supported by the Hungarian Ministry of Human Capacities. G. K. acknowledges support from the New National Excellence Program (UNKP) of the Ministry of Human Capacities in Hungary. V. Z. acknowledges support from the Graphene Flagship Project, the N8 Polaris service, the ARCHER National UK Supercomputer (RAP Project e547), and the Computational Shared Facility at the University of Manchester.  J. K. acknowledges the Bolyai and the Bolyai+ program of the Hungarian Academy of Sciences.

\end{acknowledgement}

\providecommand{\latin}[1]{#1}
\makeatletter
\providecommand{\doi}
  {\begingroup\let\do\@makeother\dospecials
  \catcode`\{=1 \catcode`\}=2 \doi@aux}
\providecommand{\doi@aux}[1]{\endgroup\texttt{#1}}
\makeatother
\providecommand*\mcitethebibliography{\thebibliography}
\csname @ifundefined\endcsname{endmcitethebibliography}
  {\let\endmcitethebibliography\endthebibliography}{}

\end{document}